\documentclass[fleqn,usenatbib]{mnras}

\usepackage{newtxtext,newtxmath}

\usepackage[T1]{fontenc}

\DeclareRobustCommand{\VAN}[3]{#2}
\let\VANthebibliography\thebibliography
\def\thebibliography{\DeclareRobustCommand{\VAN}[3]{##3}\VANthebibliography}

\usepackage{graphicx}	
\usepackage{amsmath}	

\title[Impact of \texttt{CLEAN}ing on strong lens modelling]{The Impact of \texttt{CLEAN}ing on Strong Gravitational Lens Modelling}
\author[J. Maresca et al.]{
Jacob Maresca$^{1}$\thanks{E-mail: jacob.maresca@nottingham.ac.uk}
\& Simon Dye$^{1}$
\\
$^{1}$School of Physics \& Astronomy, University of Nottingham, University Park, Nottingham, NG7 2RD, UK\\
}

\date{Accepted XXX. Received YYY; in original form ZZZ}

\pubyear{2015}

\begin{document}
\label{firstpage}
\pagerange{\pageref{firstpage}--\pageref{lastpage}}
\maketitle

\begin{abstract}
We present a comparison of image and uv-plane galaxy-galaxy strong lensing modelling results for simulated ALMA observations with different antenna configurations and on-source integration times. Image-plane modelling is carried out via use of the \texttt{CLEAN} algorithm, and we explore the effects of different visibility weighting schemes on the inferred lens model parameters. We find that direct modelling of the visibility data consistently outperforms image-plane modelling for both the naturally and Briggs-weighted images. We also find that the modelling of images created with Briggs weighting generally produces more accurate results than those obtained by modelling images constructed with natural weighting. We explain this by quantifying the suppression of information due to \texttt{CLEAN}ing on scales at which the modelling is sensitive, and how this differs between Briggs and natural weighting. At higher resolutions, the differences between the lens modelling techniques are much less pronounced and overall, modelling errors are significantly reduced. We also find that time-binning the visibilities by up to a factor of three makes no significant difference to the inferred lens parameters when directly modelling in the uv-plane. This work provides some guidance on navigating the many choices faced when modelling strong lens interferometric imaging data.
\end{abstract}




\section{Introduction}

Strong gravitational lensing provides a unique probe that enables a  wide array of astrophysical and cosmological measurements. The phenomenon can be exploited to not only determine the distribution of mass in the lens and give an enhanced view of lensed sources, but also constrain key cosmological parameters.

Due to conservation of surface brightness and commonly-observed large image-magnifications, strong lensing can be used to explore high redshift sources in unprecedented detail. The advantages of using strong lenses on both cluster scales \citep[e.g.,][]{bouwens2022,sun2022,salmon2020} and galaxy scales \citep[e.g.,][]{etherington2022,maresca,berta2021,dye2018} to understand the nature of lensed source galaxies has been realised and used for some time. Detailed source reconstruction allows study of both the morphology and physical conditions in high redshift galaxies on scales of $\sim 100$ pc \citep[e.g.,][]{swinbank2015,rybak_2015}, and, through observation of spectral lines, investigation of kinematical properties \citep[e.g.,][]{id141, geach2018, dye_2015}.

The occurrence of multiple images due to different sight lines through the lens' gravitational potential provides direct constraints on the lens mass distribution. This on its own and also in combination with spectroscopic measurements of the lens has dramatically improved our understanding of how elliptical galaxy mass density profiles have evolved over cosmic time \citep[e.g.,][]{etherington2023,shajib2021,dye2014,sonnenfeld2013}. These measurements help constrain the physical mechanisms that regulate galaxy density profiles, such as merger histories \citep[e.g.,][]{tan2023,wang2019,remus2017} and feedback mechanisms \citep[e.g.,][]{mukherjee2021}. Strong lensing is also sensitive to the perturbative effects of small-scale mass fluctuations, and therefore can be used to learn about dark matter substructure in galaxies \citep[e.g.,][]{ballard2023,he2023,minor2021} and even provide constraints on 
the nature of dark matter itself \citep[e.g.,][]{he2022,vegetti2018,birrer2017}.

In terms of cosmological constraints, delays between arrival times of source transient events observed in multiple images provide measurement of the Hubble constant \citep[e.g.,][]{birrer2022,treu2022,li2021,wong2020} in a manner that is competitive and complementary to methods involving use of supernovae and the cosmic microwave background. In cases where there are multiple sources at different redshifts, the ratio(s) of their image separations can provide measurement of the equation of state and matter density of the Universe \citep[e.g.][]{collett2014,collett2020}. Furthermore, strong lenses can be used to test the theory of general relativity \citep[e.g.,][]{melo2023,yang2020,collett2018}.

The typically lower systematics associated with the simpler mass distributions of galaxy-scale lenses compared to groups and clusters, especially in terms of their use in detecting substructure and in cosmology, has seen a concentration of effort on improving observations of galaxy-galaxy lenses. In particular, increases in image resolution through use of interferometry has brought significant recent advances. The Atacama Large Millimeter Array (ALMA) has proven itself invaluable in this plight, not least because of its sensitivity at sub-millimetre wavelengths, where an increasing fraction of lensed sources at higher redshifts emit their radiation. 

Observed interferometric data in its purest form comprises a set of visibilities, which are measurements of the two-dimensional Fourier transform of the surface brightness on the sky. The visibilities are evaluated at distinct points in Fourier space, on the so-called uv-plane, determined by the location of the interferometer receivers on the ground, the position of the source in the sky and the wavelength being observed. Generally, these samplings are not distributed uniformly across the uv-plane.

Obtaining a lens model that best fits the observed data presents a range of choices. Chief amongst these is whether to directly model the visibility data, or to employ a deconvolution algorithm such as \texttt{CLEAN} \citep{hogbom1974}, to form an image that can subsequently be modelled. Modelling the visibility data has the benefit of working with the purest form of the data, but this comes at the usually severe cost of requiring significantly more time and computational resources \citep[although short-cuts have been proposed; see][]{powell2021}. For this reason, it is attractive to work with a derived image-plane representation of the data which alleviates the modelling burden, but introduces artefacts caused by the discrete sampling of the uv-plane in the form of correlated noise. This is further complicated by an additional layer of subjectivity introduced by choices afforded in the production of these images, such as how to weight the visibilities during averaging in the \texttt{CLEAN}ing process.

The question of whether lens modelling should use visibility data or convert to the image plane is one faced by any user of interferometric data. Historically, studies have argued that if the visibility plane is densely sampled by observations, then modelling in the image plane should provide reliable results. \cite{dye2018} and \cite{maresca} in combination analysed a sample of 13 ALMA observations of strong lenses and compared the results obtained by modelling in both the image plane and the uv-plane. No significant differences were found, either in the lens models or the reconstructed sources, although the ALMA observations in these studies had relatively good coverage of the uv-plane.

This paper aims to provide a more quantitative assessment of the differences between modelling in the image plane and modelling the visibilities directly. Some of the questions that we seek to
answer are: Does uv-plane modelling provide a more accurate lens
model than its image plane counterparts? To what extent can visibilities be time-averaged to lower computation cost without affecting the resulting lens model? Does the choice of visibility-weighting scheme lead to the inference of a significantly different lens model? The ultimate goal is to provide users of interferometric lensing data with guidance on when image plane modelling is a viable shortcut to carrying out full visibility modelling. To achieve this, we have generated a set of synthetic ALMA visibilities of simulated lensed images and compared uv-plane and image-plane modelling.

The layout of this paper is as follows: Section \ref{sec:simulations} details the processes used to produce all the simulated data products used in this paper. Section \ref{sec:methods} outlines the lens modelling methodology we have used. The results of our lens modelling are presented in Section \ref{sec:results_p4}, and their interpretation is described in Section \ref{sec:discussion_p4}. Finally, Section \ref{sec:conclusions} presents a brief summary and our main conclusions. Throughout this paper, we assume a flat $\Lambda\mathrm{CDM}$ cosmology using the 2015 Planck results \citep{planck_2015}, with Hubble {constant $H_0 = 67.7 \: \mathrm{km s^{-1} Mpc^{-1}}$} and matter density parameter $\Omega_m = 0.307$.

\section{Simulated Data}
\label{sec:simulations}

\begin{figure}
	\centering
    \includegraphics[width=0.7\columnwidth]{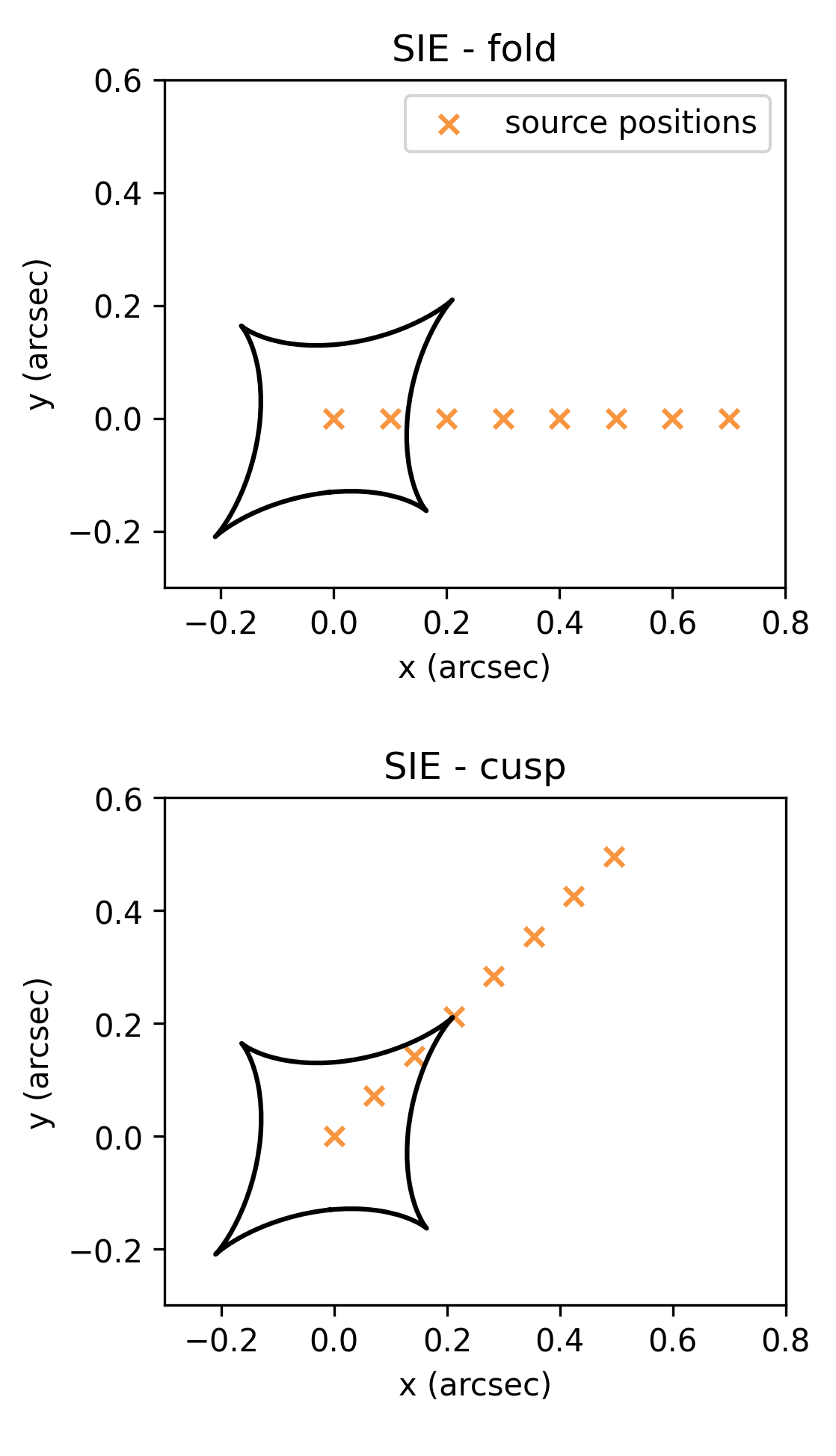}
    \vspace{-3mm}
    \caption{The source positions relative to the caustics of the two SIE datasets used in this work. A lensed image was created for each of the source positions (see Figure \ref{fig:example_sims}). The top panel shows the source stepping away from the optical axis through the caustic fold, whilst the bottom panel shows the source stepping through the caustic cusp.}
    \label{fig:caustic_diagram}
\end{figure}

This work relies on a variety of simulated data products to compare modelling results between them. Interferometer datasets, and images derived from interferometer measurements, are used. In this section, we describe the process used to create the simulated data products modelled in this paper.

Our simulated data are generated using \texttt{PyAutoLens} \citep{pyautolens} with one of three different strong lens model configurations. These comprise a singular isothermal sphere (SIS) model and two singular isothermal ellipse (SIE) models. For each model, we created eight images by positioning a background circularly symmetric Gaussian source of full-width at half-maximum of $0.24 \arcsec$ at one of eight different locations separated by $0.1 \arcsec$ along a line originating from the optical axis. In the case of the two SIE models, one line crosses a fold caustic while the other crosses a cusp caustic as shown in Figure \ref{fig:caustic_diagram}. The parameters of the lens models are given in Table \ref{tab:lens_params}. These simulated images are then used as the sky brightness distribution for our simulated interferometric observations.

\begin{table}
	\centering
	\caption{A summary of the two lens models used to create the simulated images in this work. The columns are the Einstein radius $\theta_{\rm E}$, the lens centroid $(x_{\rm c}, y_{\rm c})$, corresponding to west and north respectively, the orientation of the lens $\phi$, measured counter-clockwise from east to the lens semi-major axis, and the lens profile axis ratio.}
	\label{tab:lens_params}
	\begin{tabular}{lcccc} 
		\hline
		Lens model & $\theta_{\rm E}$ & $(x_{\rm c}, y_{\rm c})$ & $\phi$ & $q$ \\
		 & {(arcsec)} & (arcsec) & {(deg)} &  \\
		\hline
		SIS & 1 & (0, 0) & N/A & 1 \\
		SIE & 1 & (0, 0) & 45 & 0.67 \\
		\hline
	\end{tabular}
\end{table}

\begin{figure*}
	\includegraphics[width=0.98\textwidth]{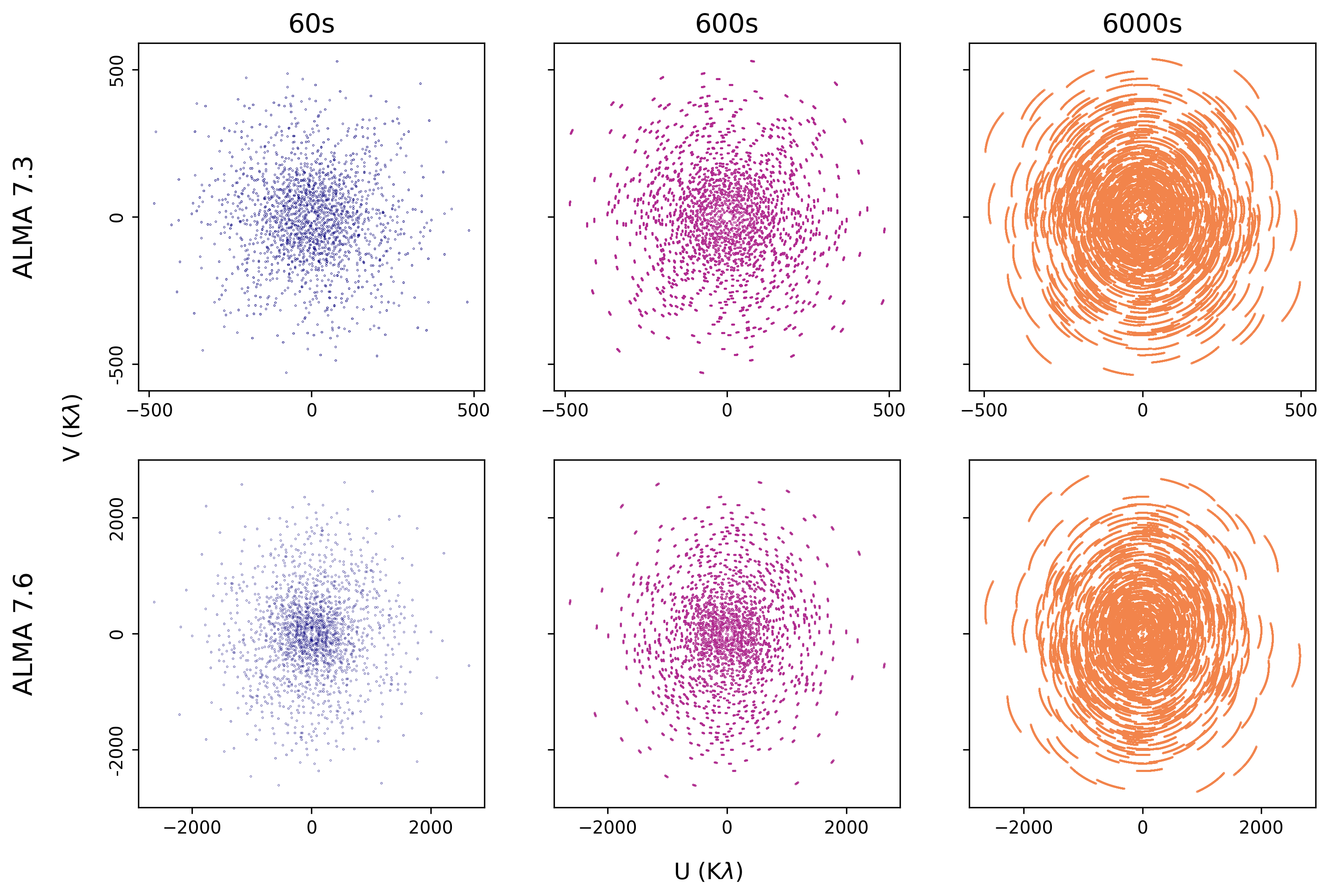}
    \caption{The uv-coverage of the different ALMA configurations and observation times used in the creation of the simulated datasets in this work. The first row shows the cycle 7.3 configuration for 60s, 600s and 6000s on-source integration times, whilst the second row shows the cycle 7.6 configuration for the same observation times.}
    \label{fig:uv_coverage}
\end{figure*}

We used the \textsc{Common Astronomy Software Application} V6.3.0.48 (\texttt{CASA}) and the \texttt{simobserve} task to produce mock ALMA measurement sets. We applied the ALMA cycle 7.3 array configuration giving a resolution of $0.41 \arcsec$ and the cycle 7.6 configuration giving a resolution of $0.089 \arcsec$. For each of these, three different on-sky exposure times (60s, 600s, 6000s) were used (see Table \ref{tab:alma_setup} for a summary of the setup). The coverage of the uv-plane for each of these observing setups can be seen in Figure \ref{fig:uv_coverage}. We assumed a constant spectral energy distribution for our lensed sources. The spectral setup in our simulated ALMA observations was identical for each system. We simulated band 7 continuum observations, comprised of four spectral windows, each with a width of 2000 MHz and centred on the frequencies 337.5 GHz, 339.5 GHz, 347.5 GHz and 349.5 GHz. The central frequency of 343.404 GHz corresponds to a wavelength of 873 $\mu$m. Each spectral window consists of 128 frequency channels, resulting in a spectral resolution of 15.6 MHz. The noise model for these simulated ALMA observations is generated by the \texttt{simobserve} task, using pre-defined values that correspond to good observing conditions (i.e. precipitable water vapour of 0.5 mm and an ambient temperature of 269 K). The three on-source integration times of 60s, 600s, and 6000s lead to a total number of 21672, 216720 and 2167200 visibilities respectively. In order to investigate the effect of reducing the number of visibilities and thus enhance the computational efficiency by time-averaging, we averaged the visibilities over three different sized time bins of 10s, 20s and 30s. Since the simulated observations use 10s exposures anyway, no averaging is performed in the 10s bin and this therefore allows us to compare the 20s and 30s bins against modelling with all the available visibilities. Time-binning reduces the number of visibilities by a factor of two for the 20s binning and three for the 30s binning.

As well as modelling the visibility data directly, we performed image-plane modelling for each system by producing images via the \texttt{tclean} algorithm. When creating images with \texttt{tclean}, one of the choices that must be made is how to weight the visibility data. A common choice is natural weighting, where visibilities are weighted according to the inverse of the noise variance of that visibility, determined during calibration, which provides good signal-to-noise and sensitivity, but relatively poor resolution. Another option is that of uniform weighting, where the weights are adjusted based on the density of points in the uv-plane, giving greater weighting to sparse regions. This achieves better resolution, but at the cost of typically poor signal-to-noise. A third option, Briggs weighting, allows for a smoothly varying compromise between the two aforementioned methods by modifying the value of the 'robustness' parameter. This parameter varies between the values of -2, which is approximately equivalent to a uniform weighting and +2, which closely resembles the natural weighting scheme \citep{briggs}. From each of our simulated measurement sets, we produced two images, one using natural weighting and another using Briggs weighting with the robustness parameter set to zero, which provides a good trade-off between resolution and sensitivity. We used an image pixel scale of $0.1 \arcsec$ pixel$^{-1}$ for the ALMA cycle 7.3 configuration simulations and a pixel scale of $0.02 \arcsec$ pixel$^{-1}$ for the ALMA cycle 7.6 configuration, providing good sampling of the minor axis of the primary beam in each case. A threshold value of twice the background noise was used as the stopping criterion for the \texttt{tclean} algorithm in all cases.

\begin{figure*}
	\includegraphics[width=0.98\textwidth]{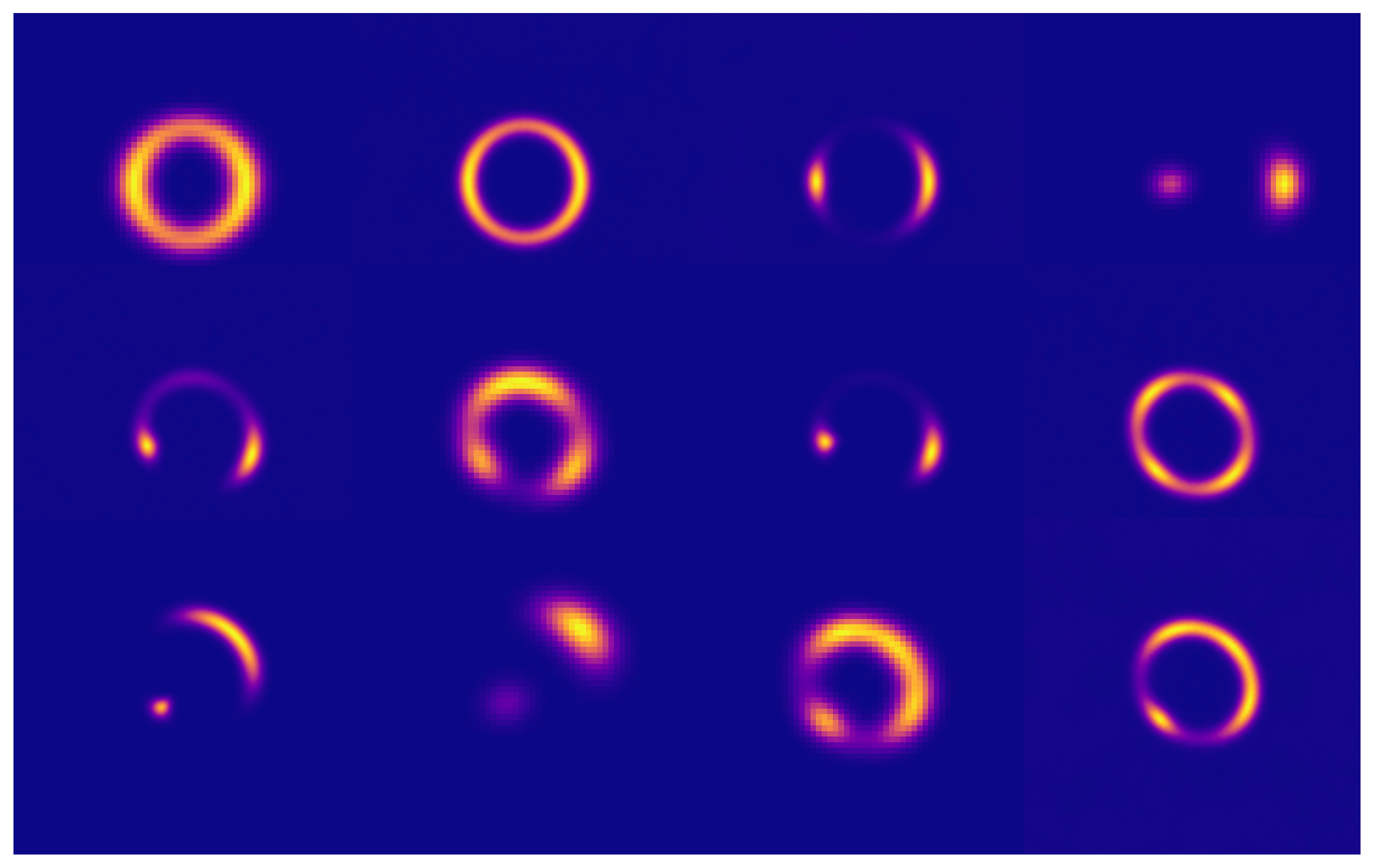}
    \caption{A selection of the images produced via the \texttt{tclean} algorithm, with a variety of on-source integration times, weighting schemes and ALMA configurations (cycle 7.3 with $0.41 \arcsec$ resolution + $0.1 \arcsec$ pixel$^{-1}$ image scale and cycle 7.6 with $0.089 \arcsec$ resolution + $0.02 \arcsec$ pixel$^{-1}$ image scale). The first row contains sources lensed by the SIS model. The second row contains sources lensed by the SIE model, with the source positions in the fold configuration. The third row contains sources lensed by the SIE model, with the source positions in the cusp configuration. All models use an Einstein radius of $1 \arcsec$ and we have displayed example lensed images across the range of source positions shown in Figure \ref{fig:caustic_diagram}.}
    \label{fig:example_sims}
\end{figure*}

\begin{table*}
	\centering
	\caption{A summary of the targeted position of each source, the number of antennae used in each observation, the continuum sensitivity (for the 60s, 600s and 6000s on-source integration times), the shortest \& longest baseline in the ALMA configuration, the angular resolution of the setup and the maximum recoverable scale.}
	\label{tab:alma_setup}
	\begin{tabular}{lccccccc} 
		\hline
		ALMA config & Position & $N_{\rm ant}$ & Sensitivity & Shortest baseline & Longest baseline & Angular resolution & Maximum recoverable scale \\
		 & {(RA DEC)} & & ($\mu$Jy/beam) & {(m)} & {(m)} & {(arcsec)} & (arcsec)\\
		\hline
		Cycle 7.3 & 19:00:00 -40:00:00 & 43 & (220, 70, 22) & 15 & 500 & 0.41  & 4.7 \\
		Cycle 7.6 & 19:00:00 -40:00:00 & 43 & (220, 70, 22) & 15 & 2500 & 0.089 & 1.2 \\
		\hline
	\end{tabular}
\end{table*}

\section{Methodology}
\label{sec:methods} 

All lens modelling in this work was carried out using \texttt{PyAutoLens} \citep{pyautolens} which uses source pixelisation based on the semi-linear inversion method of \cite{warren_dye} and includes extensions that adapt to the magnification and brightness variation across the source plane \citep{Nightingale2018}. The standard image-plane method was employed for modelling the data produced via the \texttt{tclean} algorithm, whilst modelling of the visibility data was carried out using a modified version of this technique as implemented within PyAutoLens.

\subsection{Image-plane modelling}
\label{sec:image_plane_modelling}

For modelling the image-plane data, we estimated the image noise by measuring the standard deviation of a $25 \times 25$ pixel region of each image, far away from  lensed source emission. Due to the incomplete sampling of the uv-plane, and the covariant noise that this introduces into the images, we opted to use a $1\sigma$ value of twice the measured standard deviation in our images. This is an approximation but gives values of reduced $\chi^2$ close to unity at the optimal solution.

The first stage of our modelling process was to reconstruct the source on a Voronoi grid, adapted to the magnification of the source plane. We assumed that all our observations were properly centred on the lens, and we used a Gaussian prior on the $(x_{\rm c}, y_{\rm c})$ coordinates of the lens centroid with a mean of $0\arcsec$ and standard deviation of $\sigma = 0.2\arcsec$ for both $x_{\rm c}$ and $y_{\rm c}$. A uniform prior of width $0.5 \arcsec$, centred on the true value, was used for the Einstein radius. Uniform priors covering the whole range of parameter space were used for the ellipticity components. We adopted a gradient regularisation scheme with a constant weight for the source plane. This magnification-based fit was then used to initialise a new fit, using a Voronoi grid, this time adapted to the reconstructed source light \citep{Nightingale2018}. The nested sampling algorithm \texttt{dynesty} \citep{dynesty} was used to maximise the Bayesian evidence \citep{suyu2006bayesian} and find the optimal lens model parameters. 

\subsection{The semi-linear inversion method in the uv-plane}

The image-plane implementation of the semi-linear inversion method revolves around the use of a pixelised source plane. For a realisation of lens model parameters, the image of each source pixel is formed, and the best fitting linear superposition of these images determines the source surface brightness. Similarly, when working with interferometer data, for each source pixel we produce a set of model visibilities. The best fitting linear combination of these model visibilities determines the source surface brightness for a given set of lens model parameters.

We used the operator-based source inversion method of \cite{powell} implemented within \texttt{PyAutoLens} \citep{Nightingale2021}. Interferometer datasets $\boldsymbol d$, are comprised of samples of complex visibilities. The source surface brightness vector, $\bf{s}$, contains the surface brightness of each source plane pixel. The mapping of the source light to the image-plane is described by the lensing operator $\bf L(\boldsymbol{\eta})$, where $\boldsymbol \eta$ is the parameterised projected surface mass density of the lens model. The observed sky brightness distribution can therefore be written as $\bf{L}(\boldsymbol \eta)\bf{s}$. The interferometer response is encoded into the operator $\bf D$, which Fourier transforms the pixelised source surface brightness to create a set of complex visibilities. The combination of these effects produces the observed data $\boldsymbol d$:
\begin{equation}
    \boldsymbol d = \bf D \bf{L}(\boldsymbol \eta)\bf{s} + \bf{n}.
    \label{eq:vis}
\end{equation}

The noise covariance can be represented by the diagonal matrix $\bf{C}^{-1}$, assuming uncorrelated Gaussian noise, $\bf n$, in the observed visibility data. The uncertainties on visibilities in ALMA measurement sets are often set to arbitrary values, or derived from antenna and receiver properties that are not well characterised. Accurate uncertainties are required to perform visibility data lens modelling, and so we used the \texttt{CASA} task \texttt{statwt} to empirically measure the visibility scatter over all baselines to determine the 1$\sigma$ uncertainties, as in \cite{maresca}.

Combining equation (\ref{eq:vis}) with the set of model visibilities $\bf D \bf L(\boldsymbol \eta) \bf{s}$ allows us to write the $\chi^2$ statistic as 
\begin{equation}
    \chi^2 = (\bf D \bf L \bf{s} - d)^T \bf{C}^{-1} (\bf D \bf L \bf{s} - d).
\end{equation}
\cite{powell} shows that we can write the regularised least-squares equation, with the addition of a prior on the source, $\bf R$, weighted by a regularisation strength, $\lambda_{\rm s}$, as
\begin{equation}
    \left[(\bf{D} \bf{L})^T \bf{C}^{-1}\bf{D}\bf{L} + \lambda_{s} \bf{R}^T \bf{R}\right]\boldsymbol{s}_{\mathrm{MP}} = (\bf{D} \bf{L})^T\bf{C}^{-1}\boldsymbol{d}.
    \label{eq:maps}
\end{equation}
In equation (\ref{eq:maps}), the quantity in square brackets is the solution matrix for $\boldsymbol{s}_{\mathrm{MP}}$, or the maximum a posteriori source inversion matrix. In principle, this linear system of equations is straightforward to solve, however, in practice this is an extremely memory intensive operation for large numbers of visibilities (as is the case in this work). For this reason, in place of a direct Fourier transform, a non-uniform fast Fourier transform constructed from a series of operators is used. This results in a modified version of equation (\ref{eq:maps}), with a series of operators evaluated by an iterative linear solver to find the solution for $\boldsymbol{s}_{\mathrm{MP}}$ \cite[see][for more details on this methodology and the specific implementation used here]{powell, Nightingale2021}.

Optimisation of the lens model parameters was performed using \texttt{PyAutoLens} and the nested sampling algorithm \texttt{dynesty} \citep{dynesty} to maximise the Bayesian evidence as derived within \cite{suyu2006bayesian}. A gradient regularisation scheme with a constant weight, analogous to that described in \cite{warren_dye}, was applied to the source plane. This approach lowers the number of free parameters, which is of great importance when dealing with large visibility datasets such as those treated in this work. 

The first step in our modelling procedure was to reconstruct the background source using a magnification-based pixelisation. This fit was then used to initialise a new search of parameter space, using a source plane that adapted to the reconstructed source brightness distribution \citep{Nightingale2015, Nightingale2018}.

\subsection{Lens Model}
\label{sec:lens_model}

Although the simulated data were generated with SIS and SIE lens models, for fitting the data, we used the elliptical power-law density profile. This allows us to investigate our ability to recover the extra degree of freedom caused by the slope of the density profile. The surface mass density, $\kappa$, of this profile is given by:
\begin{equation}
\centering
    \kappa(R) = \frac{3 - \alpha}{1 + q} \left(\frac{\theta_{\rm E}}{\xi}\right)^{\alpha - 1},
	\label{eq:powerlawp4}
\end{equation}
where $\theta_{\rm E}$ is the Einstein radius in arcseconds, $\alpha$ is the power-law index and $\xi$ is the elliptical radius defined by $\xi = \sqrt{x^2 + y^2 / q^2}$, where $q$ is the axis ratio \citep[e.g.][]{suyu_lens_model}. The orientation of the lens, $\phi$, is measured as the angle counter-clockwise from east to the semi-major axis of the elliptical lens profile. The centre of the lens profile is given by the image-plane coordinates $(x_{\rm c}, y_{\rm c})$. This results in six lens model parameters.

\section{Results}
\label{sec:results_p4}

In this section we present our key results, namely the effect of time-binning on inferred lens model parameters (Section \ref{sec:time_bin}) and a comparison between modelling the \texttt{CLEAN}ed image plane data and the visibility data directly (Section \ref{sec:modelling}). For the latter, we also assess the impact of total integration time of the observations and for both, we investigate the effects of resolution via the ALMA cycle 7.3 and cycle 7.6 simulated array configurations.

\subsection{Time-binning visibilities}
\label{sec:time_bin}

Our first investigation concerns the extent to which uv-plane lens modelling results are impacted by the choice of time-binning the visibility data. Time-averaging of visibilities is a means of reducing the overall data volume and thus reducing the computational cost of performing lens modelling in the uv-plane. Since the process of visibility modelling is more computationally demanding than its image plane counterpart, it is useful to know if attempts to make the task more tractable have adverse effects on modelling results. Here, we focus on time-averaging our shortest observations (60s on-source integration time) which, having the fewest visibilities, will produce the greatest impact.

\begin{figure*}
	\includegraphics[width=0.98\textwidth]{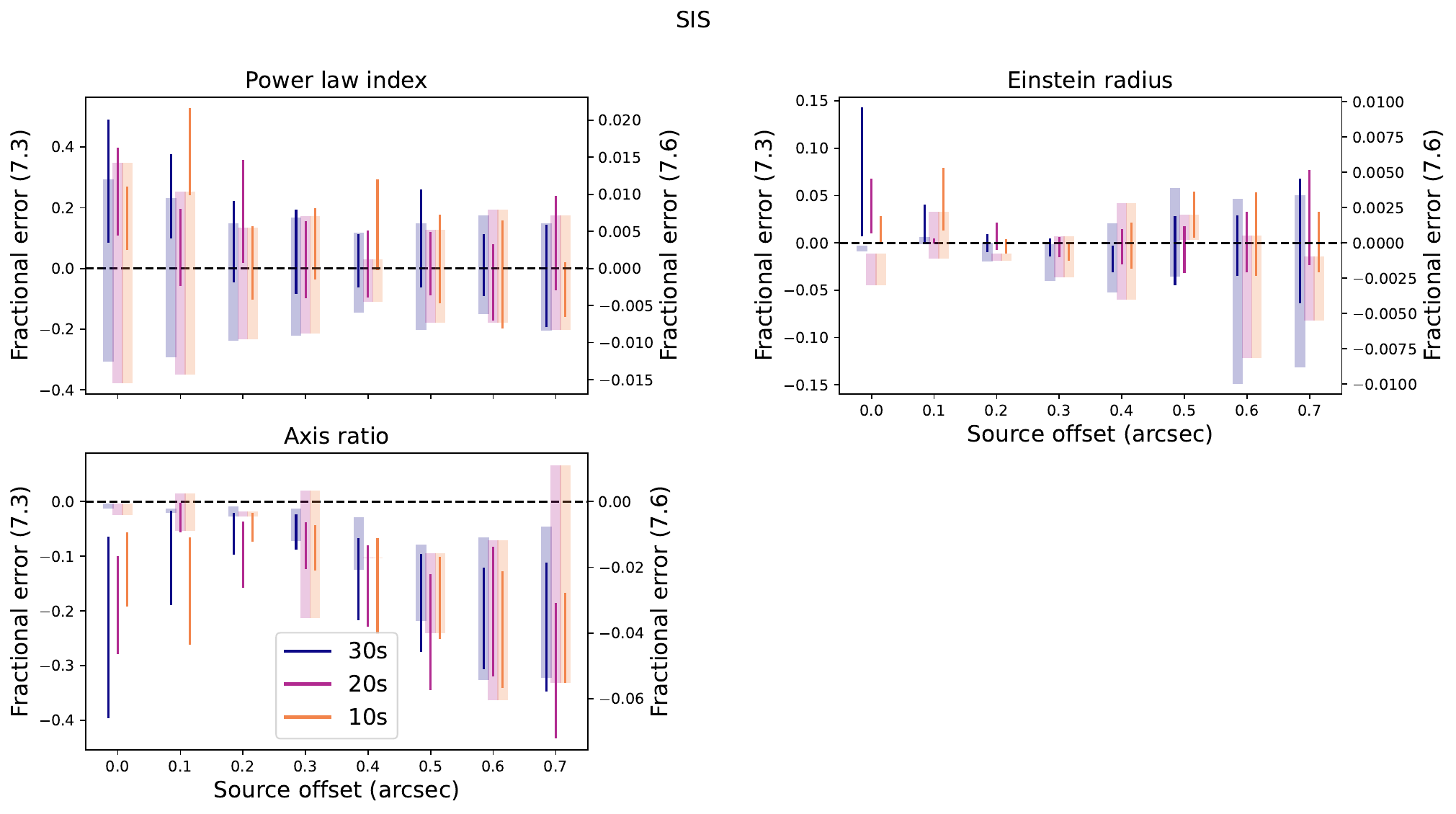}
    \caption{A comparison of uv-plane lens modelling results for three different time-binning values of visibilities. In each case, we plot the results from time-binning at 10s, 20s and 30s (shown as different colours) with the cycle 7.3 results indicated by the thin strong-coloured lines (left-hand $y$-axis) and the 7.6 results by the thick pastel-coloured lines (right-hand $y$-axis). The three panels show the 1-$\sigma$ spread in the fractional error returned by \texttt{pyAutoLens} (see Section \ref{sec:time_bin} for details) for the lens parameters of the SIS model, as a function of source offset from the lens centroid.}
    \label{fig:SIS_time_bins}
\end{figure*}

\begin{figure*}
	\includegraphics[width=0.98\textwidth]{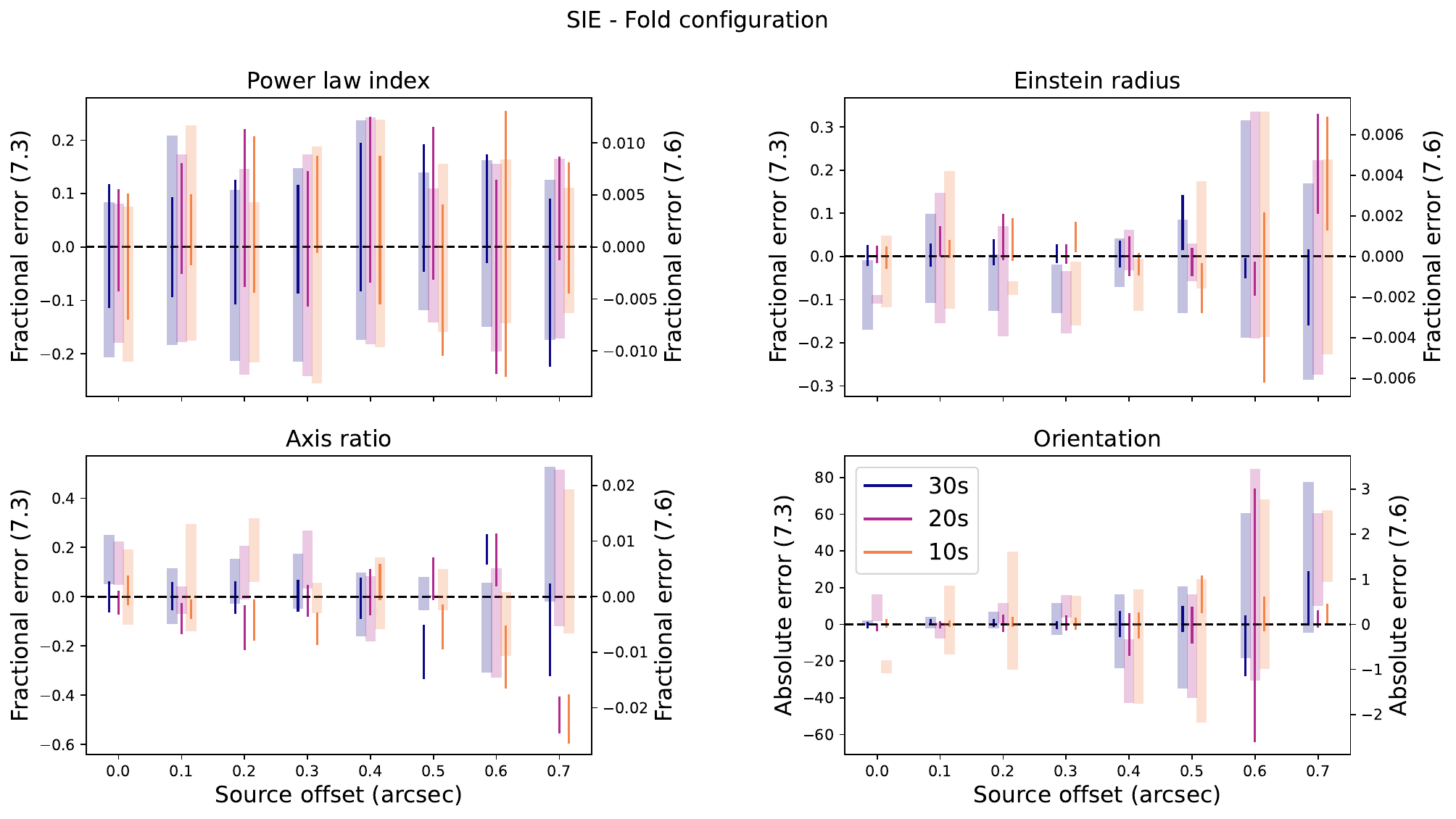}
    \caption{A comparison of uv-plane lens modelling results for three different time-binning values of visibilities. In each case, we plot the results from time-binning at 10s, 20s and 30s (shown as different colours) with the cycle 7.3 results indicated by the thin strong-coloured lines (left-hand $y$-axis) and the 7.6 results by the thick pastel-coloured lines (right-hand $y$-axis). The top-left, top-right and bottom-left panels show the 1-$\sigma$ spread in the fractional error in $\alpha$, $\theta_{\rm E}$ and $q$, and the bottom-right panel shows the 1-$\sigma$ spread in absolute error in orientation (in degrees) returned by \texttt{pyAutoLens} (see Section \ref{sec:time_bin} for details) for the SIE model (source moving through the caustic fold), as a function of source offset from the lens centroid.}
    \label{fig:SIE_fold_time_bins}
\end{figure*}

\begin{figure*}
	\includegraphics[width=0.98\textwidth]{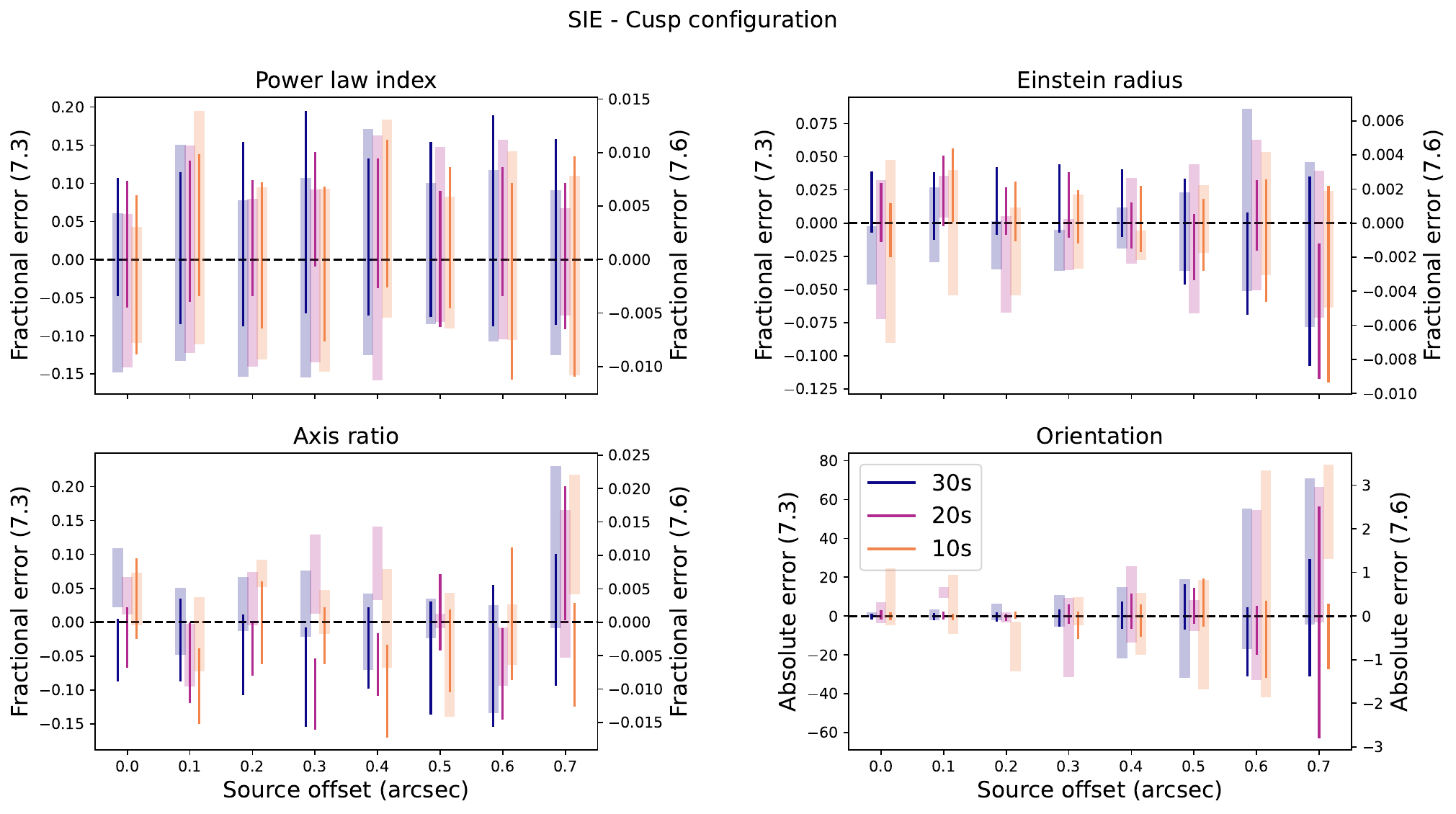}
    \caption{A comparison of uv-plane lens modelling results for three different time-binning values of visibilities. In each case, we plot the results from time-binning at 10s, 20s and 30s (shown as different colours) with the cycle 7.3 results indicated by the thin strong-coloured lines (left-hand $y$-axis) and the 7.6 results by the thick pastel-coloured lines (right-hand $y$-axis). The top-left, top-right and bottom-left panels show the 1-$\sigma$ spread in the fractional error in $\alpha$, $\theta_{\rm E}$ and $q$, and the bottom-right panel shows the 1-$\sigma$ spread in absolute error in orientation (in degrees) returned by \texttt{pyAutoLens} (see Section \ref{sec:time_bin} for details) for the SIE model (source moving through the cusp fold), as a function of source offset from the lens centroid.}
    \label{fig:SIE_cusp_time_bins}
\end{figure*}

As described in Section \ref{sec:simulations}, for each lensing configuration (the SIS and both SIE configurations, all three each having eight source positions), we produced visibility datasets with time-binning values of 10s, which retains the original number of visibilities, and 20s \& 30s which reduce the number of visibilities by a factor of two and three respectively. Time-binning is done for both the lower resolution ALMA cycle 7.3 and higher resolution 7.6 array configurations. We used \texttt{PyAutoLens} to fit the elliptical power-law lens (see Section \ref{sec:lens_model}) to each of these visibility datasets, comparing the recovered lens model parameters with the true values. The results are presented in Figures \ref{fig:SIS_time_bins}, \ref{fig:SIE_fold_time_bins}, and \ref{fig:SIE_cusp_time_bins} which show how the fractional error on each lens model parameter (i.e., (inferred parameter - true parameter)/(true parameter)) varies with source position offset. In the figures, data points are shown as vertical lines; their length and midpoint corresponds respectively to the standard deviation and peak of the posterior distribution for each parameter returned by \texttt{PyAutoLens}.

Firstly, the results show that the higher resolution data of the ALMA cycle 7.6 array configuration give nearly an order of magnitude reduction in fractional error on the lens model parameters compared to the cycle 7.3 data. Secondly, we find that for all three simulated scenarios regardless of the data resolution, the value of the time-binning plays little to no role in the overall performance of the uv-plane modelling. Except for a handful of random outliers, the results between the different time-bin values are consistent with one another and there is no significant evidence that applying shorter time-binning or indeed not using it at all (as shown by the 10s results) is worth the increased computational cost. In particular, we see no improvement to the axis ratio bias exhibited by the SIS model at larger source offsets seen in Figure \ref{fig:SIS_time_bins} by adopting shorter time bins. Likewise, the tendency to overestimate the value of the index of the power-law mass model at small source offsets in the cycle 7.3 data for the SIS model is not improved by employing shorter time-averaging. 

In light of these findings, to streamline our analysis, we have adopted 30s time-averaging of visibilities throughout the work presented hereafter.

\subsection{Image plane versus visibilities modelling}
\label{sec:modelling}

In this section, we compare the results of uv-plane lens modelling with  modelling of the \texttt{CLEAN}ed image plane data. In some instances, it is clear that our modelling has not converged upon the global solution, but rather upon an unphysical, demagnified solution \citep[see][for more discussion]{maresca2020auto}. Where this is the case (confirmed by visual inspection of the source reconstruction), we have opted to mask the data point; whilst we could have forced smaller priors on the lens model parameters for these problematic cases, this would have introduced an unfair reduction in their biases which we ultimately wish to measure, the priors being centred on the known parameter values (see Section \ref{sec:image_plane_modelling}).

\begin{figure*}
\centering
  \begin{tabular}{cc}
    \includegraphics[width=\columnwidth]{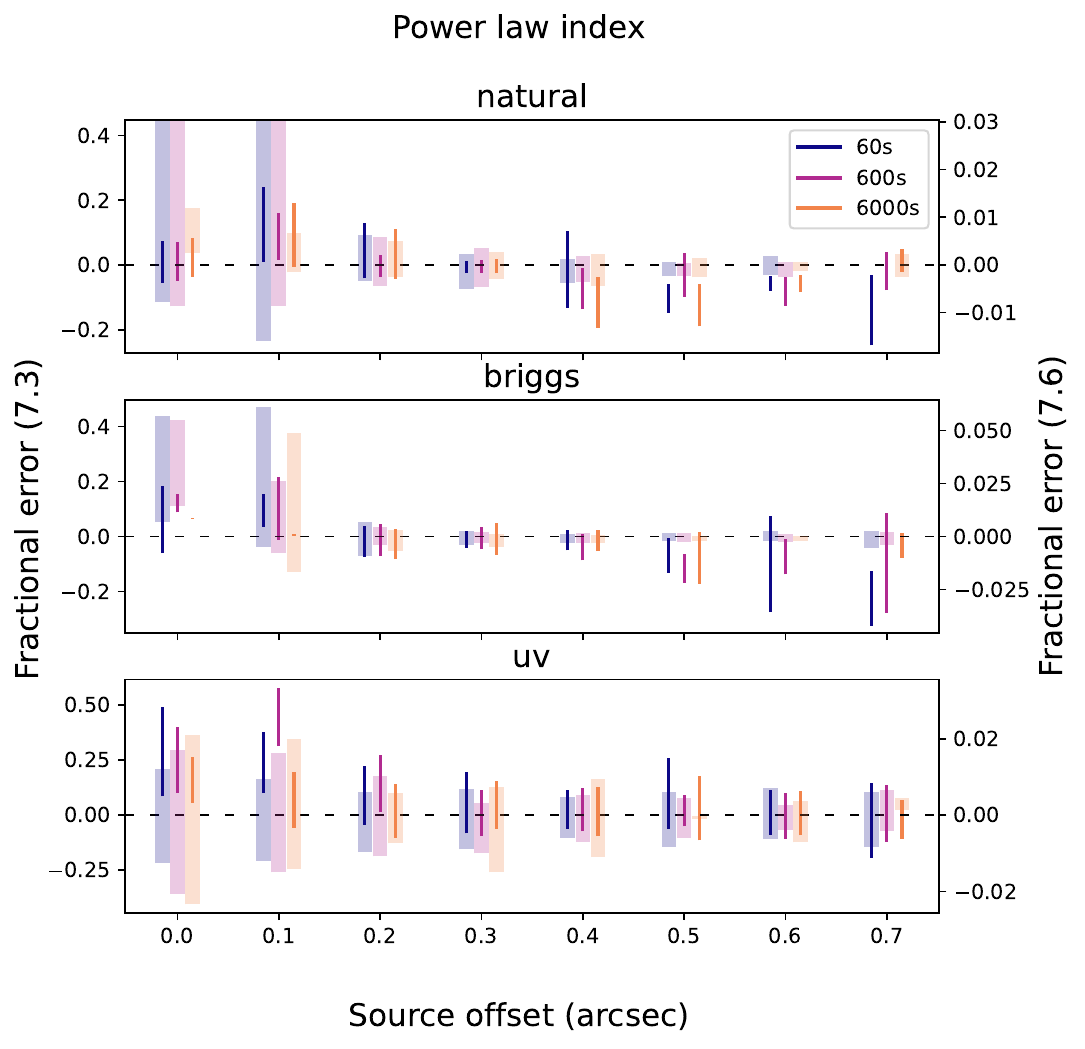} &
    \includegraphics[width=\columnwidth]{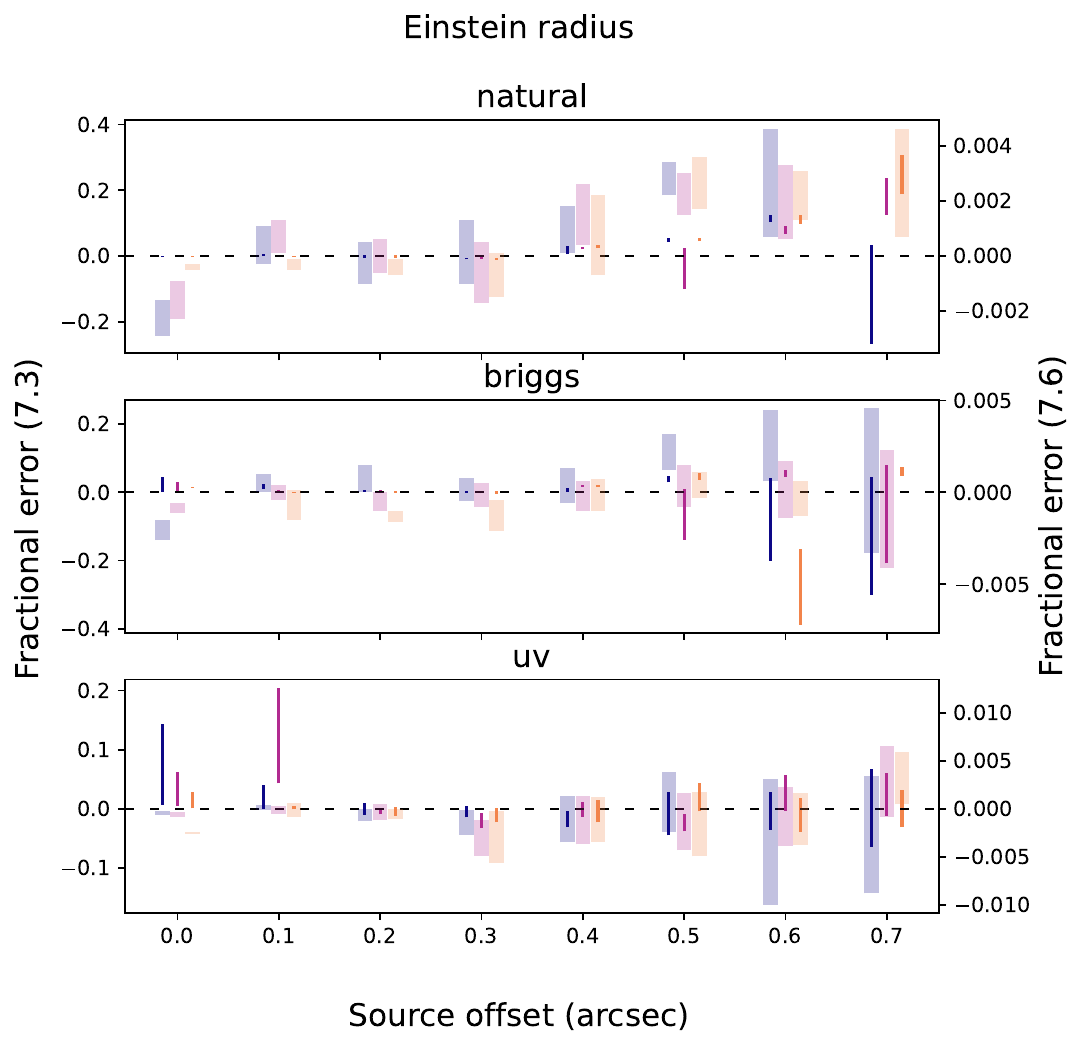} \\
    \includegraphics[width=\columnwidth]{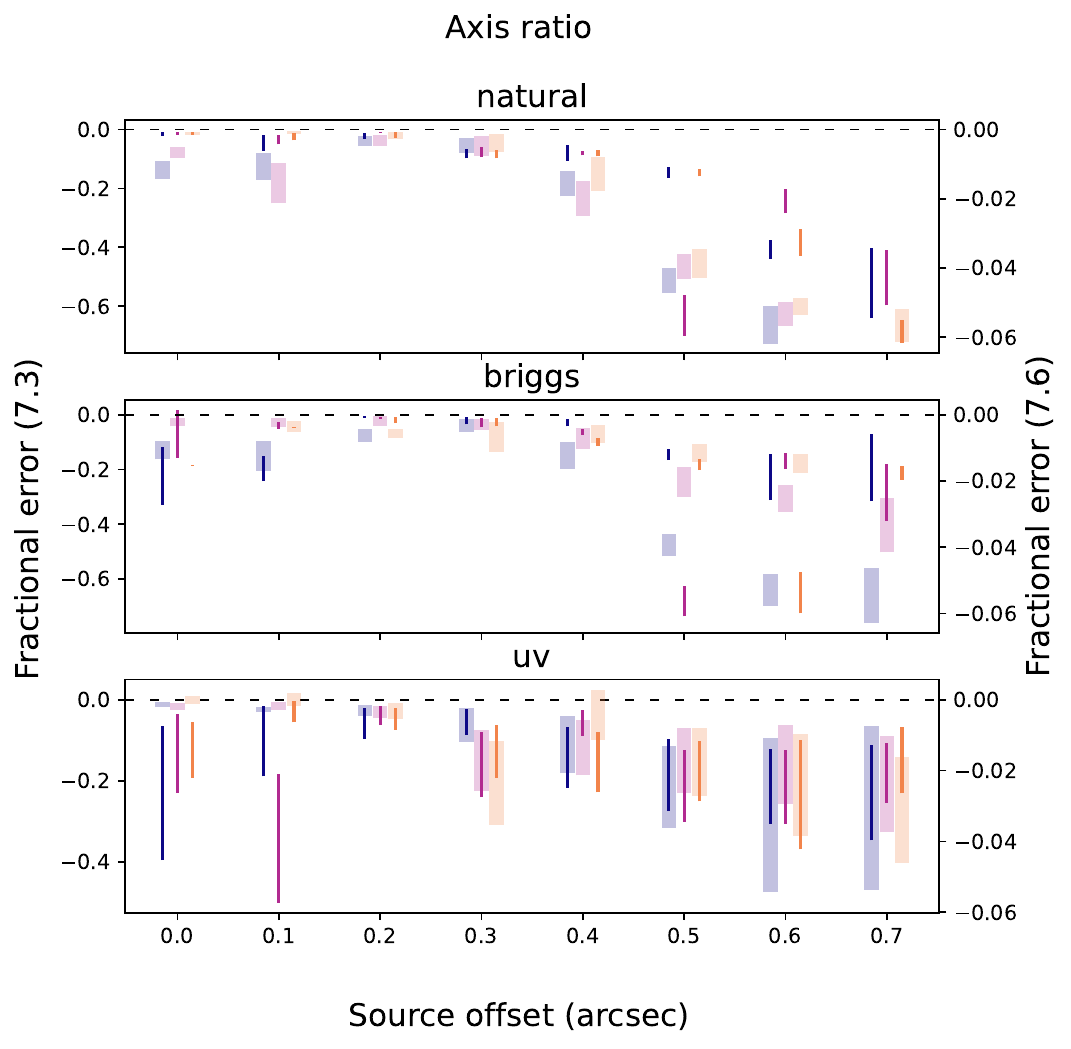} \\
  \end{tabular}
  \caption{Fractional error in SIS lens model parameters as a function of the source offset from the optical axis. The error bars represent the 1-$\sigma$ spread in the fractional error inferred by \texttt{pyAutoLens} (see Section \ref{sec:time_bin} for details). Different colours as detailed in the legend show the on-source integration time. The cycle 7.3 results are indicated by the thin strong-coloured lines (left-hand $y$-axis) and the 7.6 results by the thick pastel-coloured lines (right-hand $y$-axis). From top to bottom, the three sub-panels for each parameter show the results of modelling the \texttt{CLEAN}ed naturally weighted image, the \texttt{CLEAN}ed Briggs-weighted image and the visibilities directly.}
  \label{fig:SIS_modelling}
\end{figure*}

\begin{figure*}
\centering
  \begin{tabular}{cc}
    \includegraphics[width=\columnwidth]{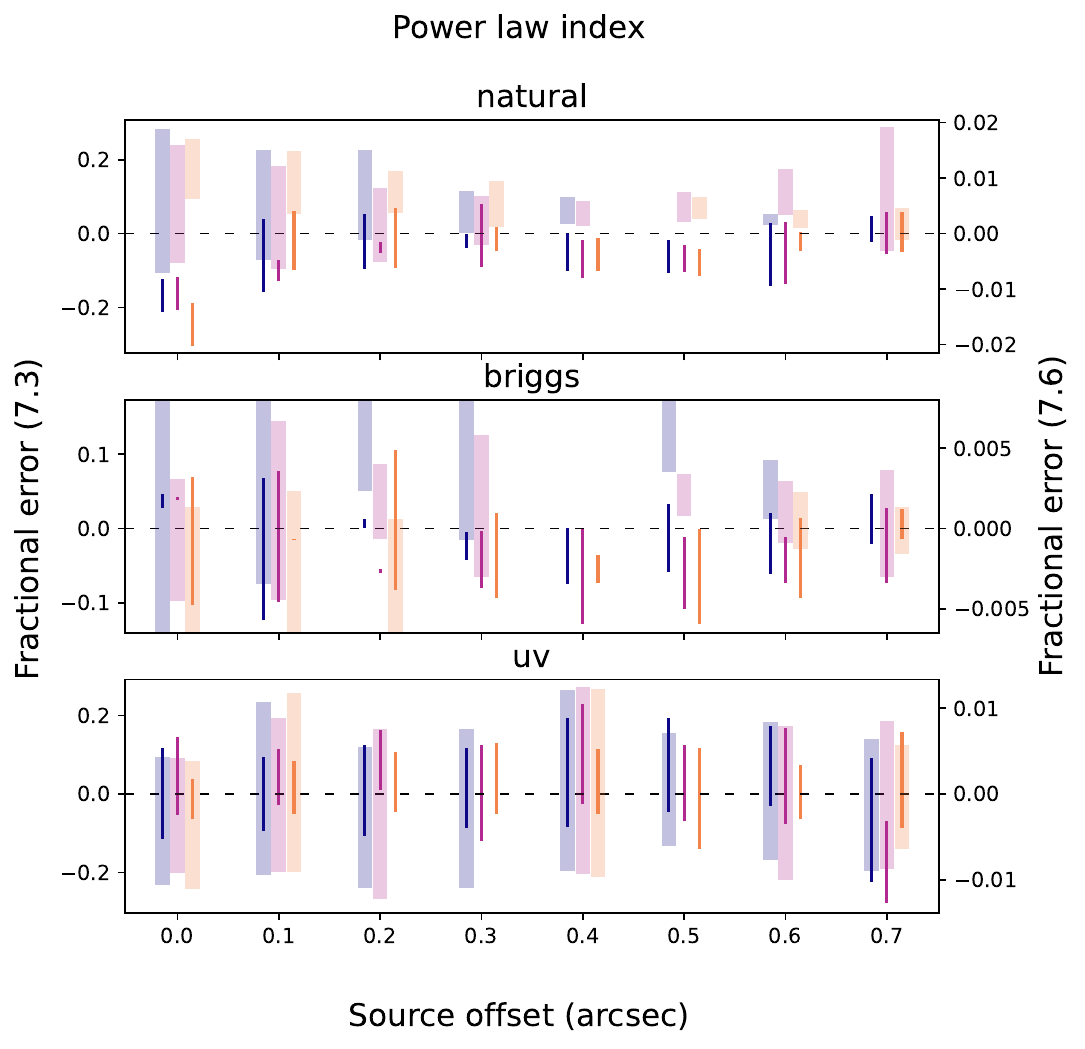} &
    \includegraphics[width=\columnwidth]{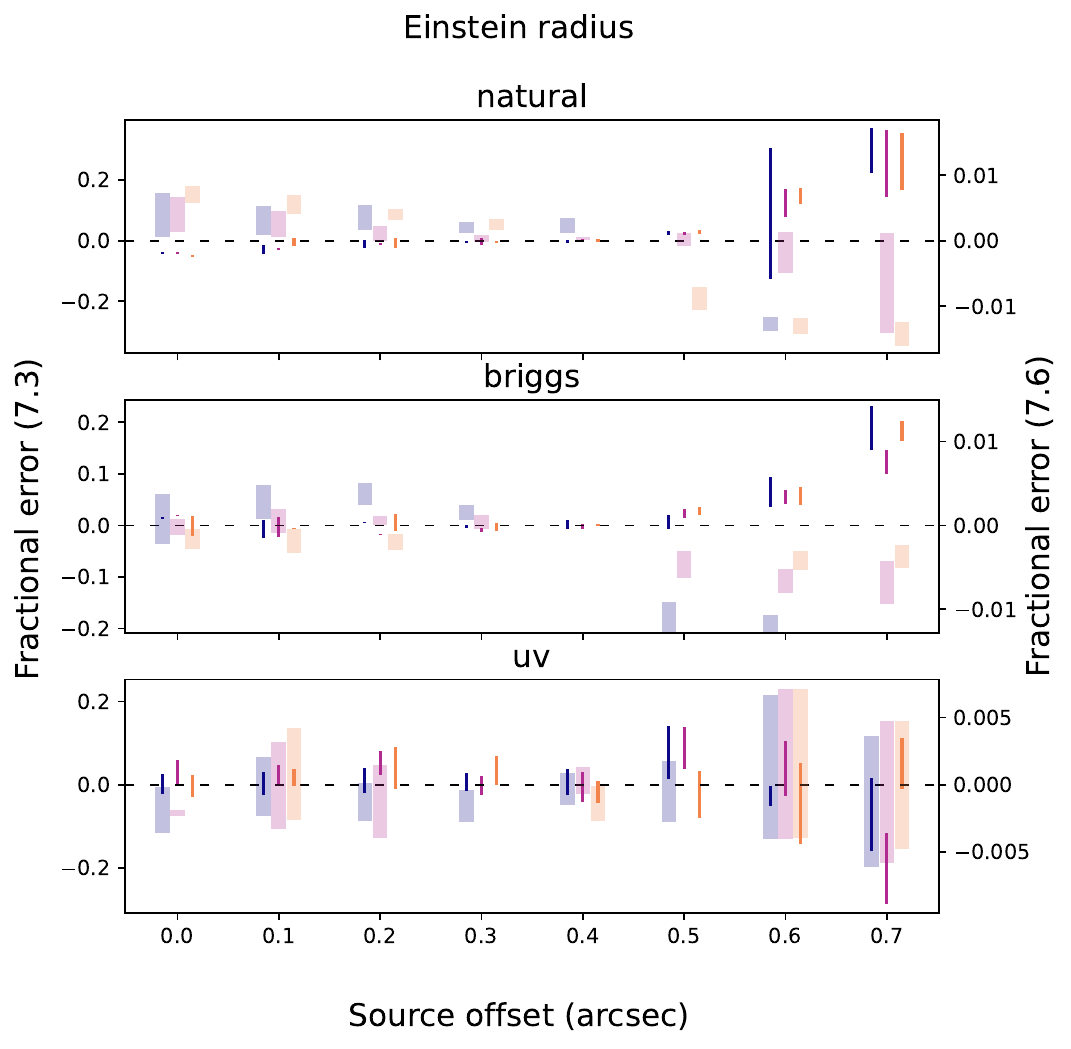} \\
    \includegraphics[width=\columnwidth]{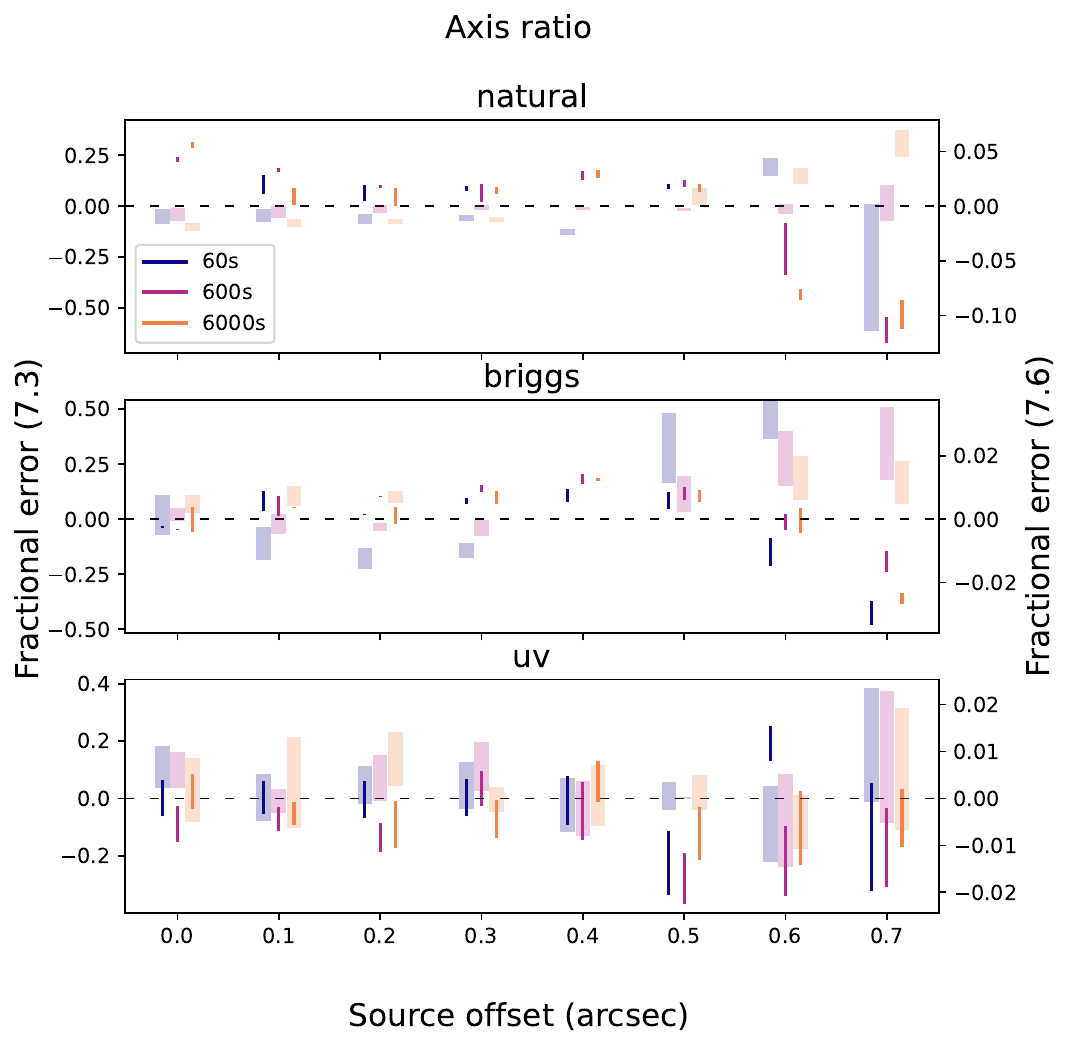} &
    \includegraphics[width=\columnwidth]{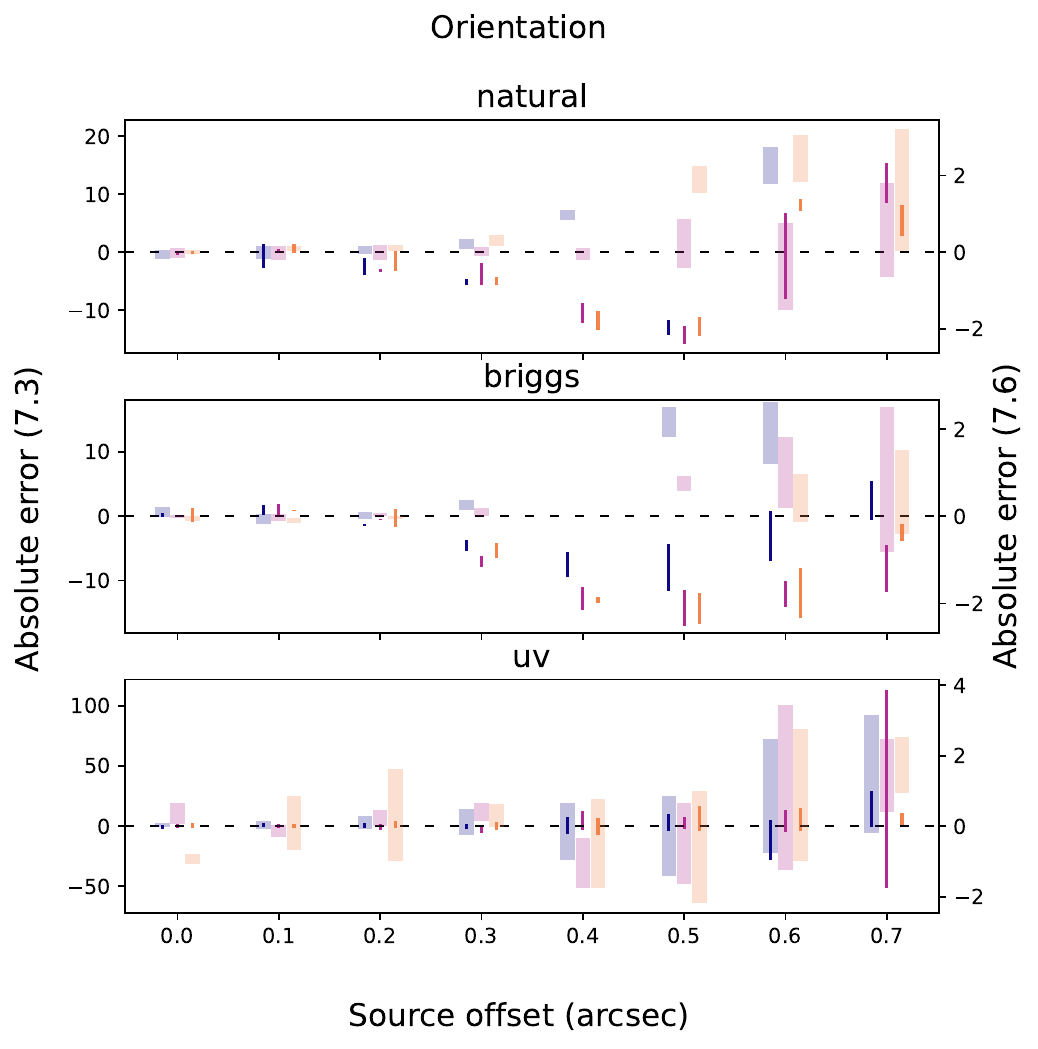} \\
  \end{tabular}
  \caption{Fractional and absolute error in SIE lens model parameters (fold configuration) as a function of the source offset from the optical axis. The error bars represent the 1-$\sigma$ spread in the fractional error for $\alpha$, $\theta_{\rm E}$ and $q$ and absolute error in orientation (in degrees) inferred by \texttt{pyAutoLens} (see Section \ref{sec:time_bin} for details). Different colours as detailed in the legend show the on-source integration time. The cycle 7.3 results are indicated by the thin strong-coloured lines (left-hand $y$-axis) and the 7.6 results by the thick pastel-coloured lines (right-hand $y$-axis). From top to bottom, the three sub-panels for each parameter show the results of modelling the \texttt{CLEAN}ed naturally weighted image, the \texttt{CLEAN}ed Briggs-weighted image and the visibilities directly.}
  \label{fig:SIE_fold_modelling}
\end{figure*}

\begin{figure*}
\centering
  \begin{tabular}{cc}
    \includegraphics[width=\columnwidth]{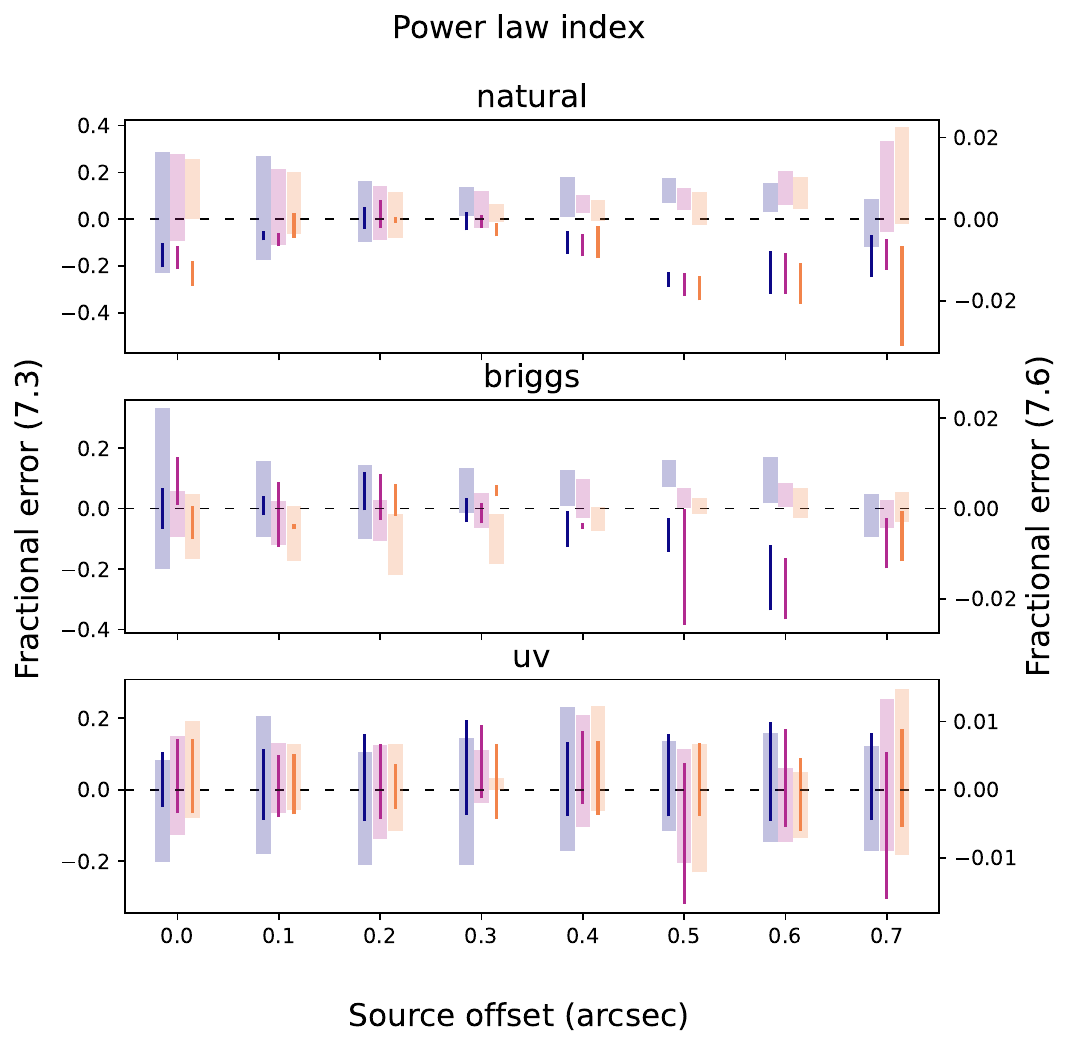} &
    \includegraphics[width=\columnwidth]{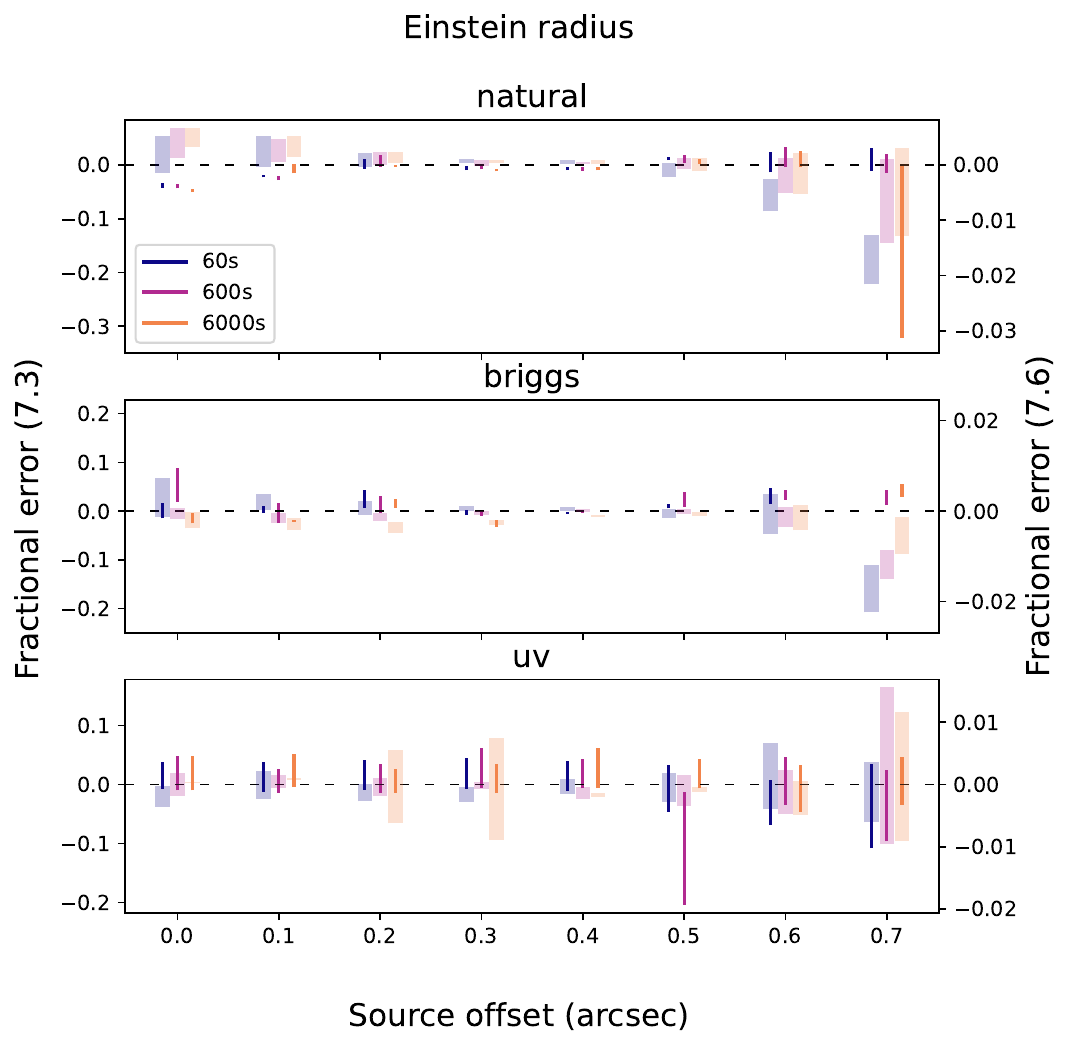} \\
    \includegraphics[width=\columnwidth]{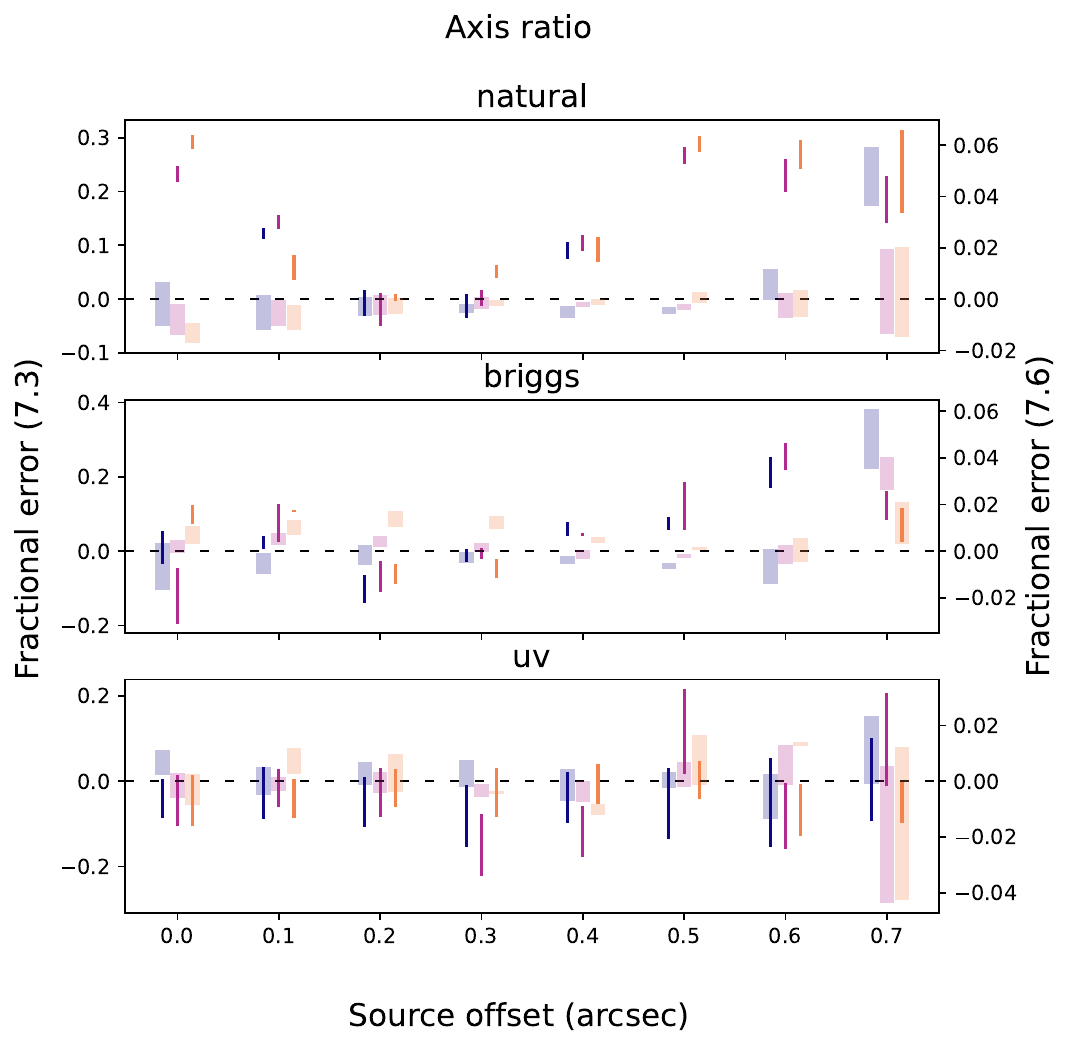} &
    \includegraphics[width=\columnwidth]{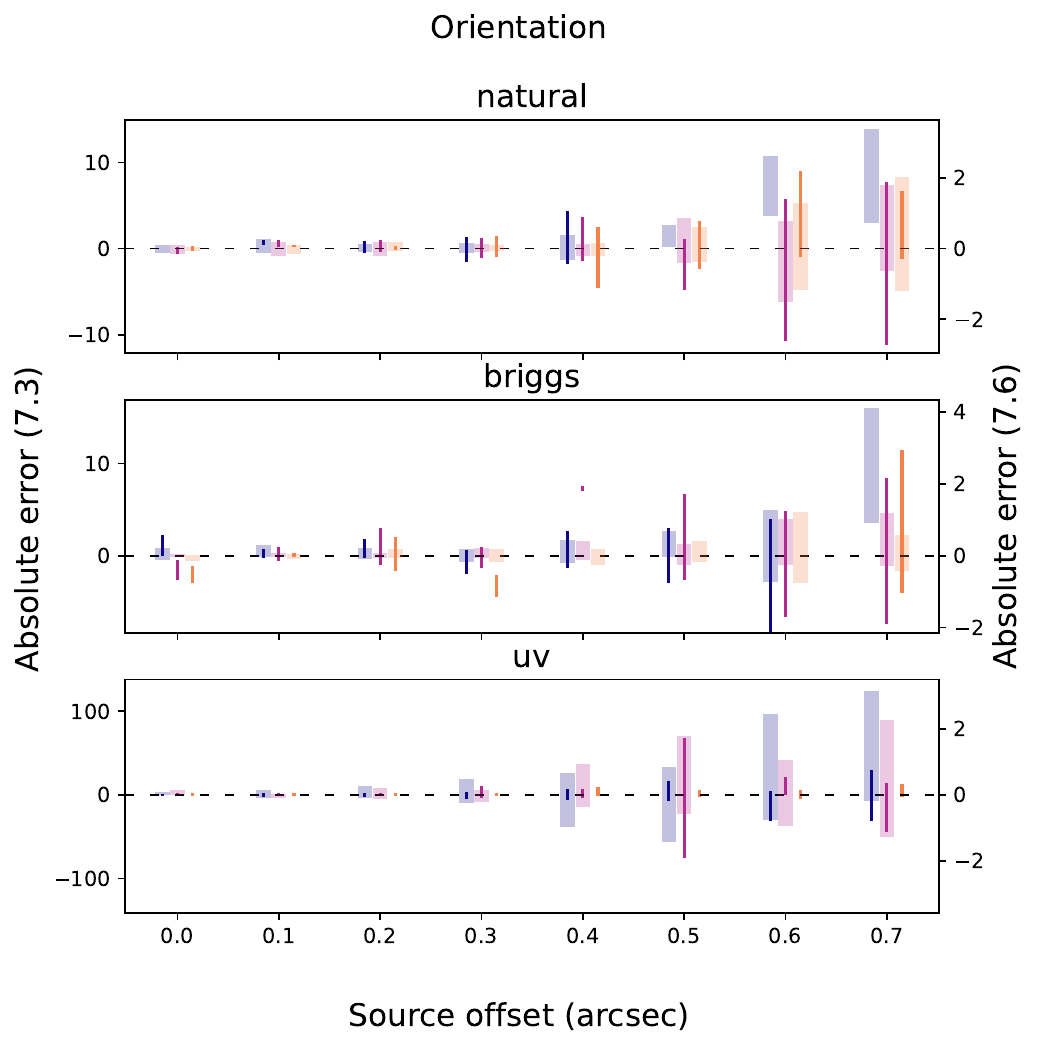} \\
  \end{tabular}
  \caption{Fractional and absolute error in SIE lens model parameters (cusp configuration) as a function of the source offset from the optical axis. The error bars represent the 1-$\sigma$ spread in the fractional error for $\alpha$, $\theta_{\rm E}$ and $q$ and absolute error in orientation (in degrees) inferred by \texttt{pyAutoLens} (see Section \ref{sec:time_bin} for details). Different colours as detailed in the legend show the on-source integration time. The cycle 7.3 results are indicated by the thin strong-coloured lines (left-hand $y$-axis) and the 7.6 results by the thick pastel-coloured lines (right-hand $y$-axis). From top to bottom, the three sub-panels for each parameter show the results of modelling the \texttt{CLEAN}ed naturally weighted image, the \texttt{CLEAN}ed Briggs-weighted image and the visibilities directly.}
  \label{fig:SIE_cusp_modelling}
\end{figure*}

Figures \ref{fig:SIS_modelling}, \ref{fig:SIE_fold_modelling} and \ref{fig:SIE_cusp_modelling} show our results. The figures show the fractional error on each lens model parameter as a function of source offset for both natural and Briggs weighting, as well as for the direct visibility modelling. Results using the lower resolution cycle 7.3 simulations are shown by the thin strongly coloured lines and the cycle 7.6 results by the thick pastel-coloured lines. Each sub-panel contains the results for the three on-source integration times as indicated by the legend. Generally, we again find that the fractional errors (see Section \ref{sec:time_bin}) on lens model parameters are reduced by an order of magnitude when using the higher resolution cycle 7.6 array configuration and that the direct visibility modelling performs better than modelling of any of the \texttt{CLEAN}ed image data. On the whole, longer integration times result in a smaller spread in the inferred parameters, but the gain is marginal; increasing from 60s to 6000s produces a $\sim 50$ per cent reduction in parameter uncertainty at best. In addition, there is little difference in the range of biases seen, independent of resolution, source offset and modelling method.

Inspecting the results in more detail, Figure \ref{fig:SIS_modelling} displays the results from modelling the SIS lens. The modelling of the naturally weighted images performs well for determining the power-law index, but significant biases in the inferred value of Einstein radius and axis ratio begin to appear at larger source offsets, regardless of integration time or resolution. The results for the Briggs-weighted images are largely similar, but the extent to which the inferred values for $\theta_{\rm E}$ and $q$ are biased at large source offsets is less than that of the naturally weighted images. The direct modelling of visibility data appears to show a small bias in the value of $\alpha$ at small source offsets for the cycle 7.3 array configuration, but not for the higher resolution cycle 7.6 configuration. The inferred values of Einstein radius for this approach do not show any evidence of being biased at large values of source offset, but there is a small bias at small source offsets, which is much lessened in the higher resolution results. There is, however, a bias at all source offsets in the axis ratio value inferred by the direct visibility modelling approach. This bias remains approximately constant over the range of offsets considered here, at approximately the level of 20 per cent for the cycle 7.3 data but only 2 per cent for the cycle 7.6 data.

Figure \ref{fig:SIE_fold_modelling} shows the results of the SIE with the fold configuration and follows the same layout as Figure \ref{fig:SIS_modelling} but with the addition of mass profile orientation. In terms of recovering the power-law index, the uv-plane modelling outperforms the two image-plane modelling approaches, both in terms of a lack of bias and smaller errors. This is the case for the cycle 7.3 and cycle 7.6 data. The Briggs-weighted image plane modelling performs slightly better than the naturally weighted image plane approach, not, for example showing the bias at low source offsets exhibited by the cycle 7.3 naturally weighted results, but both image plane methods have a tendency to underestimate (by $\sim 10$ per cent for cycle 7.3) or overestimate (by $\sim 1$ per cent for cycle 7.6) the power-law index. This is true for all three integration times. For the Einstein radius, inferred values using naturally weighted images are biased $\sim 5$ per cent low for cycle 7.3 data but only $\sim 0.5$ per cent high for cycle 7.6 data when the source is on or close to the optical axis. This is not seen with the Briggs-weighted data. At large source offsets, both image-plane approaches become biased towards higher values of $\theta_{\rm E}$ (biased higher in the cycle 7.3 data and biased lower with the cycle 7.6 data), but once again, the uv-plane results are superior with errors beneath $\sim 0.5$ per cent for the higher resolution cycle 7.6 data.

Biases in the power-law index and Einstein radius also translate to biases in the axis ratio, due to the well known degeneracy between this parameter triplet. The significant biases in $\theta_{\rm E}$ of both image plane results produce the larger biases seen in $q$ at larger source offsets. The uv-plane modelling shows some bias in $q$ at larger offsets, at the 20 per cent level for the lower resolution cycle 7.3 data but only up to 2 per cent for the cycle 7.6 data. The bias in lens orientation tends to anti-correlate with the bias in axis ratio, and this is seen most notably in the image-plane modelling results, where biases of up to 20$^\circ$ are seen. The orientation inferred from the uv-plane modelling shows insignificant bias and with the higher resolution cycle 7.6 data, the orientation is recovered to better than 4$^\circ$ at all source offsets.

Figure \ref{fig:SIE_cusp_modelling} shows the results of modelling the SIE with the cusp configuration. Broadly similar results to the fold caustic are seen, with biases generally becoming larger with the image plane modelling methods at larger source offsets where the lensed image contains less information. The uv-plane method continues to perform well at these larger source offsets. Considering the parameters in turn, for the inference of the power-law index, the visibility modelling approach significantly outperforms the image plane modelling, across the full range of source offsets. Modelling the cycle 7.6 data yields errors in power-law index of no more than $\sim 1$ per cent. Similar to the fold configuration, the results of modelling the naturally weighted images show a tendency to under-estimate $\alpha$ in the cycle 7.3 data or overestimate it in the cycle 7.6 data when the source is on or close to the optical axis. In fact, with the source on the optical axis, these images are identical except for different realisations of random noise. This behaviour is once again not seen in either the Briggs-weighted images nor the visibility modelling. At larger source offsets,  both image plane methods tend to show stronger biases in $\alpha$. The naturally weighted image modelling shows a bias towards lower values in $\theta_{\rm E}$ at small source offsets with the cycle 7.3 data but in general, the modelling of all three data sources provides a reliable estimate of the Einstein radius with errors of no more than 5 per cent at any source offset even with the lower resolution cycle 7.3 data. The visibility modelling approach provides the least biased estimate of the lens axis ratio for all source offsets. Both the naturally and Briggs-weighted datasets exhibit significant biases of up to $\sim 30$ per cent at large source offsets with the cycle 7.3 data, but this is generally much reduced with the higher resolution cycle 7.6 data. The lens orientation is recovered without any signs of bias in any of the data sources.

\section{Discussion}
\label{sec:discussion_p4}

Our results clearly show that direct visibility modelling consistently outperforms image-plane modelling. Nevertheless, image-plane modelling may still be worth consideration in some circumstances, for example, in providing an approximate lens model, perhaps to better initialise visibility modelling. Of the two image-plane weighting schemes we have explored in this work, Briggs weighting results in more reliable lens model parameters overall. The relatively good performance of the modelling of the Briggs-weighted data, compared to that of the naturally weighted data, suggests that there is important information on small scales that is being lost by using natural weighting. At least, for this simple lensing scenario, with a parametric source, any benefits from the increased sensitivity of natural weighting appear to be outweighed by the greater resolution achieved by Briggs weighting. 

In order to explore this effect, we generated a new set of lensed images using \texttt{pyAutoLens} and computed their 2D Fourier transforms to quantify the interferometric signal as a function of scale. We considered two scenarios: one where the source is on the optical axis and another where the source is located at $(0.7 \arcsec, 0.0)$ with respect to the centre of the lens. For each scenario, we produced simulated images of the source, lensed by a power-law profile with $\theta_{\rm E} = 1 \arcsec$, axis ratio varying over the interval $q = [0.4, 0.9]$, and power-law index varying over the interval $\alpha = [1.8, 2.2]$. Following the same procedure as described in Section \ref{sec:simulations}, we produced two versions of each image using the \texttt{tclean} algorithm, one using natural weighting and the other using Briggs weighting with a robustness parameter of zero. For computational efficiency and because similar (albeit reduced) behaviour is exhibited in the higher resolution data, we used the lower resolution ALMA cycle 7.3 configuration.

\begin{figure}
	\includegraphics[width=\columnwidth]{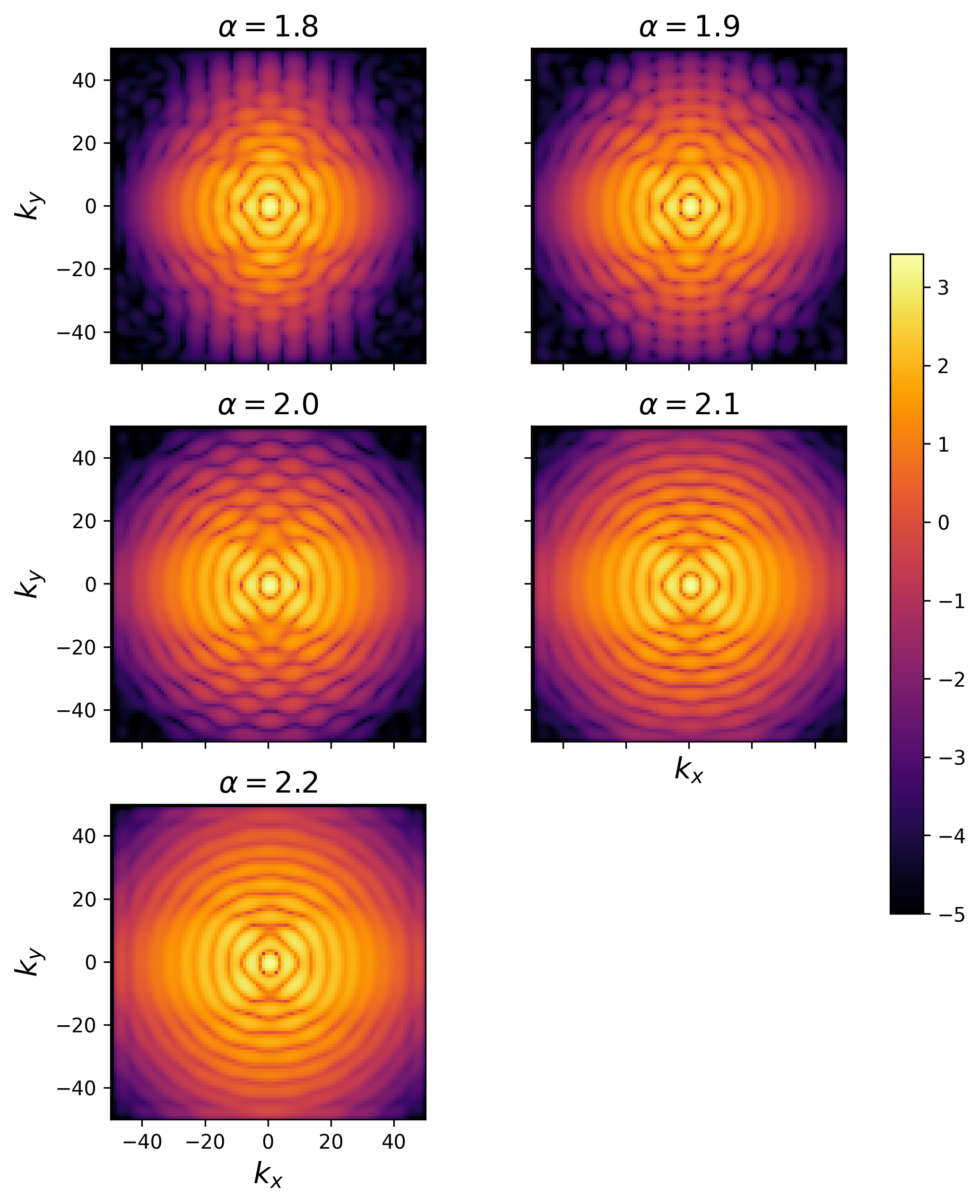}
    \caption{Fourier $q$-variance (see Section \ref{sec:discussion_p4}) of the true sky image of the source on the optical axis lensed by the power-law model. Shading corresponds to the logarithm of the variance. The five different panels correspond to a different fixed power-law index across the interval $\alpha = [1.8, 2.2]$. The true sky surface brightness images were created with a resolution of $0.1 \arcsec$ pixel$^{-1}$ to match that of the \texttt{CLEAN}ed images for the ALMA cycle 7.3 configuration.}
    \label{fig:ffts_sb_varq_lr}
\end{figure}

\begin{figure}
	\includegraphics[width=\columnwidth]{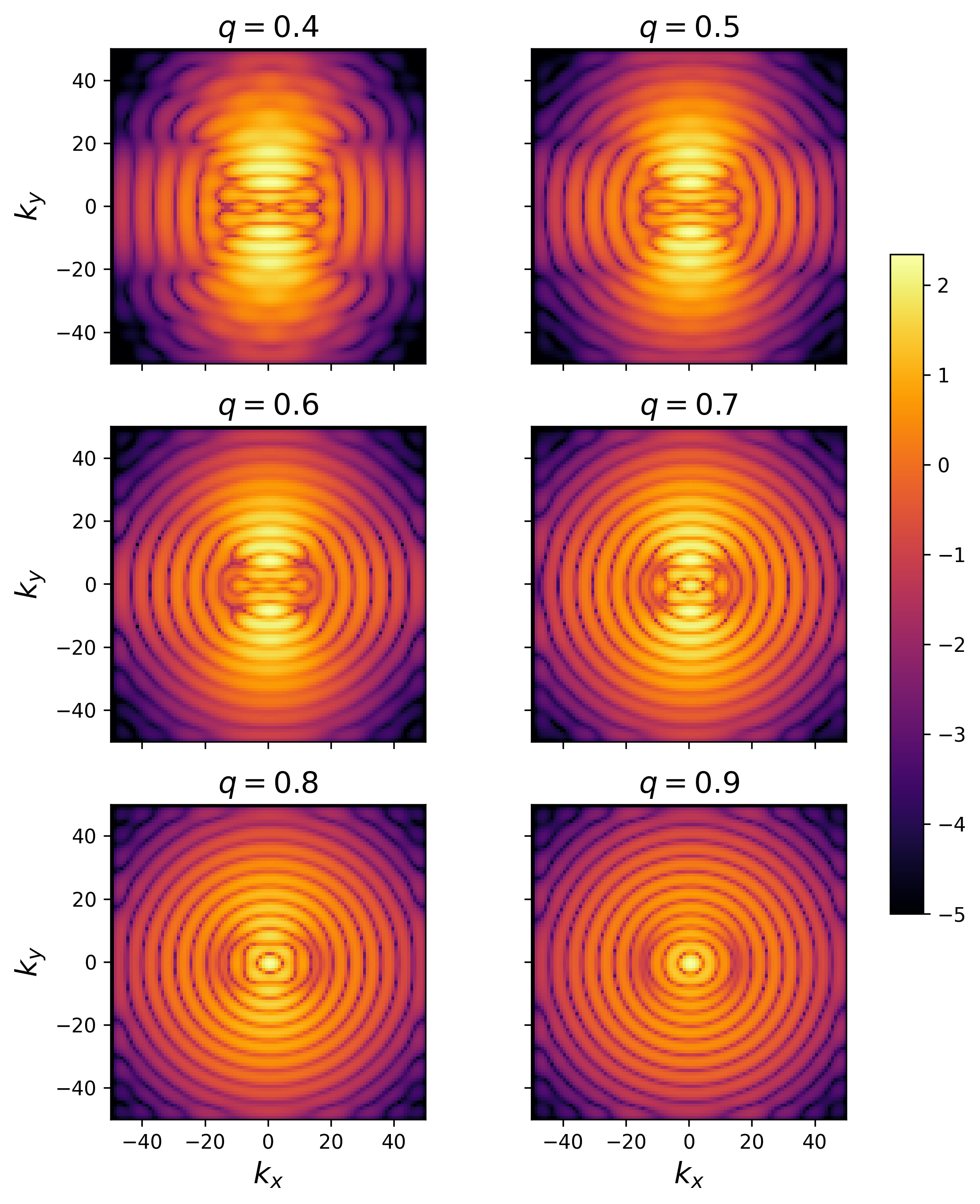}
    \caption{Fourier $\alpha$-variance (see Section \ref{sec:discussion_p4}) of the true sky image of the source on the optical axis lensed by the power-law model. Shading corresponds to the logarithm of the variance. The six different panels correspond to a different fixed axis ratio across the interval $q = [0.4, 0.9]$. The true sky surface brightness images were created with a resolution of $0.1 \arcsec$ pixel$^{-1}$ to match that of the \texttt{CLEAN}ed images for the ALMA cycle 7.3 configuration.
}
    \label{fig:ffts_sb_varalpha_lr}
\end{figure}

Our goal is to quantify the variation in signal on different spatial scales arising from varying the lens model parameters, and therefore to investigate what effect \texttt{CLEAN}ing has on them. Since we have fixed the Einstein radius and orientation for all of our models, the parameters of influence are the power-law index, $\alpha$, and the axis ratio, $q$. We therefore computed the 2D Fourier transform of 10 images, varying the axis ratio across the interval $q = [0.4, 0.9]$ for a fixed value of $\alpha$. For each point on the Fourier plane, we calculated the variance over the 10 Fourier transforms. We refer to this as the 'Fourier $q$-variance' hereafter. The logarithm of this variance for the true sky surface brightness images (i.e. those directly produced by \texttt{pyAutoLens}) when the source is on the optical axis is plotted in Figure \ref{fig:ffts_sb_varq_lr} for different values of power-law index. Similarly, for fixed $q$, we computed the variance of the Fourier transform of 10 images when the power-law index is varied across the interval $\alpha = [1.8, 2.2]$ which we refer to as the 'Fourier $\alpha$-variance' hereafter. The logarithm of this variance is plotted in Figure \ref{fig:ffts_sb_varalpha_lr} for the true sky surface brightness images when the source is on the optical axis.

As the Figures show, there is a wealth of information contained at high $k$-numbers, indicated by the intensity at large radial distances from the centre of the Fourier transforms. It can be seen from Figure \ref{fig:ffts_sb_varq_lr} that the Fourier $q$-variance at large $k$ increases with increasing power-law index. Figure \ref{fig:ffts_sb_varalpha_lr} shows that the Fourier $\alpha$-variance at high $k$ numbers has a weak radial dependence on the axis ratio, but particularly for very elliptical lenses, there is a strong directional component that is most prominent at low $k$ numbers. In both instances, there is significant variance at high $k$ numbers, indicating that variations in the lens model parameters $q$ and $\alpha$ lead to changes at high spatial frequencies, that if retained in \texttt{CLEAN}ing, could be useful in constraining the lens model.

\begin{figure}
	\includegraphics[width=\columnwidth]{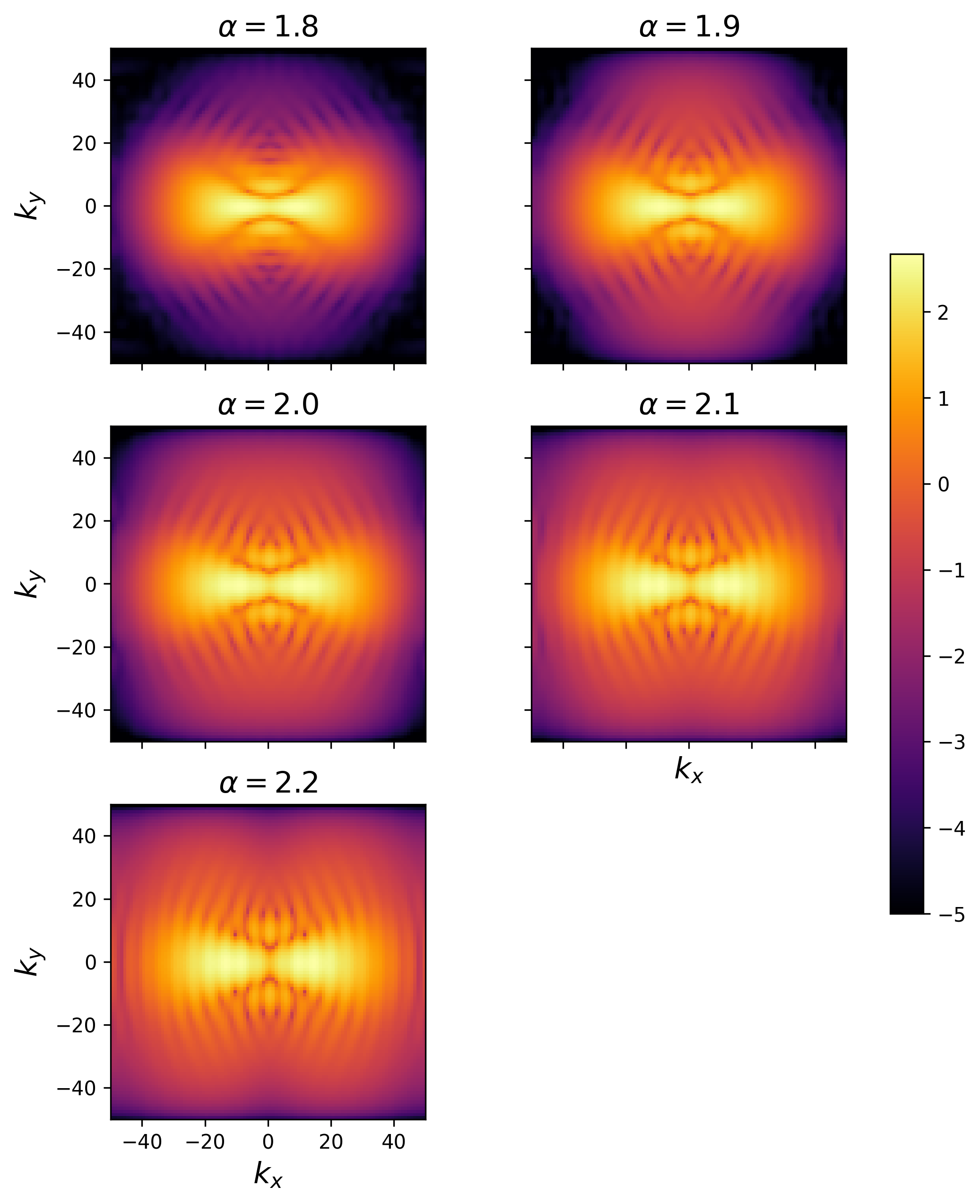}
    \caption{Fourier $q$-variance (see Section \ref{sec:discussion_p4}) of the true sky image of the source offset $(0.7\arcsec, 0.0\arcsec)$ from the optical axis lensed by the power-law model. Shading corresponds to the logarithm of the variance. The five different panels correspond to a different fixed power-law index across the interval $\alpha = [1.8, 2.2]$. The true sky surface brightness images were created with a resolution of $0.1 \arcsec$ pixel$^{-1}$ to match that of the \texttt{CLEAN}ed images for the ALMA cycle 7.3 configuration.}
    \label{fig:ffts_sb_varq_s07_lr}
\end{figure}

\begin{figure}
	\includegraphics[width=\columnwidth]{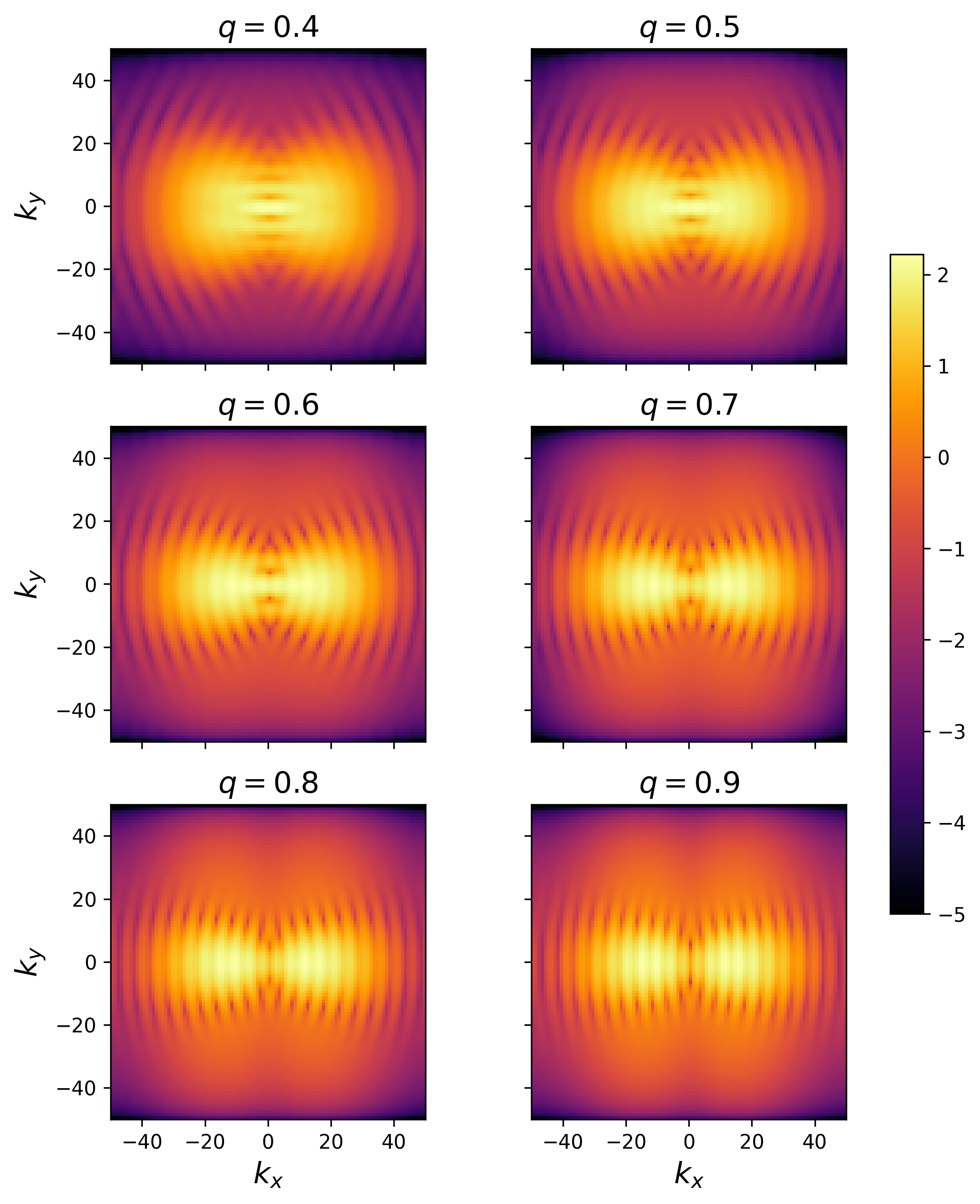}
    \caption{Fourier $\alpha$-variance (see Section \ref{sec:discussion_p4}) of the true sky image of the source offset $(0.7\arcsec, 0.0\arcsec)$ from the optical axis lensed by the power-law model. Shading corresponds to the logarithm of the variance. The six different panels correspond to a different fixed axis ratio across the interval $q = [0.4, 0.9]$. The true sky surface brightness images were created with a resolution of $0.1 \arcsec$ pixel$^{-1}$ to match that of the \texttt{CLEAN}ed images for the ALMA cycle 7.3 configuration.}
    \label{fig:ffts_sb_varalpha_s07_lr}
\end{figure}

Similar to Figures \ref{fig:ffts_sb_varq_lr} and \ref{fig:ffts_sb_varalpha_lr}, Figures \ref{fig:ffts_sb_varq_s07_lr} and \ref{fig:ffts_sb_varalpha_s07_lr} respectively show the Fourier $q$-variance and Fourier $\alpha$-variance when the source is offset from the optical axis to (0.7", 0.0). Generally,  since the images are not circularly symmetric in this case, the Fourier variance in both cases becomes less circularly symmetric; rotating the images would also result in the same rotation of the Fourier transforms. More notably, the Fourier variances become stronger at high $k$-numbers which could plausibly contribute to more tightly constraining the lens model, a postulation that we will explore.

\begin{figure}
	\includegraphics[width=\columnwidth]{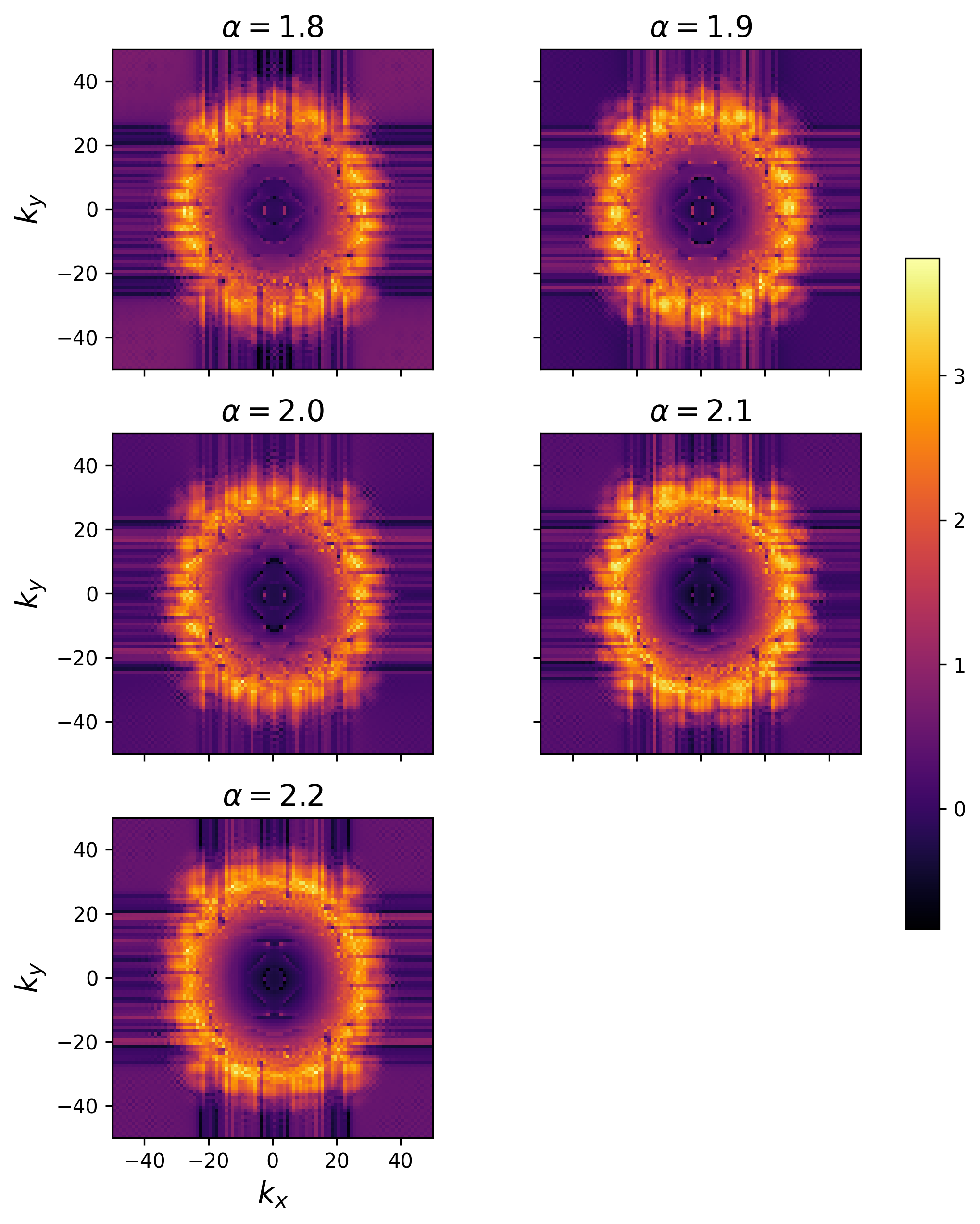}
    \caption{Ratios (Briggs:natural) of the $q$-variance (see Section \ref{sec:discussion_p4}) of the Fourier transformed \texttt{CLEAN}ed images for a range of power-law indices. The source is placed on the optical axis. A higher ratio indicates that Briggs-weighting retains more spatial information and hence gives better lens model sensitivity.}
    \label{fig:sensitivity_varq}
\end{figure}

\begin{figure}
	\includegraphics[width=\columnwidth]{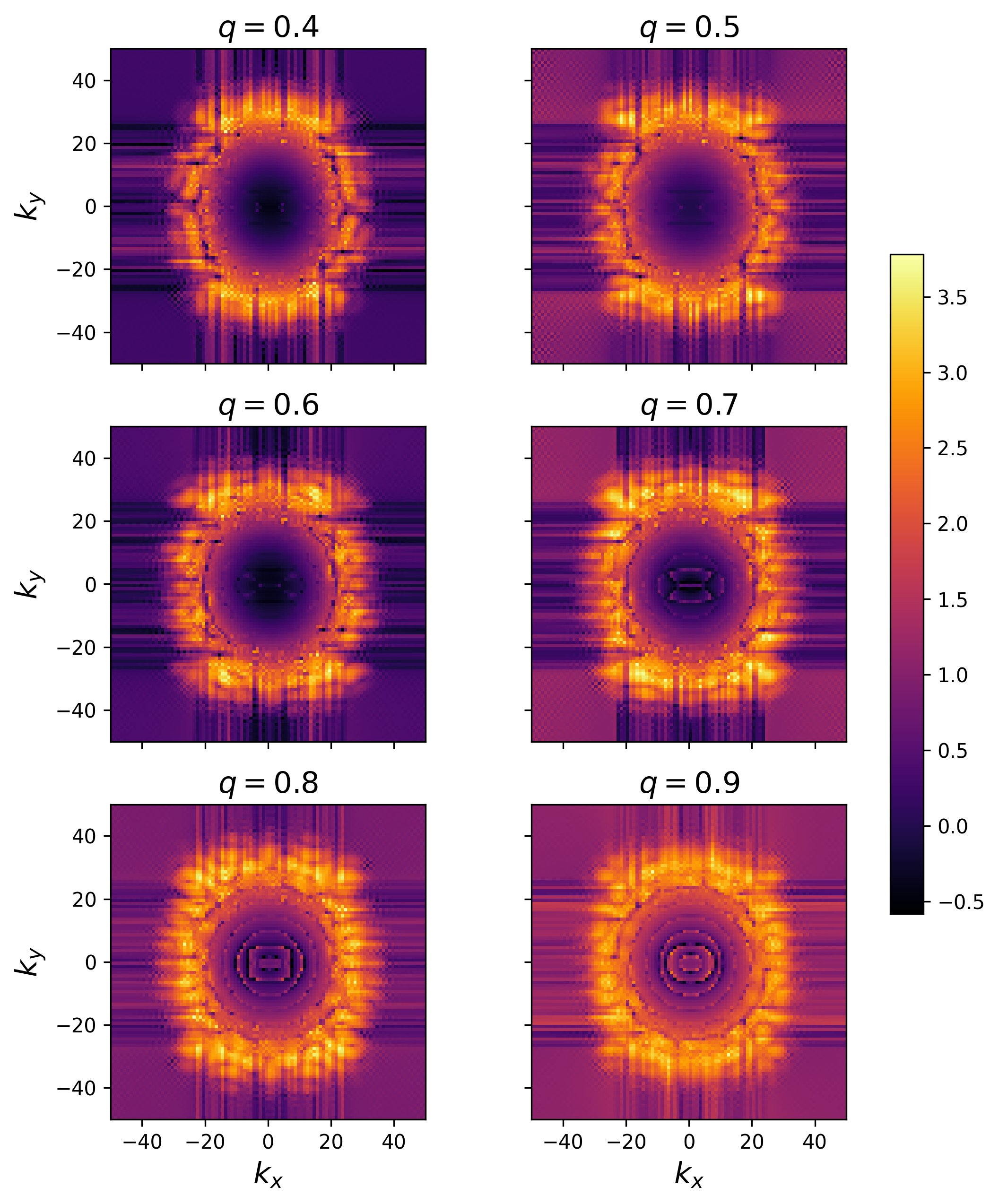}
    \caption{Ratios (Briggs:natural) of the $\alpha$-variance (see Section \ref{sec:discussion_p4}) of the Fourier transformed \texttt{CLEAN}ed images for a range of axis ratios. The source is placed on the optical axis. A higher ratio indicates that Briggs-weighting retains more spatial information and hence gives better lens model sensitivity.}
    \label{fig:sensitivity_varalpha}
\end{figure}

\begin{figure}
	\includegraphics[width=\columnwidth]{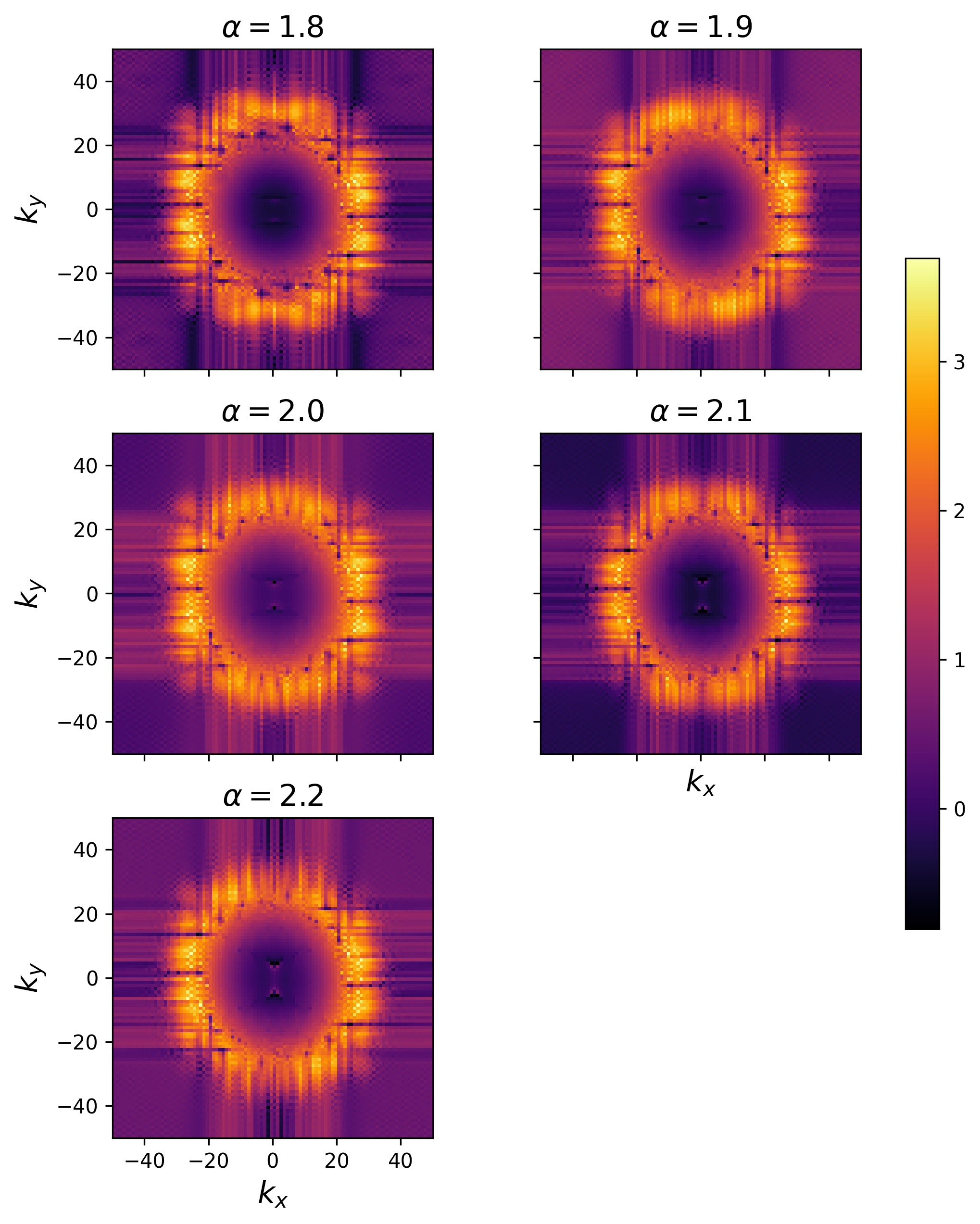}
    \caption{Ratios (Briggs:natural) of the $q$-variance (see Section \ref{sec:discussion_p4}) of the Fourier transformed \texttt{CLEAN}ed images for a range of power-law indices. The source is offset from the optical axis by $(0.7 \arcsec, 0.0\arcsec)$. A higher ratio indicates that Briggs-weighting retains more spatial information and hence gives better lens model sensitivity.}
    \label{fig:sensitivity_varq_s07}
\end{figure}

\begin{figure}
	\includegraphics[width=\columnwidth]{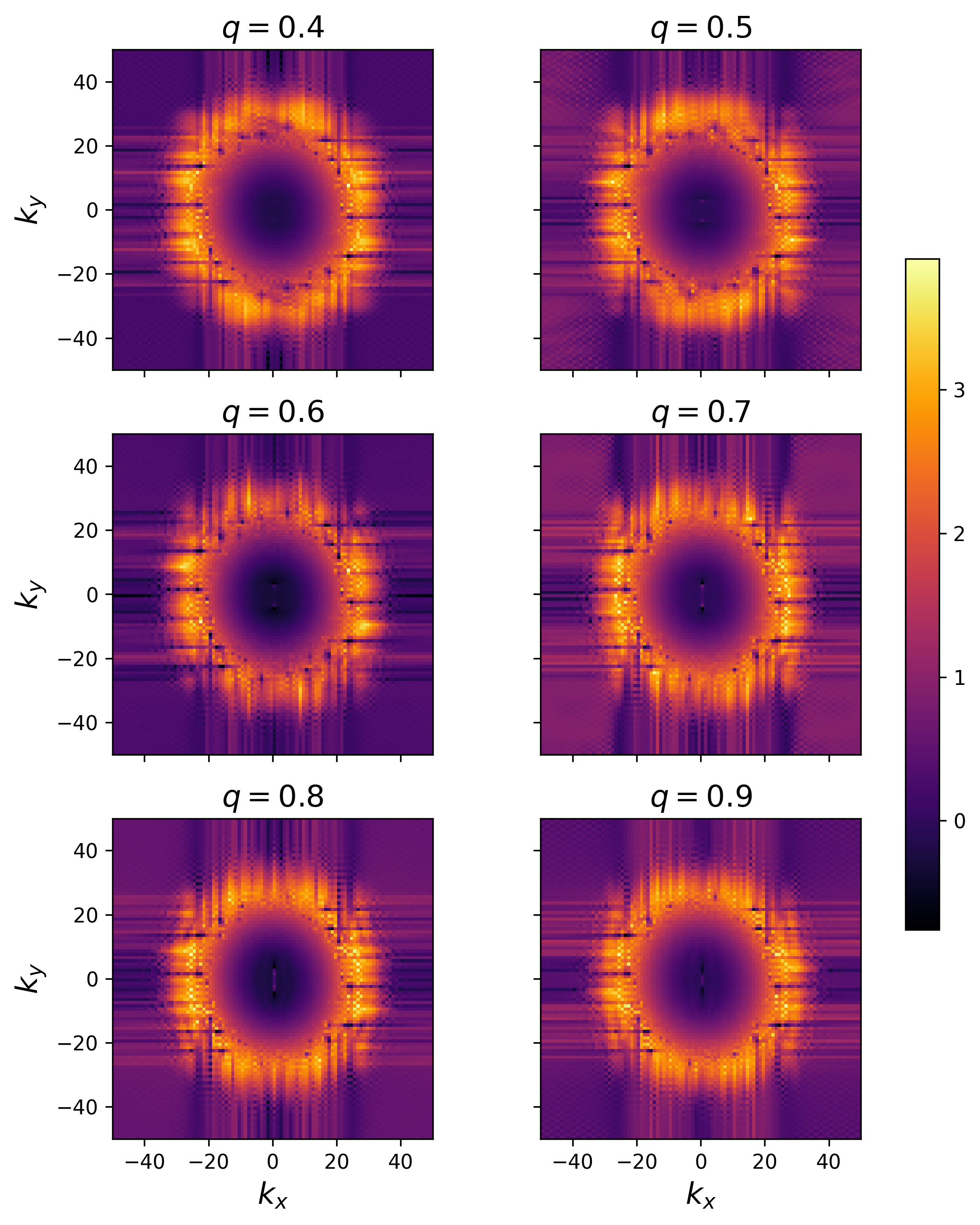}
    \caption{Ratios (Briggs:natural) of the $\alpha$-variance (see Section \ref{sec:discussion_p4}) of the Fourier transformed \texttt{CLEAN}ed images for a range of axis ratios. The source is offset from the optical axis by $(0.7 \arcsec, 0.0\arcsec)$. A higher ratio indicates that Briggs-weighting retains more spatial information and hence gives better lens model sensitivity.}
    \label{fig:sensitivity_varalpha_s07}
\end{figure}

The significance of this high-$k$ variance can be seen by looking at the relative sensitivity to these scales in the naturally and Briggs-weighted images. Due to the nature of the weighting schemes, it is expected that the naturally weighted images will contain less information on these smaller scales compared to the Briggs-weighted images. We quantified this by inspecting ratios of the Fourier $\alpha$ and $q$ variances. Figure \ref{fig:sensitivity_varq} shows the ratio of the Fourier $q$-variance of the Briggs \texttt{CLEAN}ed images to that of the naturally weighted \texttt{CLEAN}ed images for fixed values of $\alpha$ when the source is on the optical axis. Similarly, Figure \ref{fig:sensitivity_varalpha} shows the ratio of Fourier $\alpha$-variances for fixed values of $q$. It can be seen from the annular structure in these figures that the Briggs-weighted \texttt{CLEAN}ed images are capturing information at higher $k$-numbers than the naturally weighted images, in regions where the true surface brightness distributions show significant variation, as shown in Figures \ref{fig:ffts_sb_varq_lr} and \ref{fig:ffts_sb_varalpha_lr}. This increased angular resolution in the Briggs-weighted images likely plays a key role in the ability to accurately recover the lens model parameters, more so than the better point source sensitivity of the naturally weighted images. This effect becomes even more important when considering cases where the source is significantly displaced from the optical axis, such as in Figures \ref{fig:ffts_sb_varq_s07_lr} and \ref{fig:ffts_sb_varalpha_s07_lr}. Quantifying the azimuthally averaged profiles of these Fourier variance plots shows that in this case, an even greater proportion of variation and therefore higher sensitivity occurs at high $k$-values. As shown by Figures \ref{fig:sensitivity_varq_s07} and \ref{fig:sensitivity_varalpha_s07}, it is in these high $k$-regions that the Briggs-weighted images are able to reproduce more of the signal. In addition to this, there is also less information within the Einstein ring to constrain the lens model at large source offsets, which can partly explain the difficulty in recovering the true lens model parameters.

The difference in lens modelling accuracy between the Briggs and naturally weighted image datasets can, at least in part, be explained by the different angular resolutions achieved by the two weighting schemes. In order to quantify this effect, we computed the azimuthal average of the two-dimensional Fourier transforms of the true sky surface brightness images and the corresponding Briggs-weighted and naturally weighted \texttt{CLEAN}ed image. Hereafter, we refer to the azimuthal averages of these two-dimensional Fourier transforms as 'Fourier radial profiles'. 

\begin{figure*}
\centering
  \begin{tabular}{cc}
    \includegraphics[width=0.99\columnwidth]{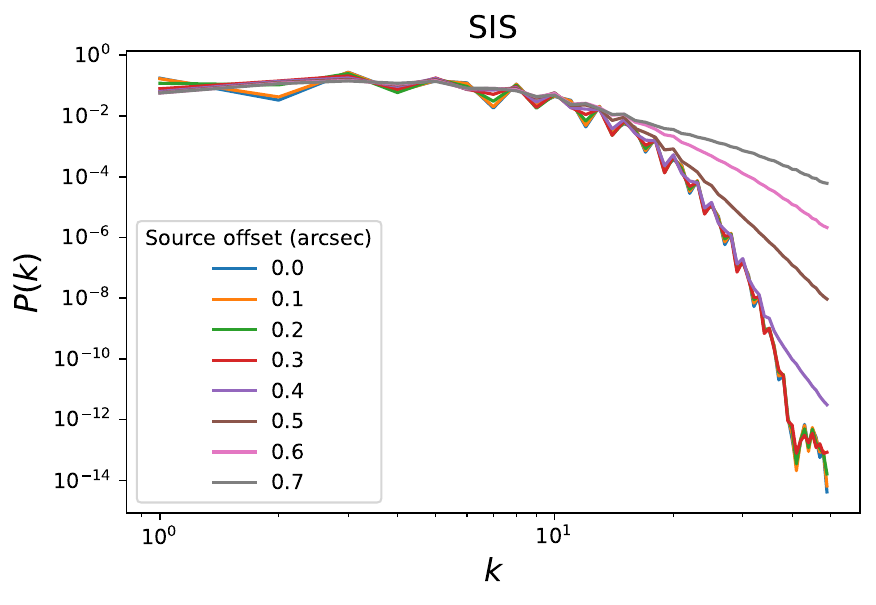} &
    \includegraphics[width=0.99\columnwidth]{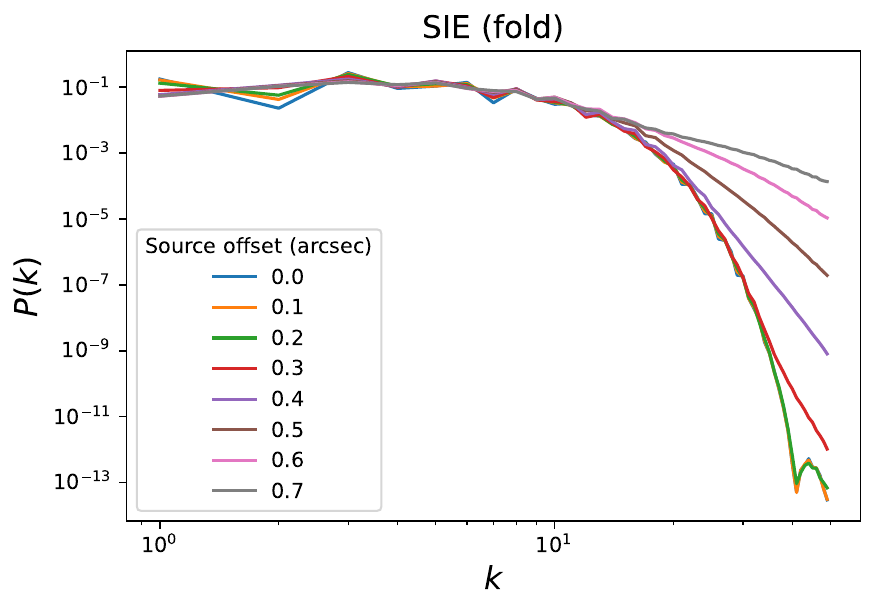} \\
    \includegraphics[width=0.99\columnwidth]{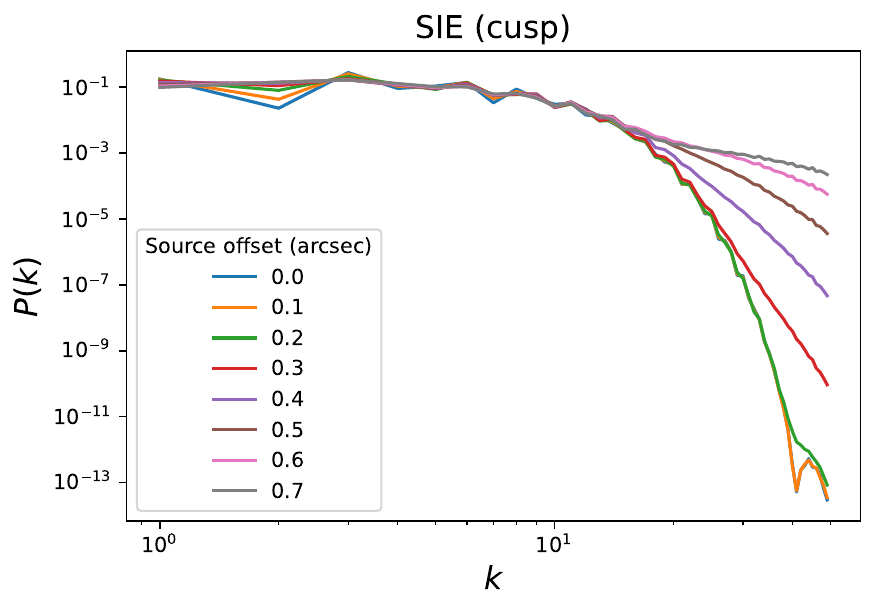} 
  \end{tabular}
  \caption{Fourier radial profile of the true sky simulated lensed images. The top left panel shows the profile for each of the source positions considered with the SIS lens model. The top right panel shows the profile for each of the source positions moving through the fold configuration of the SIE lens model. The bottom left panel shows the profile for each of the source positions moving through the cusp configuration of the SIE lens model. The legends in each plot indicate the distance between the source and lens centroid in arcseconds.}
  \label{fig:power_spectra}
\end{figure*}

\begin{figure*}
	\includegraphics[width=1.7\columnwidth]{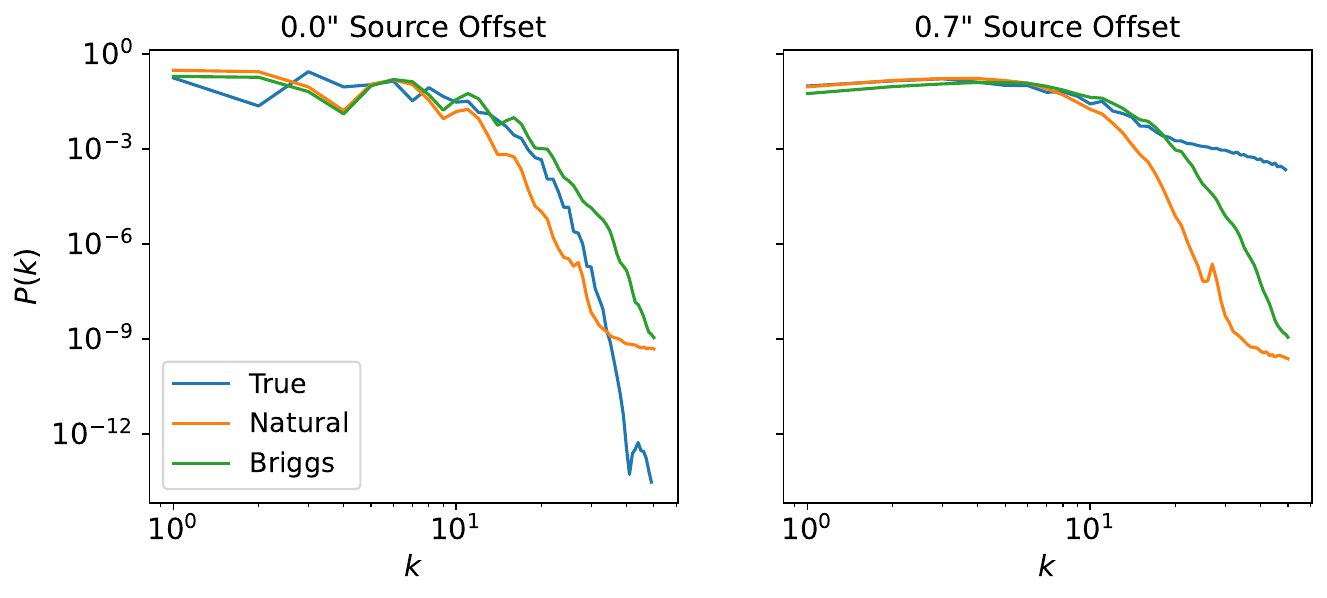}
    \caption{A comparison between the Fourier radial profiles of the simulated true sky image and the images created with Briggs and natural weighting for the two extremes in source offset position with the SIE cusp configuration.}
    \label{fig:compare_spectra}
\end{figure*}

Inspecting the Fourier radial profiles of the simulated true sky surface brightness lensed images reveals an abrupt transition in their form as a function of source position offset. As can be seen in Figure \ref{fig:power_spectra}, the profiles fall into two camps. For small source offsets ($x \leq 0.3$) the profiles are virtually identical, whilst for source offsets greater than $x \geq 0.4$ we begin to see an increasing Fourier amplitude at large $k$ numbers. This divide occurs at approximately the point where any source emission connecting the multiple images of the background source becomes statistically insignificant, and thus the image becomes dominated by smaller structure. From this, it is natural to suspect that the lower angular resolution achieved by the \texttt{CLEAN}ed images, particularly with natural weighting, is to blame for the relatively poor performance in recovery of lens model parameters at larger source offsets. 

In Figure \ref{fig:compare_spectra} we compare the Fourier radial profiles of the simulated true sky surface brightness images with those of the \texttt{CLEAN}ed images  for the two extremes of source offset in the SIE cusp configuration. With the source on the optical axis, the naturally weighted image underestimates the Fourier amplitude for $10 \leq k \leq 40$, whereas the Briggs-weighted image overestimates the Fourier amplitude in the same region. As we move the source to its extreme offset, the Fourier radial profile of the true sky image has a much shallower decline at high $k$-numbers. This serves to dramatically increase the amount by which the naturally weighted image underestimates the Fourier amplitude at these high frequencies. In contrast, the Briggs-weighted image is capable of reproducing this higher frequency structure out to larger values of $k$, but ultimately does underestimate the Fourier amplitude at the highest frequencies where the effects of noise begin to dominate.

We define the 'excess Fourier amplitude', $\epsilon$, as the difference between the integrated Fourier radial profile of the simulated true sky surface brightness image and that of the corresponding Briggs or naturally weighted \texttt{CLEAN}ed image above a threshold value of $k = 10$,
\begin{equation}
    \epsilon = \int^{50}_{10} P_C(k) dk - \int^{50}_{10} P_T(k) dk,
    \label{eq:excess}
\end{equation}
where $P_C$ and $P_T$ are the Fourier radial profiles of the \texttt{CLEAN}ed image and true surface brightness image respectively. The threshold value of $k=10$ was chosen as a reasonable approximation to where the Fourier radial profiles in all cases begin to diverge significantly from that of the true sky surface brightness image. We then use this value of $\epsilon$ to explore the biases in our lens modelling results as a function of source offset and \texttt{CLEAN}ing method.

\begin{figure*}
\centering
  \begin{tabular}{cc}
    \includegraphics[width=\columnwidth]{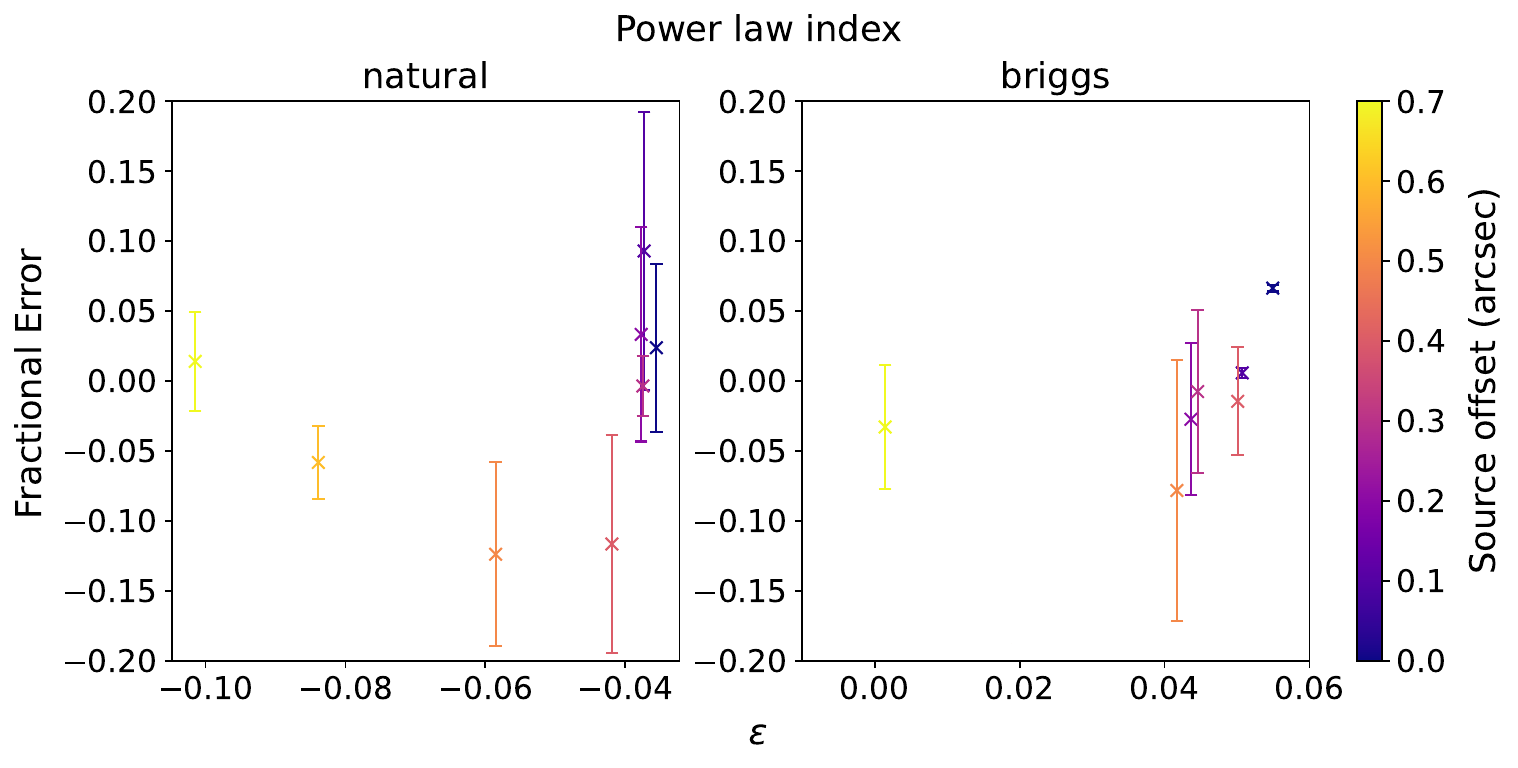} &
    \includegraphics[width=\columnwidth]{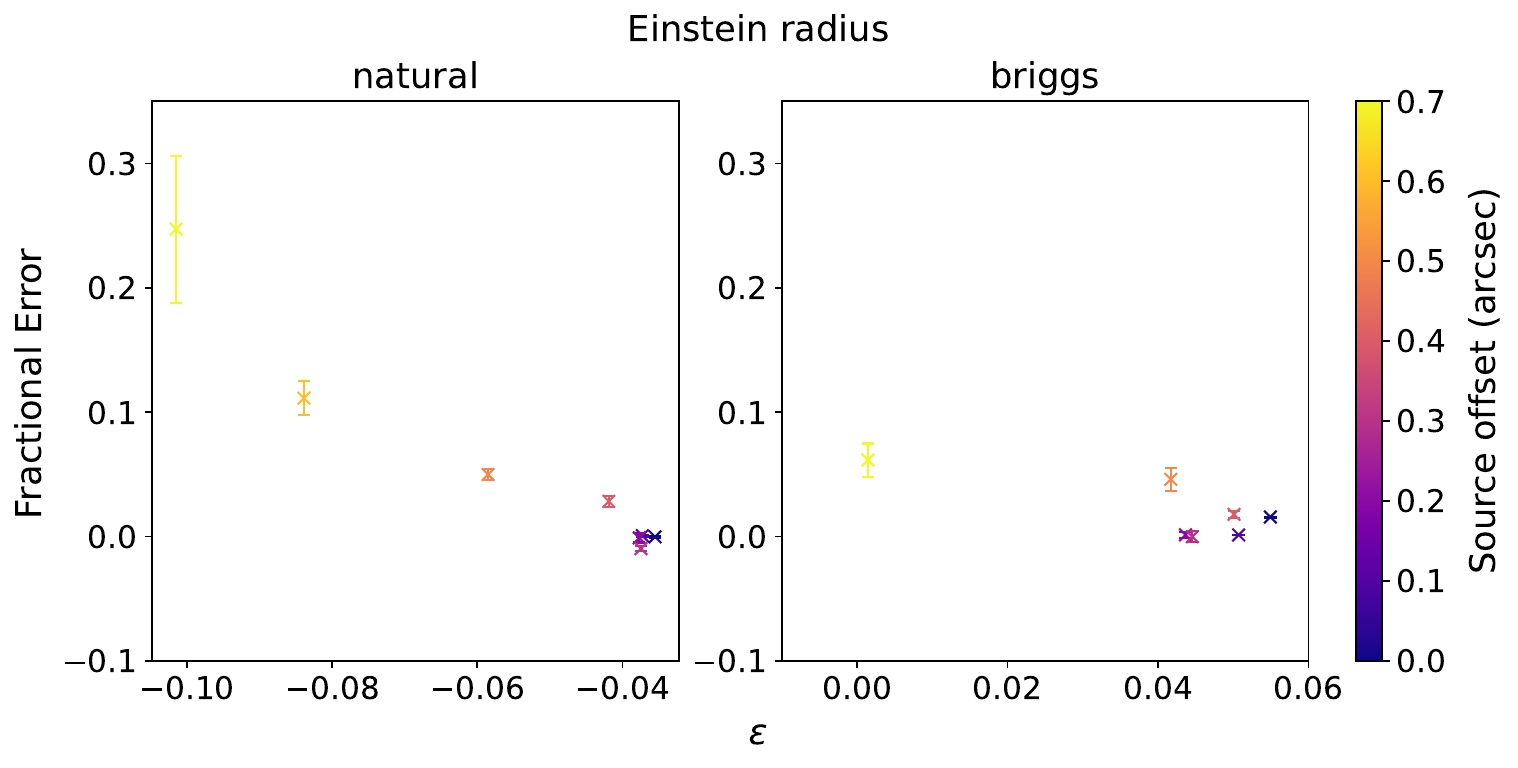} \\
    \includegraphics[width=\columnwidth]{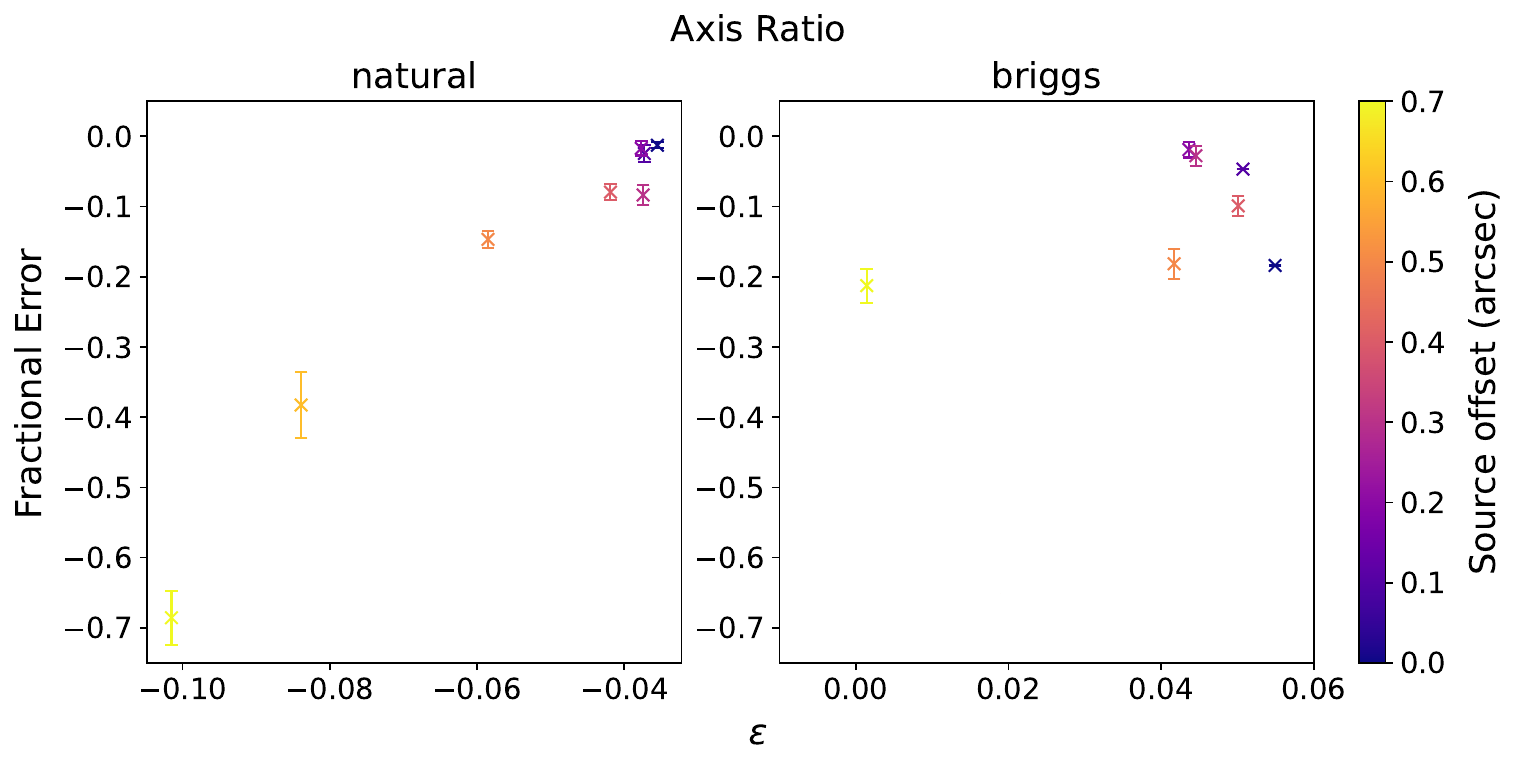} \\
  \end{tabular}
  \caption{Fractional error in SIS lens model parameters as a function of the excess Fourier amplitude, $\epsilon$, as defined in equation (\ref{eq:excess}). Plotted side by side are the same quantities for the naturally and Briggs-weighted images created from the 6000s integration time ALMA observations. The colour bar indicates the source offset in arcseconds.}
  \label{fig:power_excess_SIS}
\end{figure*}

The fractional errors on the key SIS lens model parameters as a function of excess Fourier amplitude, for the 6000s integration time ALMA cycle 7.3 observations are shown in Figure \ref{fig:power_excess_SIS}. There are several trends in these plots that can explain the behaviour seen in the modelling results. Notably, unlike the Briggs-weighted images, the naturally weighted images show a neat correlation between excess Fourier amplitude and source offset. The more negative values of $\epsilon$ at larger values of source offset indicate a progressive loss of signal on small scales in the \texttt{CLEAN}ed naturally weighted data compared to the true sky surface brightness images. Particularly in the cases of Einstein radius and axis ratio, these more negative excess Fourier amplitudes result in higher fractional errors. This suggests that the naturally weighted images are more susceptible to the degeneracy between $\theta_{\rm E}$ and $q$. Conversely, the excess Fourier amplitude for the Briggs weighted images decreases towards zero at larger source offsets, indicating a similar amount of signal on small scales compared to the true sky surface brightness image. Accordingly, the fractional errors on $\theta_{\rm E}$ and $q$ for the Briggs weighted images show a much weaker trend with source offset than that exhibited by the naturally weighted images, a reflection of the typically more robust modelling results obtained with the Briggs-weighted images.

\begin{figure*}
\centering
  \begin{tabular}{cc}
    \includegraphics[width=\columnwidth]{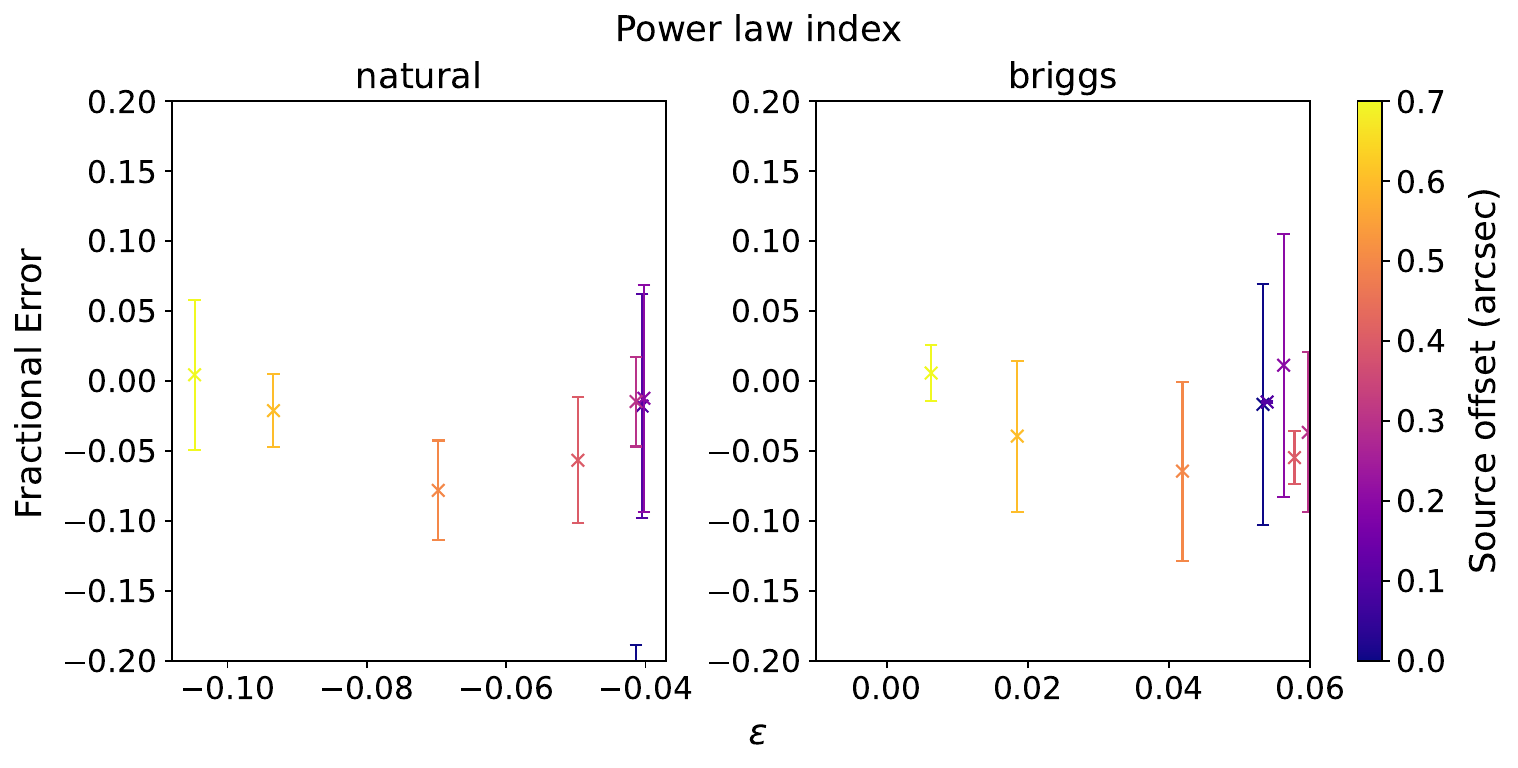} &
    \includegraphics[width=\columnwidth]{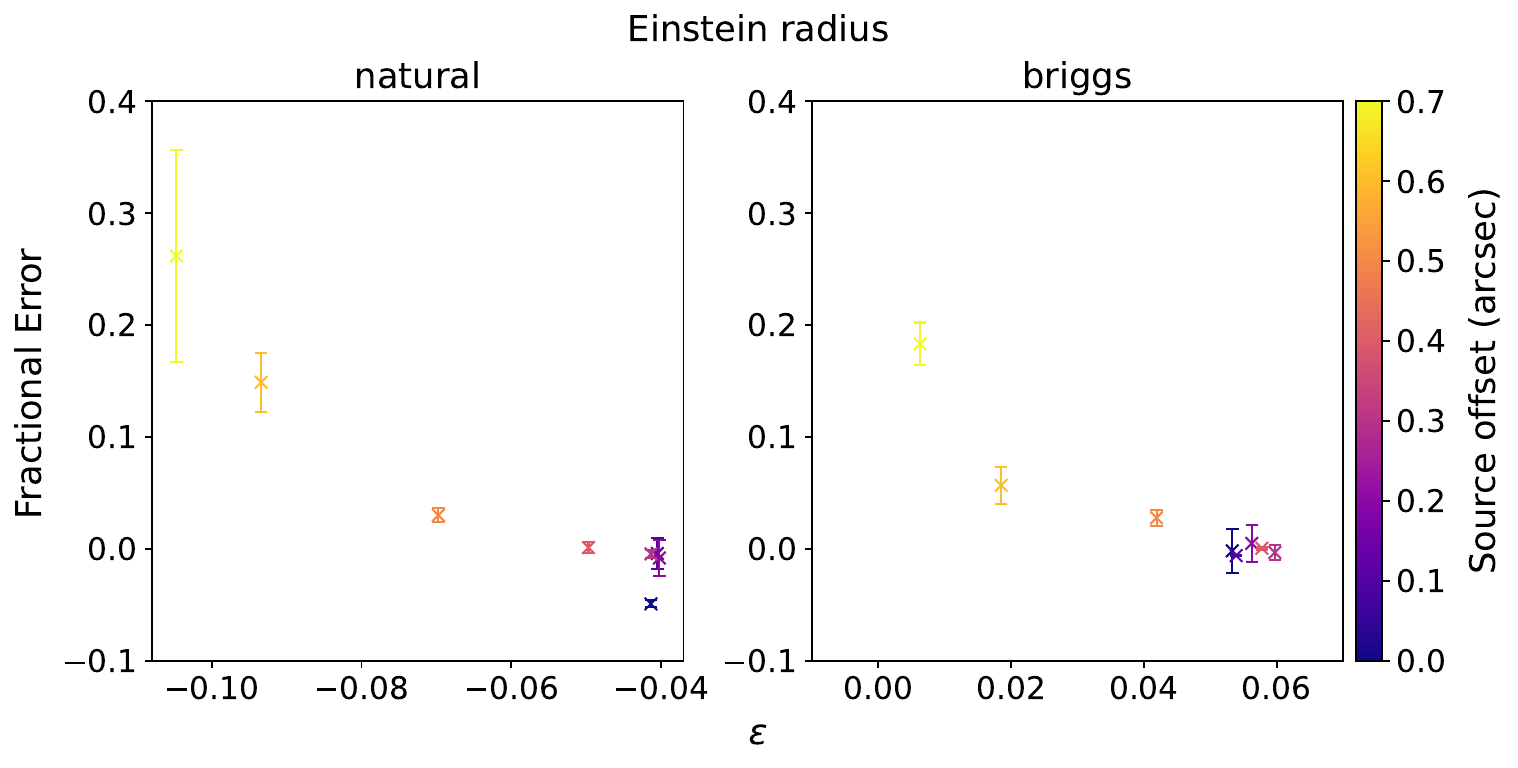} \\
    \includegraphics[width=\columnwidth]{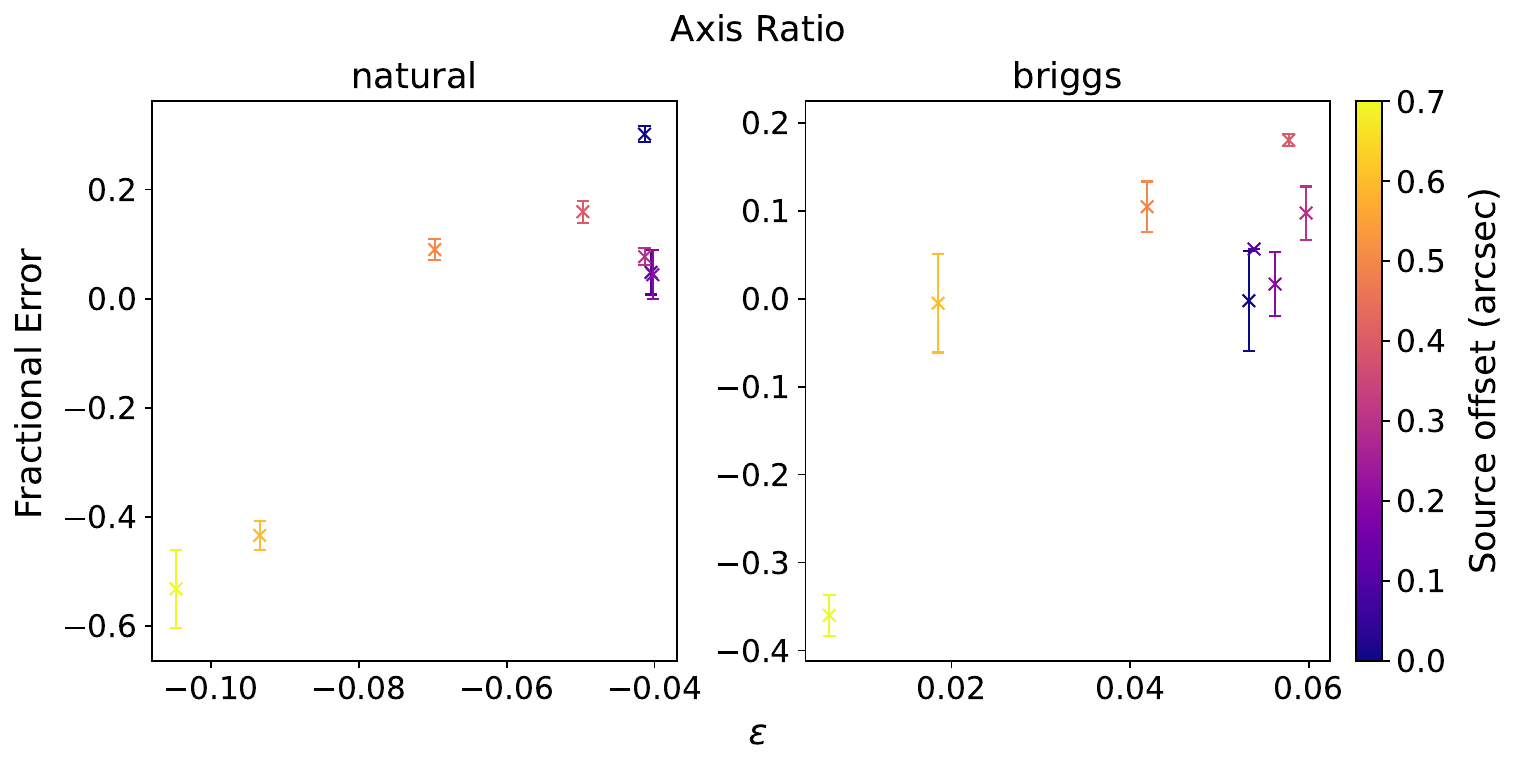} &
    \includegraphics[width=\columnwidth]{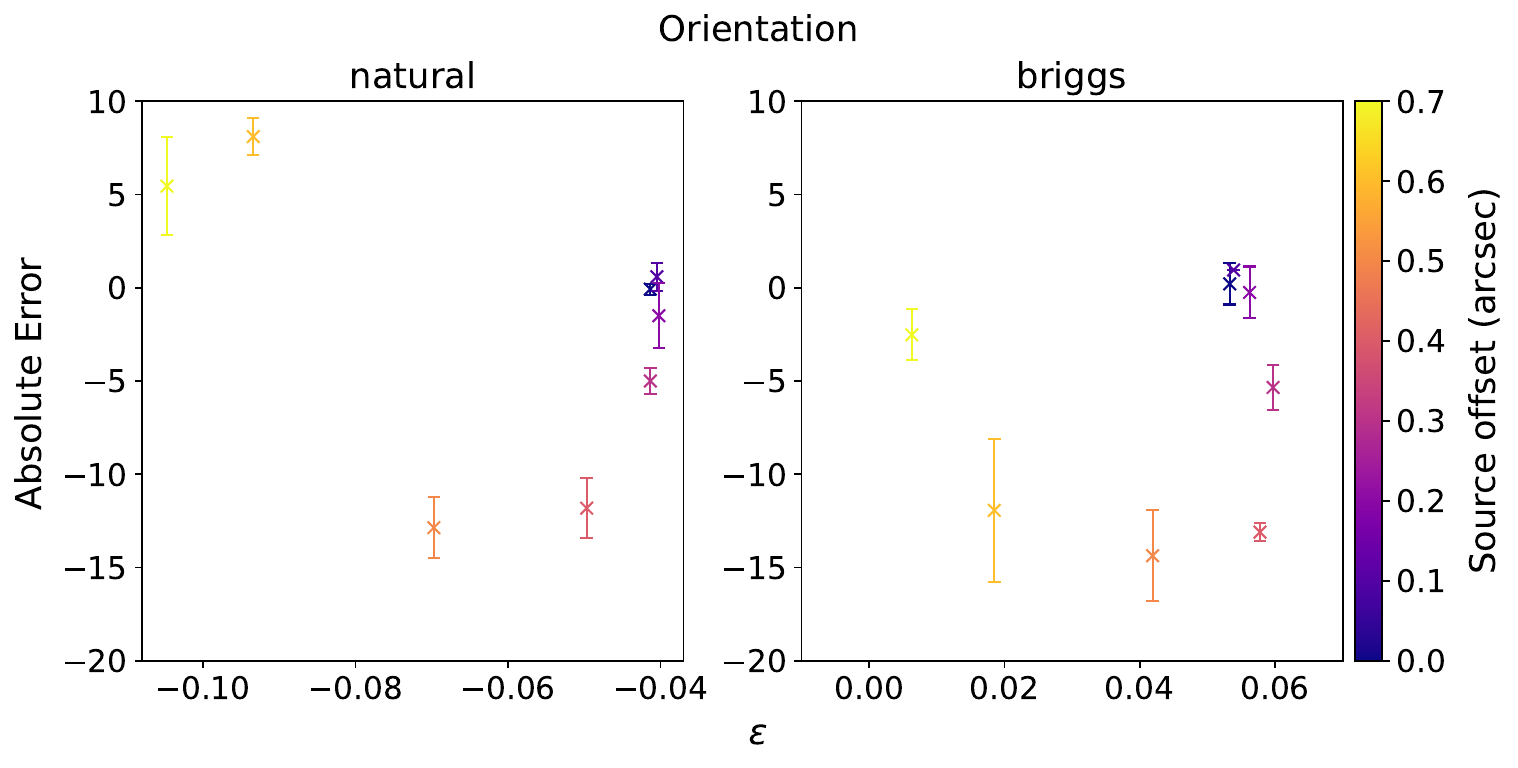} \\
  \end{tabular}
  \caption{Fractional error in SIE lens model (fold configuration) parameters as a function of the excess Fourier amplitude, $\epsilon$, as defined in equation (\ref{eq:excess}). Plotted side by side are the same quantities for the naturally and Briggs-weighted images created from the 6000s integration time ALMA observations. The colour bar indicates the source offset in arcseconds.}
  \label{fig:power_excess_SIE_fold}
\end{figure*}

The fractional errors of the SIE lens model parameters (fold configuration) as a function of excess Fourier amplitude, for the 6000s integration time ALMA observations, are shown in Figure \ref{fig:power_excess_SIE_fold}. We find the same correlation between $\epsilon$ and source offset with the naturally weighted \texttt{CLEAN}ed images and the same lack thereof with the Briggs images. We also find that the fractional errors on $\theta_{\rm E}$ and $q$ for natural weighting become higher with more negative values of $\epsilon$ at larger source offsets as the \texttt{CLEAN}ed images deviate more on small spatial scales from the true sky brightness images. Similarly, the same trend toward zero excess Fourier amplitude at larger source offsets is seen with the Briggs-weighted images, where again, the fractional errors are less severe than seen with the naturally weighted images. Recovery of the orientation is comparable between the two different \texttt{CLEAN}ed data types, and there is no clear correlation between error and $\epsilon$.

\begin{figure*}
\centering
  \begin{tabular}{cc}
    \includegraphics[width=\columnwidth]{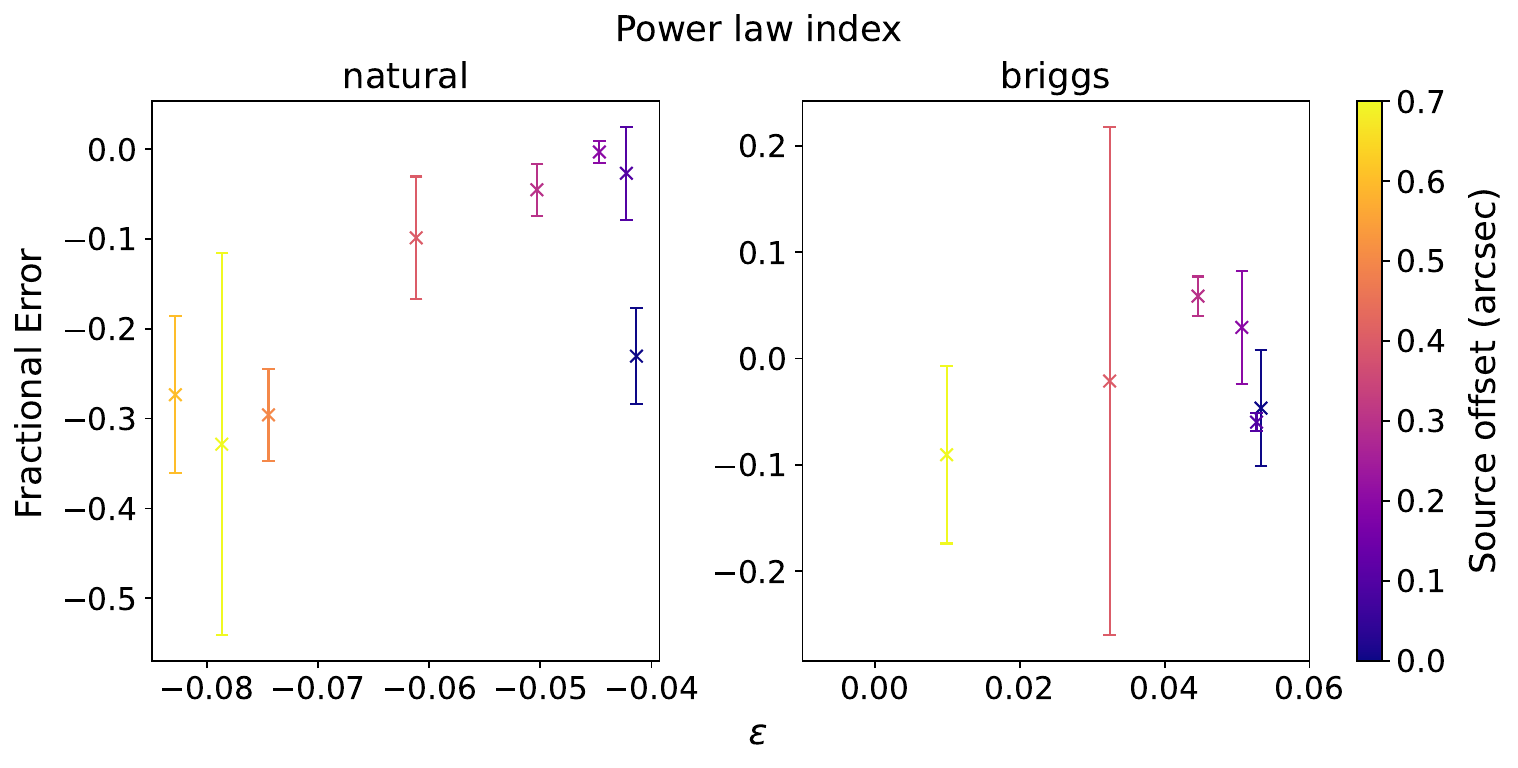} &
    \includegraphics[width=\columnwidth]{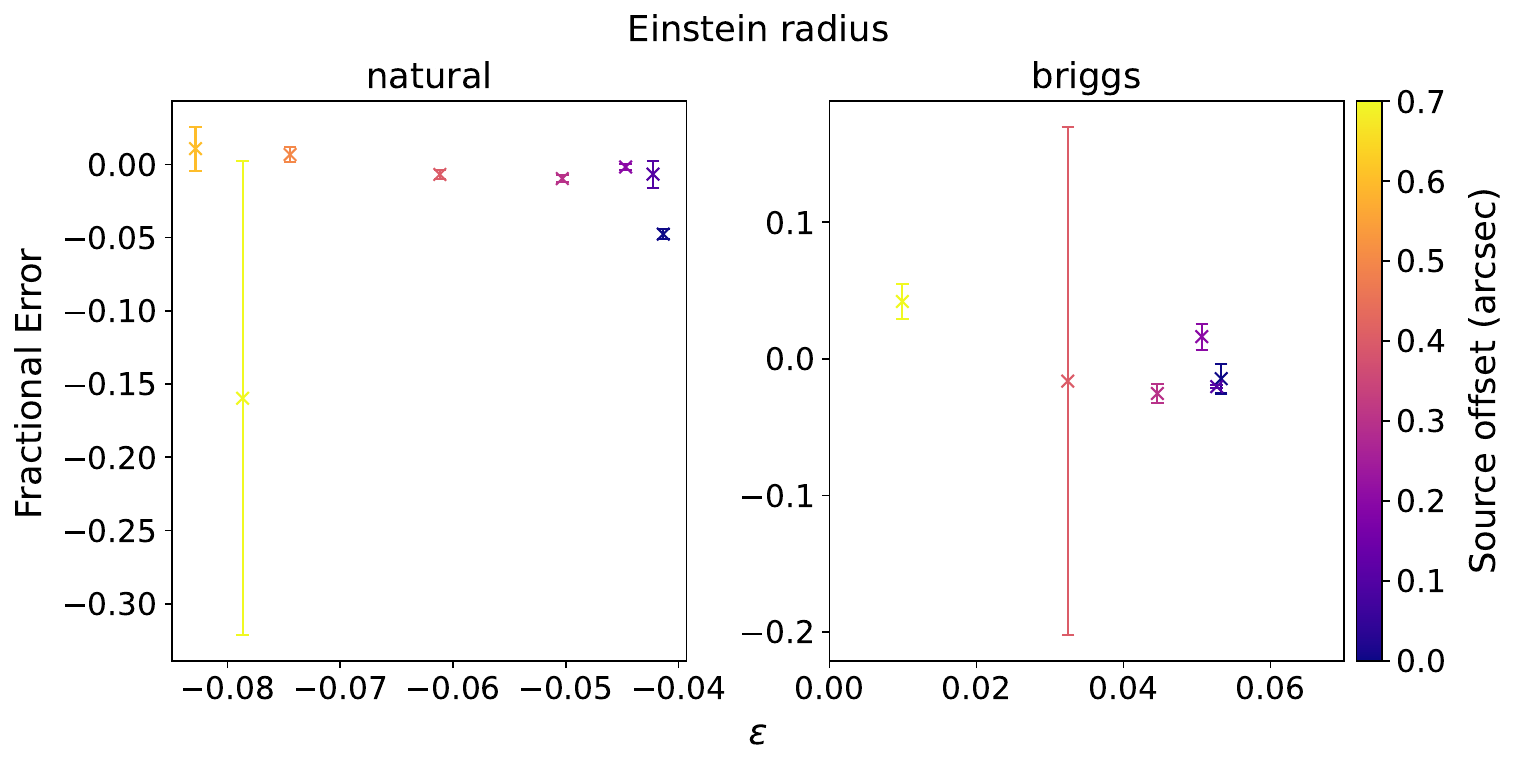} \\
    \includegraphics[width=\columnwidth]{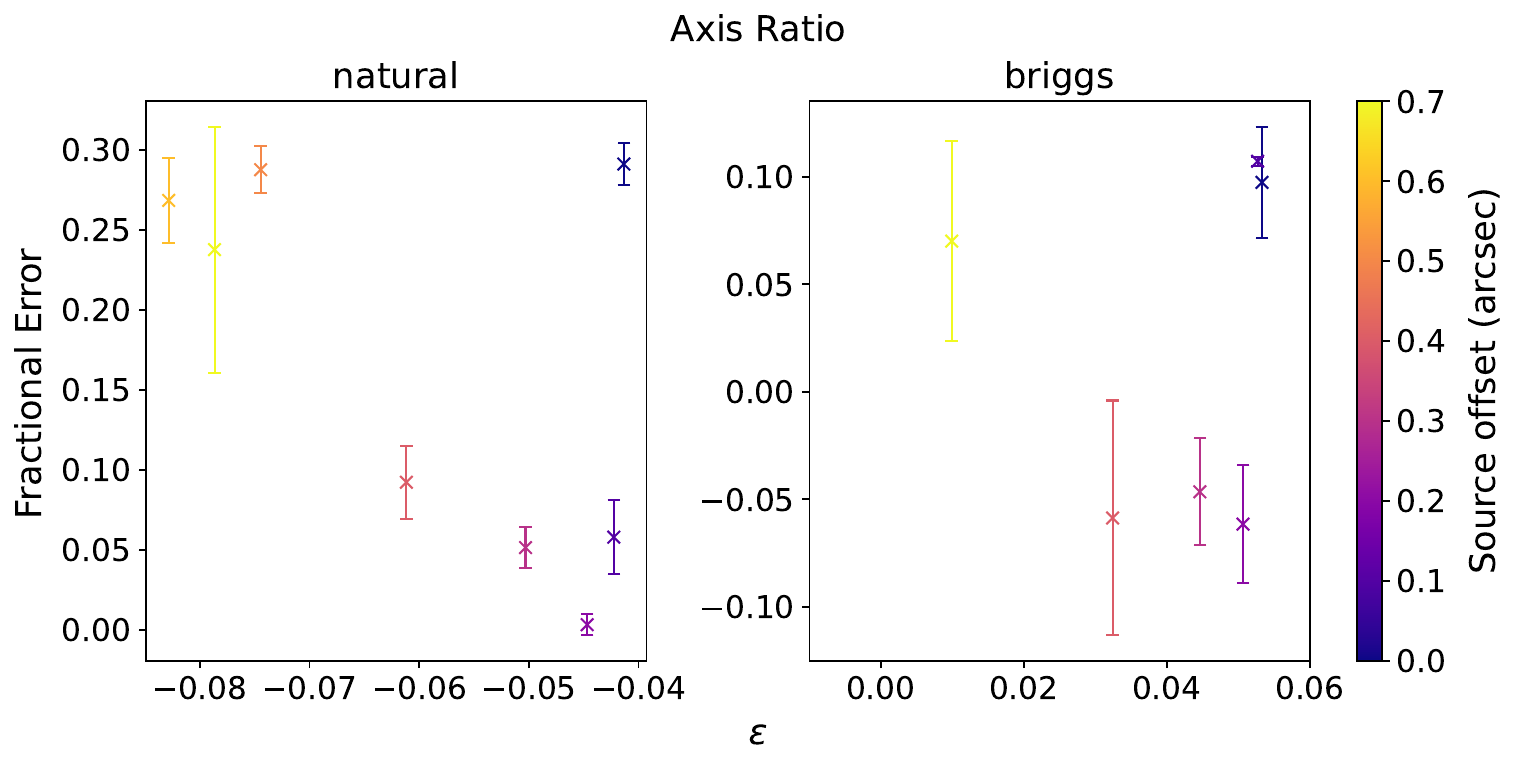} &
    \includegraphics[width=\columnwidth]{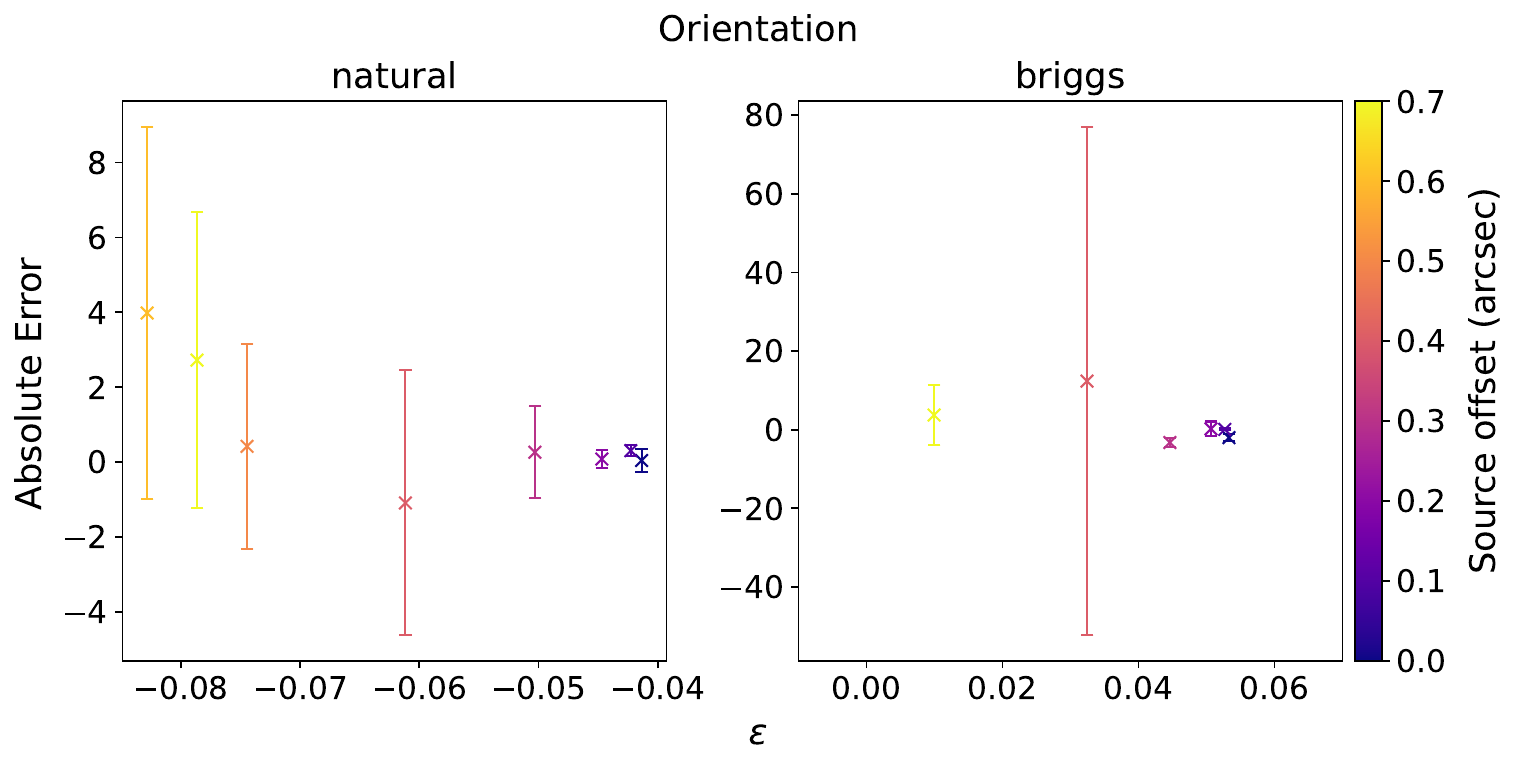} \\
  \end{tabular}
  \caption{Fractional error in SIE lens model (cusp configuration) parameters as a function of the excess Fourier amplitude, $\epsilon$, as defined in equation (\ref{eq:excess}). Plotted side by side are the same quantities for the naturally and Briggs-weighted images created from the 6000s integration time ALMA observations. The colour bar indicates the source offset in arcseconds.}
  \label{fig:power_excess_SIE_cusp}
\end{figure*}

The fractional error of the SIE lens model parameters (cusp configuration) as a function of excess Fourier amplitude, for the 6000s integration time ALMA observations, are shown in Figure \ref{fig:power_excess_SIE_cusp}. The results show similar trends to those observed in the cases of the SIS and SIE fold configuration with a few notable differences. Now the Briggs-weighted results show a clear correlation between $\epsilon$ and source offset, although again, generally, the Briggs fractional errors tend to correlate much less strongly with $\epsilon$ as seen previously. Also, there is a stronger correlation between $\epsilon$ and the bias in $\alpha$ for the naturally weighted images than seen with the SIS or SIE fold configurations, little to no correlation between $\epsilon$ and $\theta_{\rm E}$ with either \texttt{CLEAN}ed dataset and a trend in the naturally \texttt{CLEAN}ed data for $q$ to anti-correlate with $\epsilon$.

\section{Conclusions}
\label{sec:conclusions}

In this paper, we have introduced a set of realistic simulated ALMA observations, created using the \texttt{simobserve} task within \texttt{CASA}. From these simulated visibility sets, we have used the \texttt{tclean} algorithm to produce images, both with natural weighting and Briggs weighting. With these data products, we have been able to explore a number of different questions related to lens modelling methodology: How does time-binning the visibility data affect our lens model accuracy? Does the weighting scheme used in \texttt{CLEAN}ing introduce biases into our lens model? And, is the extra computational cost of performing lens modelling in the uv-plane worthwhile?

We have shown that, at least for the simple parametric source considered in this paper, significant time-binning of interferometric visibilities can be used without negatively impacting recovery of the lens model parameters. We have compared modelling of all visibilities against time-binning to reduce the number of visibilities by a factor of two and three and found that the results were all consistent with one another. We therefore conclude that the computational benefits of larger time-bins clearly outweigh any potential cost in modelling accuracy. 

Our lens modelling results indicate that the images \texttt{CLEAN}ed with Briggs weighting, which tends to leave smaller scale features intact compared to \texttt{CLEAN}ing with natural weighting, lead to slightly smaller biases on inferred lens model parameters. At large source offsets from the lens centroid, both \texttt{CLEAN}ing schemes lead to stronger biases at larger source offsets where there is less signal in the lensed image, but this effect is slightly more pronounced for the naturally weighted images. With both \texttt{CLEAN}ing schemes, increasing resolution with longer interferometric baselines lessens this effect. 

Direct modelling of the visibilities outperforms modelling of both variants of the \texttt{CLEAN}ed images considered in this work; whilst the precision of the visibility modelling (as measured from the width of the posterior distribution of the lens model parameters) is comparable to the image-plane modelling, its accuracy is consistently superior. In combination with the higher resolution (0.09 $\arcsec$ beam size) simulated ALMA data considered here, direct visibility modelling repeatedly recovers elliptical power-law lens model parameters to within a few per cent.

The on-source exposure times of 60s, 600s and 6000s investigated in this work had little impact on the accuracy and only a mild impact on the precision of our modelling results. 

In summary, whilst the computational cost of modelling the visibilities is higher, it is justified by its accuracy. This can be mitigated by time-binning visibilities, and we have shown that binning to the extent that the number of visibilities is reduced by a factor of 3 has an insignificant effect on inferred lens model parameters.

There is significant scope for further exploring the ideas broached in this work. We have, for example, only considered lensed images produced using the same simple parametric source, not more or less compact ones, nor more complex ones. Neither have we explored the effects of \texttt{CLEAN}ing on the properties of the recovered source. We leave further exploration of these questions for future work.

\section*{Acknowledgements}

JM acknowledges support from the Science and Technology Facilities Council (STFC; grant ID: ST/S505602/1). SD acknowledges support from STFC (grant ID: ST/X000982/1).

\section*{Data Availability}

The simulated data from this work can be made available upon request to the corresponding author.



\bibliographystyle{mnras}
\bibliography{paper}

\begin{thebibliography}{}
\makeatletter
\relax
\def\mn@urlcharsother{\let\do\@makeother \do\$\do\&\do\#\do\^\do\_\do\%\do\~}
\def\mn@doi{\begingroup\mn@urlcharsother \@ifnextchar [ {\mn@doi@}
  {\mn@doi@[]}}
\def\mn@doi@[#1]#2{\def\@tempa{#1}\ifx\@tempa\@empty \href
  {http://dx.doi.org/#2} {doi:#2}\else \href {http://dx.doi.org/#2} {#1}\fi
  \endgroup}
\def\mn@eprint#1#2{\mn@eprint@#1:#2::\@nil}
\def\mn@eprint@arXiv#1{\href {http://arxiv.org/abs/#1} {{\tt arXiv:#1}}}
\def\mn@eprint@dblp#1{\href {http://dblp.uni-trier.de/rec/bibtex/#1.xml}
  {dblp:#1}}
\def\mn@eprint@#1:#2:#3:#4\@nil{\def\@tempa {#1}\def\@tempb {#2}\def\@tempc
  {#3}\ifx \@tempc \@empty \let \@tempc \@tempb \let \@tempb \@tempa \fi \ifx
  \@tempb \@empty \def\@tempb {arXiv}\fi \@ifundefined
  {mn@eprint@\@tempb}{\@tempb:\@tempc}{\expandafter \expandafter \csname
  mn@eprint@\@tempb\endcsname \expandafter{\@tempc}}}

\bibitem[\protect\citeauthoryear{{Ballard}, {Enzi}, {Collett}, {Turner}  \&
  {Smith}}{{Ballard} et~al.}{2023}]{ballard2023}
{Ballard} D.~J.,  {Enzi} W. J.~R.,  {Collett} T.~E.,  {Turner} H.~C.,   {Smith}
  R.~J.,  2023, \mn@doi [arXiv e-prints] {10.48550/arXiv.2309.04535}, \href
  {https://ui.adsabs.harvard.edu/abs/2023arXiv230904535B} {p. arXiv:2309.04535}

\bibitem[\protect\citeauthoryear{{Berta} et~al.,}{{Berta}
  et~al.}{2021}]{berta2021}
{Berta} S.,  et~al., 2021, \mn@doi [\aap] {10.1051/0004-6361/202039743}, \href
  {https://ui.adsabs.harvard.edu/abs/2021A&A...646A.122B} {646, A122}

\bibitem[\protect\citeauthoryear{{Birrer}, {Amara}  \& {Refregier}}{{Birrer}
  et~al.}{2017}]{birrer2017}
{Birrer} S.,  {Amara} A.,   {Refregier} A.,  2017, \mn@doi [\jcap]
  {10.1088/1475-7516/2017/05/037}, \href
  {https://ui.adsabs.harvard.edu/abs/2017JCAP...05..037B} {2017, 037}

\bibitem[\protect\citeauthoryear{{Birrer}, {Millon}, {Sluse}, {Shajib},
  {Courbin}, {Koopmans}, {Suyu}  \& {Treu}}{{Birrer} et~al.}{2022}]{birrer2022}
{Birrer} S.,  {Millon} M.,  {Sluse} D.,  {Shajib} A.~J.,  {Courbin} F.,
  {Koopmans} L.~V.~E.,  {Suyu} S.~H.,   {Treu} T.,  2022, \mn@doi [arXiv
  e-prints] {10.48550/arXiv.2210.10833}, \href
  {https://ui.adsabs.harvard.edu/abs/2022arXiv221010833B} {p. arXiv:2210.10833}

\bibitem[\protect\citeauthoryear{{Bouwens}, {Illingworth}, {Ellis}, {Oesch}  \&
  {Stefanon}}{{Bouwens} et~al.}{2022}]{bouwens2022}
{Bouwens} R.~J.,  {Illingworth} G.,  {Ellis} R.~S.,  {Oesch} P.,   {Stefanon}
  M.,  2022, \mn@doi [\apj] {10.3847/1538-4357/ac86d1}, \href
  {https://ui.adsabs.harvard.edu/abs/2022ApJ...940...55B} {940, 55}

\bibitem[\protect\citeauthoryear{{Briggs}}{{Briggs}}{1995}]{briggs}
{Briggs} D.~S.,  1995, PhD thesis, New Mexico Institute of Mining and
  Technology, United States

\bibitem[\protect\citeauthoryear{{Collett} \& {Auger}}{{Collett} \&
  {Auger}}{2014}]{collett2014}
{Collett} T.~E.,  {Auger} M.~W.,  2014, \mn@doi [\mnras]
  {10.1093/mnras/stu1190}, \href
  {https://ui.adsabs.harvard.edu/abs/2014MNRAS.443..969C} {443, 969}

\bibitem[\protect\citeauthoryear{{Collett} \& {Smith}}{{Collett} \&
  {Smith}}{2020}]{collett2020}
{Collett} T.~E.,  {Smith} R.~J.,  2020, \mn@doi [\mnras]
  {10.1093/mnras/staa1804}, \href
  {https://ui.adsabs.harvard.edu/abs/2020MNRAS.497.1654C} {497, 1654}

\bibitem[\protect\citeauthoryear{{Collett} et~al.,}{{Collett}
  et~al.}{2018}]{collett2018}
{Collett} T.~E.,  et~al., 2018, \mn@doi [Science] {10.1126/science.aao2469},
  \href {https://ui.adsabs.harvard.edu/abs/2018Sci...360.1342C} {360, 1342}

\bibitem[\protect\citeauthoryear{{Dye} et~al.,}{{Dye} et~al.}{2014}]{dye2014}
{Dye} S.,  et~al., 2014, \mn@doi [\mnras] {10.1093/mnras/stu305}, \href
  {https://ui.adsabs.harvard.edu/abs/2014MNRAS.440.2013D} {440, 2013}

\bibitem[\protect\citeauthoryear{Dye et~al.,}{Dye et~al.}{2015}]{dye_2015}
Dye S.,  et~al., 2015, \mn@doi [Monthly Notices of the Royal Astronomical
  Society] {10.1093/mnras/stv1442}, 452, 2258

\bibitem[\protect\citeauthoryear{Dye et~al.,}{Dye et~al.}{2018}]{dye2018}
Dye S.,  et~al., 2018, \mn@doi [Monthly Notices of the Royal Astronomical
  Society] {10.1093/mnras/sty513}, 476, 4383

\bibitem[\protect\citeauthoryear{{Dye} et~al.,}{{Dye} et~al.}{2022}]{id141}
{Dye} S.,  et~al., 2022, \mn@doi [\mnras] {10.1093/mnras/stab3569}, \href
  {https://ui.adsabs.harvard.edu/abs/2022MNRAS.510.3734D} {510, 3734}

\bibitem[\protect\citeauthoryear{{Etherington} et~al.,}{{Etherington}
  et~al.}{2022}]{etherington2022}
{Etherington} A.,  et~al., 2022, \mn@doi [\mnras] {10.1093/mnras/stac2639},
  \href {https://ui.adsabs.harvard.edu/abs/2022MNRAS.517.3275E} {517, 3275}

\bibitem[\protect\citeauthoryear{{Etherington} et~al.,}{{Etherington}
  et~al.}{2023}]{etherington2023}
{Etherington} A.,  et~al., 2023, \mn@doi [\mnras] {10.1093/mnras/stad582},
  \href {https://ui.adsabs.harvard.edu/abs/2023MNRAS.521.6005E} {521, 6005}

\bibitem[\protect\citeauthoryear{{Geach}, {Ivison}, {Dye}  \& {Oteo}}{{Geach}
  et~al.}{2018}]{geach2018}
{Geach} J.~E.,  {Ivison} R.~J.,  {Dye} S.,   {Oteo} I.,  2018, \mn@doi [\apjl]
  {10.3847/2041-8213/aae375}, \href
  {https://ui.adsabs.harvard.edu/abs/2018ApJ...866L..12G} {866, L12}

\bibitem[\protect\citeauthoryear{{He} et~al.,}{{He} et~al.}{2022}]{he2022}
{He} Q.,  et~al., 2022, \mn@doi [\mnras] {10.1093/mnras/stac191}, \href
  {https://ui.adsabs.harvard.edu/abs/2022MNRAS.511.3046H} {511, 3046}

\bibitem[\protect\citeauthoryear{{He} et~al.,}{{He} et~al.}{2023}]{he2023}
{He} Q.,  et~al., 2023, \mn@doi [\mnras] {10.1093/mnras/stac2779}, \href
  {https://ui.adsabs.harvard.edu/abs/2023MNRAS.518..220H} {518, 220}

\bibitem[\protect\citeauthoryear{{H{\"o}gbom}}{{H{\"o}gbom}}{1974}]{hogbom1974}
{H{\"o}gbom} J.~A.,  1974, \aaps, \href
  {https://ui.adsabs.harvard.edu/abs/1974A&AS...15..417H} {15, 417}

\bibitem[\protect\citeauthoryear{{Li}, {Becker}  \& {Dye}}{{Li}
  et~al.}{2021}]{li2021}
{Li} N.,  {Becker} C.,   {Dye} S.,  2021, \mn@doi [\mnras]
  {10.1093/mnras/stab984}, \href
  {https://ui.adsabs.harvard.edu/abs/2021MNRAS.504.2224L} {504, 2224}

\bibitem[\protect\citeauthoryear{Maresca, Dye  \& Li}{Maresca
  et~al.}{2020}]{maresca2020auto}
Maresca J.,  Dye S.,   Li N.,  2020, arXiv preprint arXiv:2012.04665

\bibitem[\protect\citeauthoryear{Maresca et~al.,}{Maresca
  et~al.}{2022}]{maresca}
Maresca J.,  et~al., 2022, \mn@doi [Monthly Notices of the Royal Astronomical
  Society] {10.1093/mnras/stac585}

\bibitem[\protect\citeauthoryear{{Melo-Carneiro}, {Furlanetto}  \&
  {Chies-Santos}}{{Melo-Carneiro} et~al.}{2023}]{melo2023}
{Melo-Carneiro} C.~R.,  {Furlanetto} C.,   {Chies-Santos} A.~L.,  2023, \mn@doi
  [\mnras] {10.1093/mnras/stad162}, \href
  {https://ui.adsabs.harvard.edu/abs/2023MNRAS.520.1613M} {520, 1613}

\bibitem[\protect\citeauthoryear{{Minor}, {Gad-Nasr}, {Kaplinghat}  \&
  {Vegetti}}{{Minor} et~al.}{2021}]{minor2021}
{Minor} Q.,  {Gad-Nasr} S.,  {Kaplinghat} M.,   {Vegetti} S.,  2021, \mn@doi
  [\mnras] {10.1093/mnras/stab2247}, \href
  {https://ui.adsabs.harvard.edu/abs/2021MNRAS.507.1662M} {507, 1662}

\bibitem[\protect\citeauthoryear{{Mukherjee}, {Koopmans}, {Metcalf}, {Tortora},
  {Schaller}, {Schaye}, {Vernardos}  \& {Bellagamba}}{{Mukherjee}
  et~al.}{2021}]{mukherjee2021}
{Mukherjee} S.,  {Koopmans} L. V.~E.,  {Metcalf} R.~B.,  {Tortora} C.,
  {Schaller} M.,  {Schaye} J.,  {Vernardos} G.,   {Bellagamba} F.,  2021,
  \mn@doi [\mnras] {10.1093/mnras/stab693}, \href
  {https://ui.adsabs.harvard.edu/abs/2021MNRAS.504.3455M} {504, 3455}

\bibitem[\protect\citeauthoryear{Nightingale \& Dye}{Nightingale \&
  Dye}{2015}]{Nightingale2015}
Nightingale J.~W.,  Dye S.,  2015, \mn@doi [Monthly Notices of the Royal
  Astronomical Society] {10.1093/mnras/stv1455}, 452, 2940

\bibitem[\protect\citeauthoryear{Nightingale \& Hayes}{Nightingale \&
  Hayes}{2020}]{pyautolens}
Nightingale J.,  Hayes R.,  2020, PyAutoLens: Open-source Strong Gravitational
  Lensing, \url {https://github.com/Jammy2211/PyAutoLens}

\bibitem[\protect\citeauthoryear{Nightingale, Dye  \& Massey}{Nightingale
  et~al.}{2018}]{Nightingale2018}
Nightingale J.~W.,  Dye S.,   Massey R.~J.,  2018, \mn@doi [Monthly Notices of
  the Royal Astronomical Society] {10.1093/mnras/sty1264}, 478, 4738

\bibitem[\protect\citeauthoryear{Nightingale et~al.,}{Nightingale
  et~al.}{2021}]{Nightingale2021}
Nightingale J.~W.,  et~al., 2021, \mn@doi [Journal of Open Source Software]
  {10.21105/joss.02825}, 6, 2825

\bibitem[\protect\citeauthoryear{{Planck Collaboration} et~al.,}{{Planck
  Collaboration} et~al.}{2016}]{planck_2015}
{Planck Collaboration} et~al., 2016, \mn@doi [\aap]
  {10.1051/0004-6361/201525830}, \href
  {https://ui.adsabs.harvard.edu/abs/2016A&A...594A..13P} {594, A13}

\bibitem[\protect\citeauthoryear{Powell, Vegetti, McKean, Spingola, Rizzo  \&
  Stacey}{Powell et~al.}{2020}]{powell}
Powell D.,  Vegetti S.,  McKean J.~P.,  Spingola C.,  Rizzo F.,   Stacey H.~R.,
   2020, \mn@doi [Monthly Notices of the Royal Astronomical Society]
  {10.1093/mnras/staa2740}, 501, 515

\bibitem[\protect\citeauthoryear{{Powell}, {Vegetti}, {McKean}, {Spingola},
  {Rizzo}  \& {Stacey}}{{Powell} et~al.}{2021}]{powell2021}
{Powell} D.,  {Vegetti} S.,  {McKean} J.~P.,  {Spingola} C.,  {Rizzo} F.,
  {Stacey} H.~R.,  2021, \mn@doi [\mnras] {10.1093/mnras/staa2740}, \href
  {https://ui.adsabs.harvard.edu/abs/2021MNRAS.501..515P} {501, 515}

\bibitem[\protect\citeauthoryear{{Remus}, {Dolag}, {Naab}, {Burkert},
  {Hirschmann}, {Hoffmann}  \& {Johansson}}{{Remus} et~al.}{2017}]{remus2017}
{Remus} R.-S.,  {Dolag} K.,  {Naab} T.,  {Burkert} A.,  {Hirschmann} M.,
  {Hoffmann} T.~L.,   {Johansson} P.~H.,  2017, \mn@doi [\mnras]
  {10.1093/mnras/stw2594}, \href
  {https://ui.adsabs.harvard.edu/abs/2017MNRAS.464.3742R} {464, 3742}

\bibitem[\protect\citeauthoryear{Rybak, Vegetti, McKean, Andreani  \&
  White}{Rybak et~al.}{2015}]{rybak_2015}
Rybak M.,  Vegetti S.,  McKean J.~P.,  Andreani P.,   White S. D.~M.,  2015,
  \mn@doi [Monthly Notices of the Royal Astronomical Society: Letters]
  {10.1093/mnrasl/slv092}, 453, L26

\bibitem[\protect\citeauthoryear{{Salmon} et~al.,}{{Salmon}
  et~al.}{2020}]{salmon2020}
{Salmon} B.,  et~al., 2020, \mn@doi [\apj] {10.3847/1538-4357/ab5a8b}, \href
  {https://ui.adsabs.harvard.edu/abs/2020ApJ...889..189S} {889, 189}

\bibitem[\protect\citeauthoryear{{Shajib}, {Treu}, {Birrer}  \&
  {Sonnenfeld}}{{Shajib} et~al.}{2021}]{shajib2021}
{Shajib} A.~J.,  {Treu} T.,  {Birrer} S.,   {Sonnenfeld} A.,  2021, \mn@doi
  [\mnras] {10.1093/mnras/stab536}, \href
  {https://ui.adsabs.harvard.edu/abs/2021MNRAS.503.2380S} {503, 2380}

\bibitem[\protect\citeauthoryear{{Sonnenfeld}, {Treu}, {Gavazzi}, {Suyu},
  {Marshall}, {Auger}  \& {Nipoti}}{{Sonnenfeld} et~al.}{2013}]{sonnenfeld2013}
{Sonnenfeld} A.,  {Treu} T.,  {Gavazzi} R.,  {Suyu} S.~H.,  {Marshall} P.~J.,
  {Auger} M.~W.,   {Nipoti} C.,  2013, \mn@doi [\apj]
  {10.1088/0004-637X/777/2/98}, \href
  {https://ui.adsabs.harvard.edu/abs/2013ApJ...777...98S} {777, 98}

\bibitem[\protect\citeauthoryear{Speagle}{Speagle}{2020}]{dynesty}
Speagle J.~S.,  2020, \mn@doi [Monthly Notices of the Royal Astronomical
  Society] {10.1093/mnras/staa278}, 493, 3132

\bibitem[\protect\citeauthoryear{{Sun} et~al.,}{{Sun} et~al.}{2022}]{sun2022}
{Sun} F.,  et~al., 2022, \mn@doi [\apj] {10.3847/1538-4357/ac6e3f}, \href
  {https://ui.adsabs.harvard.edu/abs/2022ApJ...932...77S} {932, 77}

\bibitem[\protect\citeauthoryear{Suyu}{Suyu}{2012}]{suyu_lens_model}
Suyu S.~H.,  2012, \mn@doi [Monthly Notices of the Royal Astronomical Society]
  {10.1111/j.1365-2966.2012.21661.x}, 426, 868

\bibitem[\protect\citeauthoryear{Suyu, Marshall, Hobson  \& Blandford}{Suyu
  et~al.}{2006}]{suyu2006bayesian}
Suyu S.~H.,  Marshall P.,  Hobson M.,   Blandford R.,  2006, Monthly Notices of
  the Royal Astronomical Society, 371, 983

\bibitem[\protect\citeauthoryear{{Swinbank} et~al.,}{{Swinbank}
  et~al.}{2015}]{swinbank2015}
{Swinbank} A.~M.,  et~al., 2015, \mn@doi [\apjl] {10.1088/2041-8205/806/1/L17},
  \href {https://ui.adsabs.harvard.edu/abs/2015ApJ...806L..17S} {806, L17}

\bibitem[\protect\citeauthoryear{{Tan} et~al.,}{{Tan} et~al.}{2023}]{tan2023}
{Tan} C.~Y.,  et~al., 2023, \mn@doi [arXiv e-prints]
  {10.48550/arXiv.2311.09307}, \href
  {https://ui.adsabs.harvard.edu/abs/2023arXiv231109307T} {p. arXiv:2311.09307}

\bibitem[\protect\citeauthoryear{{Treu}, {Suyu}  \& {Marshall}}{{Treu}
  et~al.}{2022}]{treu2022}
{Treu} T.,  {Suyu} S.~H.,   {Marshall} P.~J.,  2022, \mn@doi [\aapr]
  {10.1007/s00159-022-00145-y}, \href
  {https://ui.adsabs.harvard.edu/abs/2022A&ARv..30....8T} {30, 8}

\bibitem[\protect\citeauthoryear{{Vegetti}, {Despali}, {Lovell}  \&
  {Enzi}}{{Vegetti} et~al.}{2018}]{vegetti2018}
{Vegetti} S.,  {Despali} G.,  {Lovell} M.~R.,   {Enzi} W.,  2018, \mn@doi
  [\mnras] {10.1093/mnras/sty2393}, \href
  {https://ui.adsabs.harvard.edu/abs/2018MNRAS.481.3661V} {481, 3661}

\bibitem[\protect\citeauthoryear{{Wang} et~al.,}{{Wang}
  et~al.}{2019}]{wang2019}
{Wang} Y.,  et~al., 2019, \mn@doi [\mnras] {10.1093/mnras/stz2907}, \href
  {https://ui.adsabs.harvard.edu/abs/2019MNRAS.490.5722W} {490, 5722}

\bibitem[\protect\citeauthoryear{{Warren} \& {Dye}}{{Warren} \&
  {Dye}}{2003}]{warren_dye}
{Warren} S.~J.,  {Dye} S.,  2003, \mn@doi [\apj] {10.1086/375132}, \href
  {https://ui.adsabs.harvard.edu/abs/2003ApJ...590..673W} {590, 673}

\bibitem[\protect\citeauthoryear{{Wong} et~al.,}{{Wong}
  et~al.}{2020}]{wong2020}
{Wong} K.~C.,  et~al., 2020, \mn@doi [\mnras] {10.1093/mnras/stz3094}, \href
  {https://ui.adsabs.harvard.edu/abs/2020MNRAS.498.1420W} {498, 1420}

\bibitem[\protect\citeauthoryear{{Yang}, {Birrer}  \& {Hu}}{{Yang}
  et~al.}{2020}]{yang2020}
{Yang} T.,  {Birrer} S.,   {Hu} B.,  2020, \mn@doi [\mnras]
  {10.1093/mnrasl/slaa107}, \href
  {https://ui.adsabs.harvard.edu/abs/2020MNRAS.497L..56Y} {497, L56}

\makeatother
\end{thebibliography}








\bsp	
\label{lastpage}
\end{document}